\documentclass[11pt,a4paper]{article}

\usepackage{jheppub} 

\usepackage[utf8]{inputenc}
\usepackage{graphicx,dcolumn,subcaption,mathtools,epsfig}

\usepackage{braket}
\usepackage{siunitx}
\usepackage{appendix}

\usepackage{amsmath}
\usepackage{amssymb}
\usepackage{amsbsy}
\usepackage[mathscr]{eucal}
\usepackage{mathrsfs}
\usepackage{bbm}
\usepackage{cancel}

\usepackage{footnotebackref}

\usepackage{rotating}

\usepackage[]{float}
\usepackage[font=small, labelfont=bf]{caption}

\usepackage{slantsc}
\usepackage{xspace}

\usepackage{setspace}
\allowdisplaybreaks

\usepackage{scalefnt,pstricks}

\usepackage{multirow}
\usepackage{tabularray}
\usepackage{tabularx}
\UseTblrLibrary{booktabs}

\usepackage{scalerel}
\usepackage{listings}

\usepackage[tikz]{bclogo}
\usepackage{tikz-feynman}
\usepackage{verbatim}
\usetikzlibrary{arrows,shapes}
\usepackage{afterpage}
\usepackage{enumitem}

\providecommand{\href}[2]{#2}

\newcommand{\ttH}{\ensuremath{t \bar t H}}
\newcommand{\ttHJ}{\ensuremath{t \bar t H J}}

\newcommand{\X}{{\rm X}}
\newcommand{\XJ}{{\rm XJ}}

\newcommand{\QQF}{Q\bar{Q} F}
\newcommand{\QQ}{Q\bar{Q}}

\newcommand{\nf}{n_{\mathrm{f}}}
\newcommand{\nl}{n_{\mathrm{\ell}}}
\newcommand{\nh}{n_{\mathrm{h}}}

\newcommand{\pt}{{p_{\text{\scalefont{0.77}T}}}}

\newcommand{\ptrad}{{p_{\text{\scalefont{0.77}T,rad}}}}
\newcommand{\ptH}{{p_{\text{\scalefont{0.77}T,$H$}}}}
\newcommand{\ptTT}{{p_{\text{\scalefont{0.77}T,$t \bar t$}}}}

\newcommand{\ptTTH}{{p_{\text{\scalefont{0.77}T,$t \bar t H$}}}}

\newcommand{\yH}{{y_{\text{\scalefont{0.77}$H$}}}}

\newcommand{\muF}{{\mu_{\text{\scalefont{0.77}F}}}}
\newcommand{\muR}{{\mu_{\text{\scalefont{0.77}R}}}}

\newcommand{\KF}{{K_{\text{\scalefont{0.77}F}}}}
\newcommand{\KR}{{K_{\text{\scalefont{0.77}R}}}}
\newcommand{\KQ}{{K_{\text{\scalefont{0.77}Q}}}}

\newcommand{\noun}[1]{{\scshape #1}}

\newcommand{\MADSPIN}{\noun{MadSpin}}
\newcommand{\POWHEG}{\noun{Powheg}}

\newcommand{\POWHEGBOX}{\noun{Powheg-Box}}
\newcommand{\POWHEGBOXRES}{\noun{Powheg-Box-Res}}

\newcommand{\minlo}{{\noun{MiNLO$^{\prime}$}}\xspace}
\newcommand{\minnlo}{{\noun{MiNNLO$_{\rm PS}$}}\xspace}

\newcommand{\Matrix}{{\noun{Matrix}}\xspace}

\newcommand{\Recola}{{\noun{Recola}}\xspace}
\newcommand{\PYTHIA}[1]{\noun{Pythia{#1}}\xspace}

\newcommand{\citere}[1]{Ref.\,\cite{#1}}

\newcommand{\citeres}[1]{Refs.\,\cite{#1}}

\newcommand{\eqn}[1]{Eq.\,(\ref{#1})}

\newcommand{\neqn}[1]{Eqs.\,(\ref{#1})}
\newcommand{\fig}[1]{Figure\,\ref{#1}}

\newcommand{\tab}[1]{Table\,\ref{#1}}
\newcommand{\sct}[1]{Section~\ref{#1}}

\newcommand{\M}{\mathcal M}

\newcommand{\muIR}{{\mu_{\text{\scalefont{0.77}IR}}}}
\newcommand{\SA}{\mathrm{SA}}
\newcommand{\MA}{\mathrm{MA}}
\newcommand{\CA}{\mathrm{CA}}
\newcommand{\MSbar}{\mathrm{\overline{MS}}}

\newcommand{\Mfin}{{\M^{\rm fin}_{c\bar c}}}
\newcommand{\HardVirt}{{\mathbb{H}}}

\usepackage{etoolbox}
\makeatletter
\patchcmd{\@sect}{#8}{\boldmath #8}{}{}
\let\ori@chapter\@chapter
\def\@chapter[#1]#2{\ori@chapter[\boldmath#1]{\boldmath#2}}
\makeatother

\title{
Next-to-next-to-leading order event generation for $\boldsymbol{\ttH}$ production with approximate two-loop amplitude
}
\preprint{
  \begin{flushright}
    CERN-TH-2026-031 \\
    MPP-2026-22 \\
    TUM-HEP-1595/26\\
  \end{flushright}
}

\author[a]{Christian Biello,}
\author[b]{Chiara Savoini,}
\author[c]{Chiara Signorile-Signorile}
\author[d]{and Marius Wiesemann}

\affiliation[a]{Institute for Theoretical Physics, ETH Zurich, 8093 Z\"urich, Switzerland}
\affiliation[b]{Technical University of Munich, TUM School of Natural Sciences, Physics Department, James-Franck-Stra{\ss}e 1, 85748 Garching, Germany}
\affiliation[c]{Theoretical Physics Department, CERN, 1211 Geneva 23, Switzerland}
\affiliation[d]{Max-Planck-Institut f\"ur Physik, Boltzmannstra{\ss}e 8, 85748 Garching, Germany}

\emailAdd{cbiello@phys.ethz.ch}
\emailAdd{chiara.savoini@tum.de}
\emailAdd{chiara.signorile-signorile@cern.ch}
\emailAdd{marius.wiesemann@mpp.mpg.de}

\date{Received: date / Accepted: \today}

\abstract{
We study Higgs-boson production in association with a top-quark pair (\ttH{}) at hadron colliders
and present the first matching of next-to-next-to-leading order (NNLO) QCD 
corrections to parton showers using the \minnlo{} method. 
For the two-loop amplitude, we employ two established approximations, 
based on the soft Higgs-boson and high-energy limits, respectively.
For the first time, we also construct the latter in full colour and propose 
a pointwise combination of the two approximations across phase space. 
By assigning a conservative uncertainty estimate, which remains well below the perturbative uncertainties, 
we ensure robust and reliable differential predictions, explicitly validated at the one-loop level.
Apart from the two-loop amplitude, all remaining ingredients of the \minnlo{} calculation are included exactly.
After thorough validation, we present a series of phenomenological 
results illustrating the impact of NNLO corrections and parton-shower effects. 
We consider fiducial predictions for the Higgs-boson decay into photons and include 
off-shell top-quark decays with tree-level spin correlations in both the dilepton and semileptonic channels.

Our \ttH{} \minnlo{} generator is publicly available within the \POWHEG{} framework.
}

\arxivnumber{2603.06143 }

\begin{document}

\maketitle
\flushbottom

%============================================
\section{Introduction}
\label{sec:intro}

The discovery of the Higgs boson a decade ago \cite{ATLAS:2012yve,CMS:2012qbp}
marked a historic milestone in collider physics.
Since then, the characterisation of its properties has become a central goal of
the particle physics community, with Higgs-boson measurements 
being a cornerstone of the rich physics programme at the Large Hadron Collider (LHC). 
To date, experimental data have established a precise picture of many Higgs properties, including its couplings to top ($t$) and bottom ($b$) quarks, to $W$ and $Z$ bosons, and to $\tau$ leptons, all of which are found to be consistent with the Standard-Model (SM) hypothesis \cite{ATLAS:2022vkf,CMS:2022dwd}.
Nevertheless, the Higgs sector remains among the least explored parts of the SM and, therefore, it is particularly sensitive to potential deviations and signals of new physics.

Given that Higgs interactions with SM particles scale with their masses, the Higgs coupling to top quarks plays a particularly important role. 
Owing to the large top-quark mass, it is the strongest Yukawa interaction in the SM and among the best-measured Higgs couplings. 
Any significant deviation from its SM value would constitute a clear signal of new physics. 
While several LHC processes, such as Higgs-boson production in gluon fusion, top-quark pair production, and four-top production, probe the top-quark Yukawa coupling indirectly, the associated production of a Higgs boson with a top-quark pair (\ttH{}) provides the most direct and model-independent determination. 
Moreover, the \ttH{} process constitutes an important background to Higgs-pair searches.

Accounting for just about 1\% of the total Higgs-boson production rate, the \ttH{} channel poses a significant experimental challenge and, as a result, was observed by the ATLAS and CMS collaborations only in 2018~\cite{ATLAS:2018mme,CMS:2018uxb}.
Nowadays, the corresponding signal strength is measured with an uncertainty of about $20\%$, which is expected to improve to the level of approximately $2\%$ by the end of the high-luminosity phase of the LHC~\cite{Cepeda:2019klc}.\footnote{For a comprehensive discussion about the state-of-the-art and future projections see for instance \citeres{Altmann:2025feg,ATLAS:2025eii,deBlas:2025gyz,CMS:2026nce} and references therein.}
Therefore, accurate theoretical simulations for \ttH{} production are indispensable to provide precise predictions for future
LHC measurements. 
This requires the inclusion of next-to-next-to-leading order (NNLO) QCD corrections, together with their consistent matching to parton-shower event generators, thereby enabling reliable fully differential event simulations suitable for direct use in experimental analyses.

The first calculation of the \ttH{} process at leading order (LO) was performed more than forty years ago~\cite{Ng:1983jm,Kunszt:1984ri}. 
The next-to-leading order (NLO) QCD corrections for on-shell top quarks were obtained about twenty years later \cite{Beenakker:2001rj,Beenakker:2002nc,Reina:2001sf,Reina:2001bc,Dawson:2002tg,Dawson:2003zu}. 
Subsequently, NLO electroweak (EW) corrections were computed \cite{Frixione:2014qaa,Zhang:2014gcy,Frixione:2015zaa,Frederix:2018nkq}, and the full off-shell calculations including top-quark decays became available at NLO in QCD \cite{Denner:2015yca,Stremmer:2021bnk} and at NLO in the EW  \cite{Denner:2016wet}. 
Owing to the sensitivity of \ttH{} production to near-threshold dynamics, soft-gluon resummation has also been investigated extensively over the past decade \cite{Kulesza:2015vda,Broggio:2015lya,Broggio:2016lfj,Kulesza:2017ukk,Broggio:2019ewu,Ju:2019lwp,Kulesza:2020nfh}.

More recently, NNLO QCD predictions have become available, first for the inclusive \ttH{} cross section~\cite{Catani:2021cbl,Catani:2022mfv}, where the two-loop amplitude was treated within a soft-Higgs approximation, and afterwards for differential distributions \cite{Devoto:2024nhl}, which additionally relied on a high-energy expansion of the two-loop amplitude in the small top-quark mass limit.
These NNLO results were later combined with soft-gluon resummation and subleading LO and NLO EW effects \cite{Balsach:2025jcw}.
While these approximations yield phenomenologically reliable predictions, the computation of the full two-loop amplitude for \ttH{} production remains an important open challenge.
Significant progress has been reported recently \cite{FebresCordero:2023pww,Agarwal:2024jyq,Wang:2024pmv}, but a complete result is still elusive.
On the other hand, parton-shower simulations for \ttH{} production have seen little development over the past decade, since the first NLO+PS predictions appeared \cite{Frederix:2011zi,Hartanto:2015uka}.

In this work, we present the first NNLO QCD calculation matched to parton showers (NNLO+PS) for \ttH{} production.
We employ the \minnlo{} method~\cite{Monni:2019whf,Monni:2020nks}, in particular its extension to heavy-quark pair production~\cite{Mazzitelli:2020jio} in association with a colour-singlet final state~\cite{Mazzitelli:2024ura}. 
Except for the two-loop amplitude, all ingredients of our NNLO+PS calculation are included exactly.
Following \citere{Devoto:2024nhl}, we employ two approximations for the two-loop amplitude, valid in the soft-Higgs and high-energy limits, respectively.
We then construct a novel pointwise combination of these approximations and assign a conservative systematic uncertainty to it, reflecting the absence of the exact two-loop result.
A detailed comparison with exact one-loop and NLO predictions demonstrates that this approach yields robust and reliable differential results throughout phase space.
After validation against fixed-order predictions, we use our NNLO+PS generator for phenomenological studies, highlighting the benefits of fully exclusive Monte Carlo simulations.
We present fiducial predictions for the Higgs-boson decay into photons and include off-shell top-quark decays with tree-level spin correlations in both the dilepton and semileptonic channels.

The remainder of this paper is organised as follows.
In \sct{sec:MiNNLO}, we outline our \ttH{} NNLO+PS calculation and summarise the \minnlo{} methodology.
\sct{sec:2loop_approx} reviews the approximations for the two-loop amplitude, introduces their pointwise combination, and validates them extensively at first order in the strong-coupling expansion.
In \sct{sec:pheno_results}, we present phenomenological predictions obtained with our \ttH{} NNLO+PS generator, including comparisons to fixed-order results (\sct{sec:MATRIX_validation}), on-shell predictions (\sct{sec:pheno_on-shell_results}), fiducial cross sections for the Higgs-boson decay into photons (\sct{sec:pheno_Higgs-decay_results}), and top-quark decays in the dilepton and semileptonic channels (\sct{sec:pheno_top-decay_results}).
We conclude in \sct{sec:conclusions}.

%============================================
\section{Outline of the calculation}
\label{sec:MiNNLO}
\begin{figure}[b]
  \begin{center}
\begin{subfigure}[b]{.3\linewidth}
  \centering
\begin{tikzpicture}
  \begin{feynman}
	\vertex (a1) at (0,0) {\( g\)};
	\vertex (a2) at (0,-1.7) {\( g\)};
	\vertex (a3) at (1.53,0);
	\vertex (a32) at (1.53,-0.85);
	\vertex (a33) at (3,-0.85){\( H\)};
	\vertex (a4) at (1.53,-1.7);
	\vertex (a5) at (3,0){\( t\)};
	\vertex (a6) at (3,-1.7){\(\bar t\)};
        \diagram* {
          {[edges=fermion]
            (a6)--[fermion, ultra thick](a4)--[fermion, ultra thick](a32)--[fermion, ultra thick](a3)--[fermion, ultra thick](a5),
          },
          (a2)--[gluon,thick](a4),
          (a3)--[gluon,thick](a1),
          (a32)--[scalar,thick](a33),
        };
  \end{feynman}
\end{tikzpicture}\vspace{0.15cm}
\caption{$t$-channel $gg$ diagram}
        \label{subfig:gg}
\end{subfigure}
\begin{subfigure}[b]{.3\linewidth}
  \centering
\begin{tikzpicture}
  \begin{feynman}
	\vertex (a1) at (0,0) {\( g\)};
	\vertex (a2) at (0,-2) {\( g\)};
	\vertex (a3) at (1.4,-1);
	\vertex (a4) at (2.7,-1);
	\vertex (a42) at (3.26,-0.6);
	\vertex (a43) at (4.05,-1.15);
	\vertex (a99) at (4.3,-1.15) {\( H\)};	
	\vertex (a5) at (4.1,0){\( t\)};
	\vertex (a6) at (4.1,-2){\(\bar t\)};
        \diagram* {
          {[edges=fermion]
            (a6)--[fermion, ultra thick](a4)--[fermion, ultra thick](a42)--[fermion, ultra thick](a5),
          },
          (a3) -- [gluon,thick] (a4),
          (a2)--[gluon,thick](a3)--[gluon,thick](a1),
          (a42) -- [scalar,thick] (a43),
        };
  \end{feynman}
\end{tikzpicture}
\caption{$s$-channel $gg$ diagram}
        \label{subfig:gg}
\end{subfigure}%
\quad
\begin{subfigure}[b]{.3\linewidth}
      \centering
\begin{tikzpicture}
\begin{feynman}
	\vertex (a1) at (0,0) {\( q\)};
	\vertex (a2) at (0,-2) {\(\bar q\)};
	\vertex (a3) at (1.4,-1);
	\vertex (a4) at (2.7,-1);
	\vertex (a42) at (3.26,-0.6);
	\vertex (a43) at (4.05,-1.15);
	\vertex (a99) at (4.3,-1.15) {\( H\)};
	\vertex (a5) at (4.1,0){\( t\)};
	\vertex (a6) at (4.1,-2){\(\bar t\)};
        \diagram* {
          {[edges=fermion]
            (a1)--[thick](a3)--[thick](a2),
            (a6)--[fermion, ultra thick](a4)--[fermion, ultra thick](a42)--[fermion, ultra thick](a5),
          },
          (a3) -- [gluon,thick] (a4),
          (a42) -- [scalar,thick] (a43),
        };
      \end{feynman}
\end{tikzpicture}
\caption{$s$-channel $q\bar{q}$ diagram}
        \label{subfig:qq}
\end{subfigure}%
\end{center}
\caption{\label{fig:diagrams} Sample of Feynman diagrams for 
  the process $pp\to t\bar{t}H$ at LO.}
\end{figure}
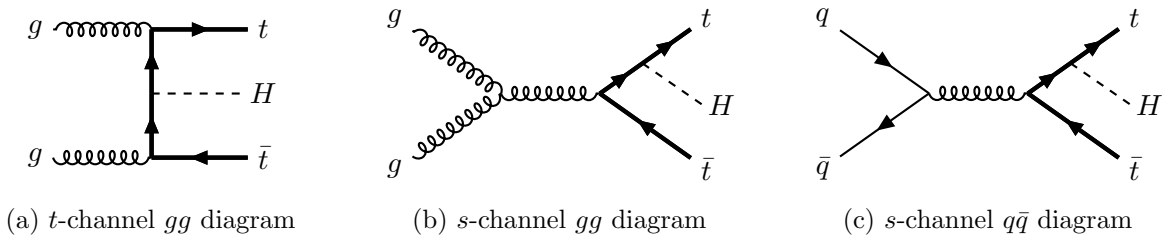

We consider the process 
\begin{equation}
	pp \rightarrow t\bar{t}H + \mathcal{X} 
\end{equation}
 inclusive over additional radiation $\mathcal{X}$. The relevant Born-level partonic scattering is defined as 
\begin{equation}
\label{eq:process}
c(p_1)+\bar{c}(p_2) \rightarrow t(p_3)+\bar{t}(p_4)+H(p_5) \; ,
\end{equation}
where the initial-state partons $c$ and $\bar{c}$ can be either quarks of same flavour 
or two gluons. A sample of the Feynman diagrams
contributing to this process at LO is shown in \fig{fig:diagrams}.
We compute NNLO QCD corrections to this process and consistently match them
to parton showers using the \minnlo{} method \cite{Monni:2019whf, Monni:2020nks}. 
% historical development 
Originally developed for colour-singlet ($F$) production and first applied to single-boson processes \cite{Monni:2019whf, Monni:2020nks}, \minnlo{} was later extended to multi-boson reactions~\cite{Lombardi:2020wju}
and subsequently reformulated to describe heavy-quark pair ($\QQ$) production \cite{Mazzitelli:2020jio}. 
In the present work, we employ the most recent extension of the method, which is applicable to the associated production of a heavy-quark pair with a generic colour-singlet system ($\QQF$), developed in \citere{Mazzitelli:2024ura}.

Within the \minnlo{} framework, NNLO QCD accuracy is achieved for observables inclusive over QCD radiation, while maintaining fully exclusive event generation and resumming logarithmically enhanced contributions through the parton shower.
The method is computationally efficient and flexible, which has led to the design of \minnlo{} generators for a wide range of LHC processes by now \cite{Lombardi:2020wju,Lombardi:2021rvg,Mazzitelli:2020jio,Mazzitelli:2021mmm,Buonocore:2021fnj,Lombardi:2021wug,Zanoli:2021iyp,Gavardi:2022ixt,Haisch:2022nwz,Lindert:2022qdd,Mazzitelli:2023znt,Mazzitelli:2024ura,Biello:2024vdh,Niggetiedt:2024nmp,Biello:2024pgo}, many of which are publicly available within the \POWHEGBOXRES{} framework \cite{Jezo:2015aia}. Most notably, the approach has been successfully applied to a variety of phenomenologically relevant heavy-quark processes, including $t\bar t$~\cite{Mazzitelli:2020jio, Mazzitelli:2021mmm} and $b\bar b$~\cite{Mazzitelli:2023znt} production, as well as $b\bar b Z$~\cite{Mazzitelli:2024ura} and $b\bar b H$~\cite{Biello:2024pgo} final states. 
 
The \minnlo{} method relies on the fact that 
the infrared (IR) singular structure of the cross section can be organised through the analytic resummation of a suitable jet-resolution variable, chosen here as the transverse momentum ($\pt$) of the Born-level system.\footnote{Extensions of the framework to different jet-resolution variables have been explored in \citere{Ebert:2024zdj}.} 
This choice is compatible with matching to traditional transverse-momentum ordered parton showers in order to preserve their leading logarithmic accuracy.
The singular (and constant) terms in $\pt{}$ are combined 
with a \POWHEG{} \cite{Frixione:2007vw} calculation for the production of 
a final-state system \X{} and an extra parton ($\XJ$), which generates the 
second emission, while subsequent emissions are added by the parton shower.
In our case, $\mathrm{X} = t\bar t H$. Exploiting the exponential suppression in $\pt$, this procedure avoids the introduction of any explicit phase-space slicing parameter to regulate the IR region, thereby preserving power corrections.

In practice, \minnlo{} is implemented by modifying the standard \POWHEG{} $\bar{B}$ 
function such that:
\vspace{-0.3cm}
\begin{itemize}
	\item[(i)] NNLO accuracy is achieved for observables inclusive over QCD emissions, 
	\item[(ii)] NLO accuracy is retained for observables sensitive to one resolved jet, 
	\item[(iii)] the (leading) logarithmic structure of traditional parton showers is preserved.
\end{itemize}
\vspace{-0.3cm}
The \minnlo{} master formula can be symbolically written as~\footnote{For simplicity, we 
keep the more involved colour structure of heavy-quark processes implicit here.}
\begin{equation}
\label{eq:minnlo_cross_section}
	{\rm d} \sigma_{\X}^{ \text{\minnlo{}}} = {\rm d}\Phi_{\XJ} \; 
	\bar{B}^{\text{\minnlo{}}} \times \bigg\{
	\Delta_{\rm pwg} (\Lambda_{\rm pwg})
	+ {\rm d}\Phi_{\rm rad} \Delta_{\rm pwg}(\ptrad) \, \frac{R_{\XJ}}{B_{\XJ}}
	\bigg\} \; ,
\end{equation}
where 
\begin{align}
\label{eq:Bbar}
	\bar{B}^{\text\minnlo{}}
	\sim
	e^{-\widetilde S(\pt)} 
	&\bigg\{\! \frac{\alpha_s(\pt)}{2\pi}\, {\rm d}\sigma_{\XJ}^{(1)}
	\bigg(1+\frac{\alpha_s(\pt)}{2\pi}\,
	\widetilde S^{(1)}(\pt)
	\bigg) \notag \\
	&+
	\left(\frac{\alpha_s(\pt)}{2\pi}\right)^2{\rm d}\sigma_{\XJ}^{(2)}
	+ D^{(\geq 3)}(\pt)\times F^{\rm corr} \! \bigg\} \; 
\end{align}
is the modified \POWHEG{} function.
In Eq.~\eqref{eq:minnlo_cross_section}, ${\rm d}\Phi_{\XJ}$ denotes the \XJ{} phase space, while ${\rm d}\Phi_{\rm rad}$ and $\ptrad$ stand for the phase space and transverse momentum of the second radiation. The Born and real contributions to the NLO $\XJ$ cross section are $B_{\XJ}$ and $R_{\XJ}$, respectively,  
while $\Delta_{\rm pwg}$ is the \POWHEG{} Sudakov form factor and \mbox{$\Lambda_{\rm pwg} = 0.89\,$GeV} is the default cutoff. 
In Eq.~\eqref{eq:Bbar}, ${\rm d}\sigma_{\XJ}^{(i)}$ denotes the $i$-th order correction to the $\XJ$ cross section, $\widetilde S$ is the exponent of the Sudakov form factor for $\pt$ resummation of the final-state system \X{}, and $F^{\rm corr}$ is a spreading factor that distributes the NNLO corrections over the full phase space.\footnote{We denote with $A^{(k)}$ the $k$-th coefficient of the expansion $A= \sum_k \big( \alpha_s(\pt)/(2\pi)  \big)^k \,A^{(k)}$ in the strong coupling. 
Note that the $D^{(\geq 3)}$ function includes, \textit{a priori}, all the perturbative coefficients $D^{(k)}$ times $\big( \alpha_s(\pt)/(2\pi)  \big)^k$, starting from the third order, see Eq.~\eqref{eq:D_geq_3}.}
The term $D^{(\geq 3)}$ collects all singular and constant contributions in $\pt{}$ required to achieve NNLO accuracy for observables inclusive over radiation, and it is derived from the $\pt{}$ resummation formula. Without $D^{(\geq 3)}$, inclusive observables would only be NLO accurate, as in the original \minlo{} approach~\cite{Hamilton:2012rf, Hamilton:2012np}. At large $\pt{}$, where Sudakov effects become negligible, both \minlo{} and \minnlo{} predictions smoothly approach the NLO result of the $\XJ$ generator, while the NNLO corrections added through $D^{(\geq 3)}$ contribute at small $\pt{}$.
A key advantage of this construction is that the Sudakov suppression renders the cross section integrable down to vanishing values of the resolution variable, eliminating the need for an explicit slicing cutoff and thereby improving both numerical stability and efficiency in event generation.

For $\QQF$ production, the structure of the large logarithmic corrections at small transverse momentum remains universal \cite{Zhu:2012ts,Li:2013mia,Catani:2014qha}, but is enriched by the presence of coloured particles in the final state. In addition, the availability of the exact two-loop virtual amplitudes for $2 \to 3$ processes involving massive quarks can be a limiting factor. 
These contributions enter \eqn{eq:Bbar} through the $D^{(\geq 3)}$ terms, defined as
 \begin{equation}
 \label{eq:D_geq_3}
D^{(\geq 3)}(\pt) = D(\pt) 
- \frac{\alpha_s(\pt)}{2\pi} \, D^{(1)}(\pt) 
-\bigg(\frac{\alpha_s(\pt)}{2\pi} \bigg)^2 \, D^{(2)} (\pt) \; , 
\end{equation}
where, for brevity, we have omitted the dependence on the Born phase space $\Phi_{\X}$. 
From \eqn{eq:D_geq_3}, it is clear that $D^{(\geq 3)}$ starts formally at order $\alpha_s^3(\pt)$. However, after
integration over $\pt$ at fixed $\Phi_\X$, $D^{(\geq 3)}$ contributes starting at relative order $\alpha_s
^2(M)$, i.e.\ it is relevant for NNLO accuracy, where $M$ denotes the invariant mass of the system \X. 
The all-order function $D$, derived from the $\QQF{}$ resummation formula, depends on both the Sudakov form factor and a luminosity factor, whose derivation is presented in \citere{Mazzitelli:2021mmm}.
In particular, these contributions are constructed to properly account for the richer colour structure relative to the colour-singlet case.
Here, we focus on the luminosity factor, which can be written as~\cite{Mazzitelli:2021mmm}
\begin{equation}
\label{eq:luminosity}
{\mathcal{L}}_{c\bar{c}} (\pt)
\equiv
\frac{\big| \mathcal{M}_{c\bar{c}}^{(0)}\big|^2}{2 M^2}
\sum_{ij} \Big[
{\rm Tr}
\big(
\mathbf{\widetilde{H}}_{c\bar{c}} \; 
\mathbf{{D}}
\big) \; 
\big(
\tilde{C}_{ci}
\otimes f_i 
\big)
\; 
\big(
\tilde{C}_{\bar{c}j}
\otimes
f_j
\big)
\Big]_{\phi} \; ,
\end{equation}
where $\mathcal{M}_{c\bar{c}}^{(0)}$ denotes the LO matrix element for the partonic $c\bar c$ channel in \eqn{eq:process}.
Note that in Eq.~\eqref{eq:luminosity} we have specified the initial-state flavour indices, 
and use bold font to indicate operators that act in colour space. 
The convolutions involve the collinear coefficient functions $\tilde{C}_{ij}$ and the parton-distribution functions $f_i$,
while the operator $\mathbf{D}$ captures azimuthal correlations of the $\QQF$ system in the small-$\pt$ limit\,\cite{Catani:2014qha}. 
The azimuthal average over these correlations is denoted by $\big[\dots\big]_\phi$, with $\big[\mathbf{D} \big]_\phi = \mathbb{I}$. 
After computing the azimuthal average, the hard-virtual operator $\mathbf{\widetilde{H}}_{c\bar{c}}$
leads to the contribution
\begin{equation}
\label{eq:hard_virtual}
\begin{split}
\widetilde{H}_{c\bar{c}} \equiv 
{\rm Tr}
\big(
\mathbf{\widetilde{H}}_{c\bar{c}}
\big) 
={}&
\frac{\bra{\Mfin} \bar{\bf{h}}^\dagger \, \bar{\bf{h}} \ket{\Mfin}}
{
\langle \M_{c\bar{c}}^{(0)}
\ket{\M_{c\bar{c}}^{(0)}}}
=
1 
+
\frac{\alpha_s(\pt)}{2\pi} H_{c\bar{c}}^{(1)}
+
\left(
\frac{\alpha_s(\pt)}{2\pi} \right)^{\!2}  \widetilde{H}_{c\bar{c}}^{(2)}
+ \mathcal{O}(\alpha_s^3) \; .
\end{split}
\end{equation}
The ket $\ket{\dots}$ denotes a vector in colour space and $\ket{\Mfin} = \mathbf{Z}^{-1}| \mathcal{M}_{c\bar c} \rangle$ is the finite remainder of the
UV-renormalised virtual amplitude $\mathcal{M}_{c\bar c}$ obtained after subtracting the IR divergences through the operator $\mathbf{Z}$~\cite{Becher:2009cu,Becher:2009qa,Becher:2009kw,Ferroglia:2009ii}.
The normalisation satisfies
$\langle \M_{c\bar{c}}^{(0)} |
\M_{c\bar{c}}^{(0)} \rangle
=
|\mathcal{M}^{(0)}_{c\bar c}|^{2}$, while the operator $\bar{\mathbf{h}}$ encodes additional finite terms arising from soft-parton contributions.
The latter are known up to NNLO accuracy for $\QQ$ production~\cite{Catani:2023tby}, while for generic $\QQF$ kinematics we rely on the numerical results of~\citere{Devoto:2025eyc}. Recently, these contributions have also been derived analytically for $\QQF$ processes, both in leptonic~\cite{Liu:2024hfa} and hadronic collisions~\cite{Liu:2025ldi}. We stress that the second-order coefficient $\widetilde{H}_{c\bar{c}}^{(2)}$ in \eqn{eq:hard_virtual} incorporates a resummation-scheme dependent shift as explained in \citere{Mazzitelli:2021mmm}.

The purely virtual process-dependent contributions to the hard function in \eqn{eq:hard_virtual} are encoded by the
minimally subtracted finite remainder $\langle \Mfin | \Mfin \rangle$. 
Upon normalisation to the squared LO amplitude, this quantity defines the \textit{hard-virtual coefficient function} $\HardVirt_{c\bar c}$, whose expansion in $\alpha_s(M)/(2\pi)$ reads
\begin{align}
	\HardVirt^{(n)}_{c\bar c}(\muIR) \equiv \left.\frac{2{\mathrm{Re}}\left(\M_{c\bar{c}}^{(n), \rm fin}(\muIR,\muR)\M_{c\bar{c}}^{(0)*}\right)}{|\M_{c\bar{c}}^{(0)}|^2}\right\vert_{\muR=M}\,, ~~~~~~~~~~n=1,2,\dots,
	\label{eq:Hn}
\end{align}
where $\M_{c\bar{c}}^{(n), \rm fin}$ is the finite $n$-loop amplitude contributing to $\Mfin$ and $\muIR$ the scale at which the IR poles are subtracted.\footnote{Note that \eqn{eq:hard_virtual} and \eqn{eq:Hn} differ by the choice of the renormalisation and IR subtraction scales. Moreover, resummation shifts and scheme-dependent terms are implicitly contained in $\bf{\bar{h}}$, and therefore absent in \eqn{eq:Hn}. We refer the reader to \citere{Biello:2024pgo} and the references therein for further details.} From here on, we omit the subscript $c\bar c$ for brevity, unless relevant for the discussion.

For many involved collider reactions, two-loop amplitudes are not yet available.
In some cases, NNLO accuracy can nevertheless be achieved by introducing suitable approximations for the virtual contributions.
This strategy has already enabled first NNLO predictions for processes such as $t\bar t H$~\cite{Catani:2022mfv, Devoto:2024nhl}, $t\bar t W$~\cite{Buonocore:2023ljm},
$b\bar{b}H$~\cite{Biello:2024pgo}, $b\bar{b}W$~\cite{Buonocore:2022pqq} and $b\bar{b}Z$~\cite{Mazzitelli:2024ura}.
In a favourable scenario, these approximations capture the dominant virtual effects in the kinematic regions relevant to LHC phenomenology and can be applied to fully differential NNLO+PS predictions.
When combined with exact real-emission contributions and full-colour dependence, the resulting simulations provide a realistic and reliable description of physical final states.
Since the two-loop finite remainder is not known for $\ttH{}$ production, we employ the approximations introduced in \sct{sec:2loop_approx}.
Up to NNLO, the only component affected by these approximations is the second-order hard-virtual coefficient function defined in \eqn{eq:Hn} for $n=2$, while all other contributions are retained exactly.

We conclude this section by highlighting several technical aspects of the \minnlo{} method that are relevant for the phenomenological studies presented later.
First, in kinematic regions dominated by relatively hard emissions, the \minnlo{} prediction is expected to smoothly approach the corresponding fixed-order result.
This behaviour is achieved by suppressing resummation effects when $\pt \sim M$, through the introduction of a modified logarithm $\log(Q/\pt) \mapsto L$ and a resummation scale $Q = \KQ\, M$ in both the Sudakov radiator and the luminosity factor.
We employ the piecewise modified logarithm $L$ as defined in Eq.~(4.15) of \citere{Mazzitelli:2021mmm}.
This prescription ensures a smooth transition from the resummation to the fixed-order regime, with logarithmically enhanced contributions turned off above $Q$. 
For a detailed discussion of the dependence of the \minnlo{} cross section on the perturbative scales and their respective modifiers, including $\KQ$ for the resummation scale, $\KR$ for the renormalisation scale ($\muR$), and $\KF$ for the factorisation scale ($\muF$), we refer to \citere{Mazzitelli:2021mmm}.
In addition, the \minnlo{} method allows for adapting the renormalisation and factorisation scales at large transverse momentum.
This is implemented through the parameter \texttt{largeptscales}, which can modify the scale choice to capture the hardness of the underlying scattering.
Finally, in the low-$\pt$ region, we employ the profiled renormalisation and factorisation scales introduced in \citere{Monni:2020nks} to avoid the Landau singularity. Specifically, we set \mbox{$Q_0 = 2\,\mathrm{GeV}$} to ensure that the perturbative expansion remains well defined.

%============================================
\section{Approximation of the two-loop amplitude}
\label{sec:2loop_approx}
Considerable effort is currently devoted to the computation of the exact two-loop QCD amplitude for \ttH{} production. 
However, despite impressive progress, a complete result is not yet available due to its substantial complexity. 
To overcome this limitation, we adopt an alternative strategy that has proven effective in recent fixed-order studies~\cite{Catani:2022mfv,Devoto:2024nhl}, both for total cross sections and differential distributions. 
Our approach relies on two well-established approximations that capture the dominant behaviour of the two-loop amplitude in two complementary kinematic regimes, while significantly simplifying its structure.\footnote{Alternative approximations have been proposed, for instance the description of Higgs radiation off top quarks in terms of perturbative fragmentation functions in \citere{Brancaccio:2021gcz}.}
The first one is based on the \textit{soft Higgs-boson limit} \cite{Ellis:1975ap,Shifman:1979eb,Spira:1995rr,Kniehl:1995tn,Catani:2022mfv}, in which the \ttH{} amplitude factorises into a soft eikonal factor associated with the Higgs boson and the lower-multiplicity $t\bar t$ matrix element, with the Higgs boson stripped off.
The second approximation is based on the {\it high-energy regime}, where the \ttH{} system is highly boosted and power-suppressed effects in the top-quark mass ($m_t$) become negligible.\footnote{In principle, power corrections in the Higgs-boson mass ($m_H$) could also be neglected in this limit. However, in contrast to \citere{Wang:2024pmv}, we retain the full dependence on $m_H$.}
In this limit, the two-loop amplitude is approximated through an expansion in the small top-quark mass, systematically neglecting terms suppressed by powers of $m_t$.
This expansion establishes a direct connection between the massive amplitude and its massless counterpart: up to power corrections in $m_t$, the two are related via the so-called \textit{massification procedure}~\cite{Penin:2005eh,Mitov:2006xs,Becher:2007cu,Engel:2018fsb,Wang:2023qbf}.
Throughout this manuscript, the term {\it high-energy regime} denotes the small-$m_t$ expansion.

In contrast to \citere{Devoto:2024nhl}, we evaluate the high-energy limit of the massive two-loop amplitude in full colour and introduce, for the first time, a \textit{pointwise} combination of the two approximations across phase space. We also assign a dedicated estimate of the systematic uncertainty associated with this procedure, which accounts for the absence of the exact two-loop result.
The main features and formulae underlying the two approximations are briefly summarised in \sct{sec:soft-Higgs_massification_approx}, while their pointwise combination and the corresponding uncertainty estimate are described in detail in \sct{sec:pointwise_combination}. Finally, we validate our approach at the one-loop level in \sct{sec:oneloop_validation}.

%...............................................................................
\subsection{Soft-Higgs and high-energy limits}
\label{sec:soft-Higgs_massification_approx}

%% Soft Approx
The {\it soft-Higgs approximation} of the two-loop $\ttH$ amplitude was originally introduced in \citere{Catani:2022mfv} and relies on its factorisation properties in the limit where the Higgs boson becomes soft ($p_5 \to 0$) \cite{Ellis:1975ap,Shifman:1979eb,Spira:1995rr,Kniehl:1995tn}. In this limit, its energy and mass $m_H$ are parametrically much smaller than the other scales involved in the process, such as the invariant mass $M$ of the $t \bar t H$ system and the top-quark mass $m_t$.
At the leading power, i.e.\ neglecting contributions that are less singular than $1/p_5$, the all-order \ttH{} finite remainder factorises into a product of a perturbatively computable soft-emission factor and the underlying $t \bar t$ amplitude. More precisely, it can be written as \cite{Catani:2022mfv}
\begin{equation}
	\label{eq:finrem_SA}
	\M^{\rm fin}(\{p_{i}\}; \muR, \muIR) \simeq F(\alpha_s(\muR), \muR/m_t) \,\frac{m_t}{v}\left(\frac{m_t}{p_3 \cdot p_5} + \frac{m_t}{p_4 \cdot p_5} \right) \M^{{\rm fin}, t \bar t}_{}(\{\tilde{p}_i\}; \muR, \muIR) \,,
\end{equation}
where $v = ( \sqrt{2} G_F)^{-1/2}$ is the Higgs vacuum expectation value, $G_F$ the Fermi constant, and $\muR, \muIR$ the renormalisation and IR subtraction scales, respectively.
%mapping
The set of momenta $\{p_i\}$ corresponds to the kinematics of the $t\bar{t}H$ process, while the set $\{\tilde p_i\}=\widetilde{\mathbb{P}}[\{p_i\}]$ refers to the underlying $t\bar{t}$ configuration, where the Higgs boson has been removed by the action of the projection operator $\widetilde{\mathbb{P}}$. Each set satisfies momentum conservation and on-shell conditions within its respective phase space. In order to consistently define the correspondence between the left- and right-hand sides of Eq.~\eqref{eq:finrem_SA}, it is necessary to introduce a prescription to map $\{p_i\}$ into $\{\tilde p_i\}$.
The projection is not unique: in this work, we adopt the same $q_T$-recoil prescription~\cite{Catani:2015vma} used in fixed-order studies~\cite{Catani:2022mfv,Devoto:2024nhl},\footnote{According to the $q_T$-recoil prescription, we left unchanged the momenta of the top and anti-top quarks in the \ttH{} event, while the initial-state partons equally reabsorb the transverse momentum of the Higgs boson.} ensuring a consistent comparison with existing results.

The perturbative function \mbox{$F(\alpha_s(\muR), \muR/m_t)$} in \eqn{eq:finrem_SA} can be extracted by taking the soft limit of the heavy-quark scalar form factor~\cite{Bernreuther:2005gw,Ablinger:2017hst,Fael:2022miw}, or by exploiting the \textit{low-energy theorems} \cite{Ellis:1975ap,Shifman:1979eb,Spira:1995rr,Kniehl:1995tn}. 
The explicit expression of \mbox{$F$} up to two-loop order is given in Eq.\,(2.3) of \citere{Devoto:2024nhl}.
For a detailed discussion of the soft-Higgs factorisation formula we refer the reader to Section\,(2.1) and Appendix A of \citere{Devoto:2024nhl}. 
Here, we limit ourselves to defining the soft-Higgs approximation (SA) of the hard-virtual coefficient $\HardVirt^{(n)}$, at scale $\muIR$, as
\begin{equation}
\label{eq:Hn_SA}
	\HardVirt^{(n)}_{\SA}(\muIR)=\left.\frac{2{\mathrm{Re}}\left(\M^{(n), \mathrm{fin}}_{\SA}(\muIR,\muR)\M^{(0)*}_{\SA}\right)}{|\M^{(0)}_{\SA}|^2}\right\vert_{\muR={\widetilde M}}\,, ~~~~~~~~~~n=1,2,\dots,
\end{equation}
where ${\widetilde M}=\widetilde{\mathbb{P}}[M]$ is the virtuality of the $t{\bar t}$ pair. Unless stated otherwise, $\muIR$ is set to $\widetilde M$.
The $n$-loop coefficient of the finite remainder \mbox{$\M^{(n), \mathrm{fin}}_{\SA}$} and the Born amplitude \mbox{$\M^{(0)}_{\SA}$} are obtained by applying the factorisation formula in \eqn{eq:finrem_SA}.

%% Massification
Having discussed the soft-Higgs limit, we now turn to the second approximation, which is formally valid in the high-energy regime ($M \gg m_t$) where power-suppressed corrections in the top-quark mass are numerically small. 
One can therefore systematically expand the massive amplitude around the small-mass limit, retaining only the logarithmically enhanced and constant terms in the top-quark mass. This correspondence between massive and massless amplitudes holds up to power corrections in $m_t$ and is established via the so-called {\it massification} procedure.
This method was first applied to the computation of NNLO QED corrections to large-angle Bhabha scattering~\cite{Penin:2005eh,Becher:2007cu} and was later extended to generic $2 \to n$ scattering processes in QCD~\cite{Mitov:2006xs}. More recently, the technique has been generalised to massive fermionic loops~\cite{Wang:2023qbf}. 
The massification procedure has proven particularly effective for approximating massive amplitudes in processes involving relatively light massive quarks, as demonstrated by recent NNLO results for $b \bar b W$~\cite{Buonocore:2022pqq}, $b \bar b Z$~\cite{Mazzitelli:2024ura}, and $b \bar b H$~\cite{Biello:2024pgo} production.

Exploiting the general factorisation formula \cite{Mitov:2006xs}
and including the contributions from heavy-quark loops~\cite{Wang:2023qbf}, we apply the massification procedure directly at the level of the finite remainder, following the formulation introduced in \citere{Mazzitelli:2024ura}.
Adapting the notation of \citere{Biello:2024pgo}, we write
\begin{align}
\label{eq:finrem_MA}
	|\M^{\text{fin}}(\{p_{i}\}; \mu)\rangle = \bar{\mathcal{F}}_{c\bar c}\left(\frac{m_t^2}{\mu^2},\alpha_s^{(\nf)}(\mu) \right) \bar{\boldsymbol{S}}\left(\frac{m_t^2}{\mu^2}, \{ \bar{p}_i\}, \alpha_s^{(\nf)}(\mu) \right) |\M_0^{\text{fin}}(\{\bar p_{i}\}; \mu)\rangle  + \mathcal{O}\left(\frac{m_t}{M}\right) \,,
\end{align}
where $\mu = \muR = \muIR$.\footnote{To evaluate the finite remainder at a renormalisation scale $\muR \ne \muIR$ it is sufficient to add the logarithmically enhanced terms from the renormalisation-group evolution of the strong coupling.} 
While $\M^{\text{fin}}$ denotes the finite remainder of the $t\bar t H$ amplitude with massive top quarks, $\M_0^{\text{fin}}$ refers to the corresponding $q\bar q H$ finite remainder with massless quarks, evaluated with $\nf=\nl+1$ massless flavours ($\nl = 5$ in our case).
The function $\bar{\mathcal{F}}_{c\bar c}$, which depends on the partonic channel $c \bar c$, and the operator $\bar{\boldsymbol{S}}$ acting in colour space are free of $\epsilon$ poles and they admit a perturbative expansion in $\alpha_s^{(\nf)}$. Their explicit expressions are given in \citeres{Mazzitelli:2024ura,Biello:2024pgo}.
Finally, we note that, despite the different organisation of the perturbative terms, \eqn{eq:finrem_MA} is fully equivalent to Eq.\,(2.39) of the journal version of \citere{Devoto:2024nhl}. 

Analogously to the soft-Higgs approximation, the matrix elements appearing on the two sides of \eqn{eq:finrem_MA} are evaluated using two different sets of momenta. In particular, each phase-space point $\{p_i\}$ generated with massive top quarks must be mapped, via the projection operator $\overline{\mathbb{P}}$, onto a corresponding set of momenta 
$\{\bar p_i\}=\overline{\mathbb{P}}[\{p_i\}]$ in which the top quarks are treated as massless.
Although the choice of the projection is not unique, a crucial requirement is infrared safety: the mapping must be constructed such that IR-divergent regions of the massless amplitude, which are absent in the massive one, are properly screened. 
This is particularly relevant in the $gg$ channel, where an initial-state collinear singularity (no longer protected by the top-quark mass) can arise in the massless limit when the (anti)top-quark transverse momentum vanishes.
Owing to the different leading-order momentum flows in the $q \bar q$ and $gg$ partonic channels, we employ a dedicated mapping for each channel. 
More specifically, we adopt the momentum mappings described in Appendix B of \citere{Devoto:2024nhl} to facilitate direct comparisons and cross-checks against fixed-order results.

By expanding both sides of \eqn{eq:finrem_MA} in $\alpha_s^{(\nf)}/(2\pi)$, one obtains the one- and two-loop relations between the massive and massless finite remainders. 
As a final step, in order to be consistent with a massive computation performed in an $\nl$-flavour number scheme, we decouple the top quark ($\nh =1$) with pole mass $m_t$ from the running of $\alpha_s$, by applying the finite renormalisation shift
\begingroup
\allowdisplaybreaks
\begin{align}
	\alpha_s^{(\nf)}(\mu^2) &= \alpha_s^{(\nl)}(\mu^2) \biggl\{ 1 + \frac{\alpha_s^{(\nl)}(\mu^2) }{2\pi}\frac{\nh}{3} \ln\frac{\mu^2}{m_t^2} \notag \\
	&+ \biggl( \frac{\alpha_s^{(\nl)}(\mu^2) }{2 \pi} \biggr)^{\!2} \nh \left[ \frac{15}{8}C_F -\frac{4}{9}C_A +\biggl( \frac{C_F}{2} + \frac{5}{6} C_A  \biggr) \ln\frac{\mu^2}{m_t^2}  
	+ \frac{\nh}{9} \ln^2\frac{\mu^2}{m_t^2}   \right]
	+ \mathcal{O}(\alpha_s^3) \biggr\} \, ,
	\label{eq:twoloop_decoupling}
\end{align}
\endgroup
where we have set $T_{\rm R} = 1/2$.

Another source of mass logarithms is the change of the renormalisation scheme for the top-Yukawa coupling. 
Indeed, the two-loop massless finite remainder, originally computed in the leading-colour approximation in \citere{Badger:2021ega} and recently in full colour in \citere{Badger:2024mir}, is expressed in terms of the heavy-quark Yukawa coupling renormalised in the $\mathrm{\overline{MS}}$ scheme. By contrast, in this work the top-quark Yukawa coupling is renormalised in the on-shell (OS) scheme. 
The conversion between the two schemes is given by~\cite{Chetyrkin:1999qi}
%\begingroup
%\allowdisplaybreaks
\begin{align}
	y_t^{{\MSbar}, (\nf)}(\mu) &= \frac{Z_m^{\rm OS}(\mu^2/m_t^2)}{Z_m^{\MSbar}(\mu^2/m_t^2)} \,y_t \equiv  z_m(\mu^2/m_t^2) \,y_t \notag \\
	&= 
	\biggl[ 1+  \frac{\alpha_s^{(\nf)}(\mu^2)}{2 \pi} z_m^{(1)}(\mu^2/m_t^2) 
	+ \bigg(\frac{\alpha_s^{(\nf)}(\mu^2)}{2\pi}\bigg)^{\!2} z_m^{(2)}(\mu^2/m_t^2)  
	+ \mathcal{O}(\alpha_s^3) \biggr] \,y_t \,,
	\label{OS-MS_Yuk}
\end{align}
%\endgroup
with
\begingroup
\allowdisplaybreaks
\begin{align}
	z_m^{(1)}(\mu^2/m_t^2) &= 
	- 2 C_F \left( 1 + \frac{3}{4}\ln\frac{\mu^2}{m_t^2} \right)  \,,\\
	z_m^{(2)}(\mu^2/m_t^2) &= 
	C_F^2 \left( \frac{7}{32}
	-\frac{15}{2}\zeta_2 
	-3\, \zeta_3 
	+12 \zeta_2\log2 
	+\frac{21}{8}\ln\frac{\mu^2}{m_t^2} 
	+\frac{9}{8}\ln^2\frac{\mu^2}{m_t^2} \right) \notag \\
	&
	+ C_A C_F 
	\left( -\frac{1111}{96} 
	+ 2\,\zeta_2 
	+\frac{3}{2}\zeta_3 
	-6\zeta_2\log2 
	-\frac{185}{24}\ln\frac{\mu^2}{m_t^2} 
	-\frac{11}{8}\ln^2\frac{\mu^2}{m_t^2} \right) \notag \\
	&
	+ C_F \nl \left( \frac{71}{48} 
	+ \zeta_2 
	+\frac{13}{12}\ln\frac{\mu^2}{m_t^2} 
	+ \frac{1}{4}\ln^2\frac{\mu^2}{m_t^2} \right) 
	\notag \\ 	&
	+ C_F \nh 
	\left( \frac{143}{48} 
	-2\, \zeta_2 
	+\frac{13}{12}\ln\frac{\mu^2}{m_t^2} 
	+ \frac{1}{4}\ln^2\frac{\mu^2}{m_t^2} \right)  \, .
\end{align}
\endgroup
This shift must be applied to the massless finite remainder before using \eqn{eq:finrem_MA}, in order to ensure consistency with the OS renormalisation scheme adopted for the top-quark Yukawa coupling.

Analogously to \eqn{eq:Hn_SA}, we define the hard-virtual coefficient in the {\it massified approximation} (MA) as
\begin{equation}
\label{eq:Hn_MA}
	\HardVirt^{(n)}_{\MA}(\muIR)=\left.\frac{2{\mathrm{Re}}\left(\M^{(n), \mathrm{fin}}_{\MA}(\muIR,\muR)\M^{(0)*}_{\MA}\right)}{|\M^{(0)}_{\MA}|^2}\right\vert_{\muR={\overline M}}\,, ~~~~~~~~~~n=1,2,\dots,
\end{equation}
where $\overline{M}=\overline{\mathbb{P}}[M]$ is the invariant mass of the projected massless event.
The $n$-loop coefficient of the finite remainder \mbox{$\M^{(n), \mathrm{fin}}_{\MA}$} and the Born amplitude \mbox{$\M^{(0)}_{\MA}$} are obtained by applying the massification procedure defined in \eqn{eq:finrem_MA}.
For the numerical evaluation of the massless two-loop finite remainder $\M^{(2), \mathrm{fin}}_{0}$, we rely on two complementary \texttt{C++} implementations: a public library providing the recent full-colour results of \citere{Badger:2024mir}, and a library developed by the authors of \citere{Biello:2024pgo} for the leading-colour results \cite{Badger:2021ega}.\footnote{We remind the reader that extensive cross-checks have been performed against an independent implementation of the same leading-colour massless amplitude by the authors of \citere{Devoto:2024nhl}.} 
This setup ensures numerical stability and computational efficiency in the evaluation of \eqn{eq:Hn_MA}.

We recall that the massless two-loop amplitude entering the massification procedure is included in full colour for the first time.
We have checked that the massified finite remainder in full colour satisfies the expected renormalisation-group equation (RGE), given by 
\begingroup
\allowdisplaybreaks
\begin{equation}
	|\M^{\mathrm{fin}}(\mu_2)\rangle = \exp\left( \int_{\mu_1}^{\mu_2}\frac{d\mu}{\mu} \mathbf{\Gamma}(\mu) \right)\, |\M^{\mathrm{fin}}(\mu_1)\rangle \,,
	\label{eq:finrem-mu1mu2_relation}
\end{equation}
\endgroup
where $\mu_1$ and $\mu_2$ denote the initial and final scales at which the finite remainder is evaluated. 
The anomalous-dimension matrix $\mathbf{\Gamma}$, which governs the IR pole structure of the UV-renormalised amplitude, has been computed in \citere{Ferroglia:2009ii} for processes involving heavy quarks and must be appropriately expanded in the small top-quark mass limit.
We exploit this property to improve the numerical performance of our implementation, avoiding multiple evaluations of the computationally expensive $\M^{(2),\mathrm{fin}}_{0}$ contribution at different scales $\mu$, as would otherwise be required for the systematic uncertainty estimate of our combination procedure discussed in \sct{sec:pointwise_combination}.

Finally, it is worth noting that each approximation is applied consistently to both the numerator and denominator of the two-loop hard-virtual coefficient $\HardVirt^{(2)}$, defined in \eqn{eq:Hn}.
The corresponding two-loop finite remainder is then reconstructed by multiplying the approximated $\HardVirt^{(2)}$ coefficient by the exact squared Born matrix element, thereby improving the overall reliability of the approximation.

%...............................................................................
\subsection{Pointwise combination}
\label{sec:pointwise_combination}

Having discussed the soft-Higgs approximation and massification separately, we now turn to their combination, which constitutes the main novelty of the two-loop approximation adopted in this work. 
Unlike previous fixed-order studies~\cite{Buonocore:2023ljm,Devoto:2024nhl}, where the two approximations were applied independently and combined {\it a posteriori}, i.e.\ after phase-space integration, either through an arithmetic or a weighted average, here we combine them directly at each phase-space point. 
One key advantage of this differential approach is that the two approximations can be weighted by a kinematic function that reflects their respective regions of validity.
In addition, we assign a conservative uncertainty estimate to account for the missing exact two-loop contribution.
Such a pointwise combination enables the construction of a reliable approximation of the two-loop hard-virtual contribution for each unweighted event, generated within the \minnlo{} framework.

While an \textit{a posteriori} combination is possible, it is less optimal in the context of an event generator. 
From a practical perspective, generating two independent event samples, i.e.\ one for each two-loop approximation, and merging them only afterwards would affect the flexibility and computational efficiency of a framework designed to produce unweighted events.
Moreover, since event generators provide exclusive events spanning the full phase space rather than a set of pre-selected distributions, a pointwise combination with a reliable uncertainty estimate is important to cover all kinematic regions. 
Last but not least, all Born-like contributions, including the two-loop amplitude, are not strictly confined to Born kinematics, as in fixed-order calculations.
Indeed, through the \minnlo{} matching procedure, subsequent \POWHEG{} radiation, and parton-shower evolution, Born-like configurations are smeared across events through soft and collinear emissions, yielding a more physical description of the cross section. 
As a consequence, the two-loop contribution is effectively redistributed over a broader region of phase space, and therefore
a kinematically dependent weight function cannot be defined \textit{a posteriori}, given that the underlying Born kinematics is not unambiguously reconstructable.

This can lead to situations in which a poorly approximated contribution in a certain kinematic region---for instance, outside the validity domains of the soft-Higgs or the high-energy limit---is redistributed into regions where the approximation would otherwise perform well, potentially entering with a sizeable weight. 
In such circumstances, the accuracy of an \textit{a posteriori} combination becomes difficult to assess.
Although these effects are partly mitigated by the fact that the additional radiation is typically soft or collinear---thereby constraining the mappings of the spreading function $F^{\rm corr}$ in \eqn{eq:Bbar}---they nevertheless provide a strong motivation for performing the combination directly at the level of individual phase-space points during event generation.
For completeness, we note that even in fixed-order calculations such a pointwise combination would be advantageous, as it avoids relying on integrated distributions with finite bin sizes and yields a more robust, practical, and physically sound result.

The approach proposed here is motivated by the observation that the soft-Higgs approximation and massification are expected to provide reliable descriptions of the two-loop amplitude in different, largely complementary kinematic regions.
Rather than selecting a single approximation globally, we interpolate between them using a kinematics-dependent weight function $\omega$, which smoothly modulates their relative contributions across phase space. 
More precisely, we define the hard-virtual coefficient at scale $\muIR$ in the {\it combined approximation} (CA) as
\begin{equation}
	\label{eq:Hn_CA}
	\HardVirt^{(n)}_{\CA}(\muIR) = \omega\, \HardVirt^{(n)}_{\SA}\big(\widetilde{\mathbb{P}}[\muIR]\big) + (1-\omega)\,\HardVirt^{(n)}_{\MA}\big( \overline{\mathbb{P}}[\muIR] \big) \,, ~~~~~~~~~~n=1,2,\dots,
\end{equation}
where $\HardVirt^{(n)}_{\SA}$ and $\HardVirt^{(n)}_{\MA}$ are defined in \neqn{eq:Hn_SA} and~\eqref{eq:Hn_MA}, respectively. We stress that the operators $\widetilde{\mathbb{P}}$ and $\overline{\mathbb{P}}$ project the scale $\muIR$ according to the mapping chosen for the two approximations. Setting by default $\muIR$ to the invariant mass of the event, we get
\begin{equation}
	\label{eq:Hn_CA_default}
	\HardVirt^{(n)}_{\CA}(M) \equiv \omega\, \HardVirt^{(n)}_{\SA}(\widetilde M) + (1-\omega)\,\HardVirt^{(n)}_{\MA}(\overline M) \,, ~~~~~~~~~~n=1,2,\dots,
\end{equation}
where $\widetilde M= \widetilde{\mathbb{P}}[M]$ and $\overline M= \overline{\mathbb{P}}[M] $ are the projections of the invariant mass $M$ of the original \ttH{} event in the soft-Higgs and in the high-energy limits, respectively.

%..................
\begin{figure}[t!]
\begin{center}
\includegraphics[width=0.52\textwidth]{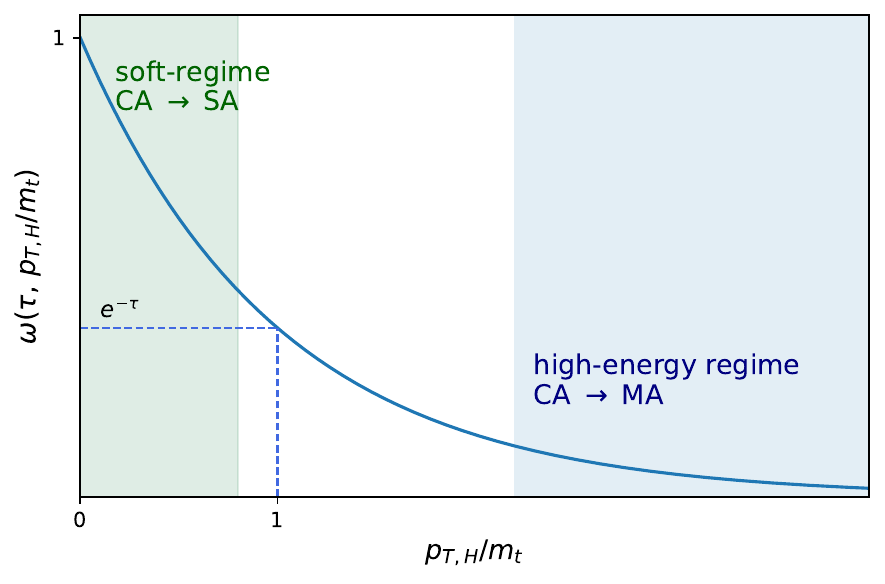}
\vspace*{1ex}
\caption{Sketch of the behaviour of the weight function $\omega$, which ensures a smooth transition between the two kinematic regimes. SA and MA refer to the soft and the massification approximation, respectively. Their pointwise combination is indicated with CA.}
\label{fig:omega_plot}
\end{center}
\end{figure}
%..................

To parametrise the transition between the two kinematic regimes, we choose $\omega$ to be a function of the dimensionless ratio $\ptH/m_t$, where $\ptH$ denotes the transverse momentum of the Higgs boson. 
This choice is physically well motivated: in the soft-Higgs limit, the Higgs transverse momentum is small compared to the top-quark mass, $\ptH \ll m_t$, whereas in the boosted regime, relevant for the massification procedure, the characteristic hard scale of the process is set by momenta much larger than $m_t$, thus corresponding to $m_t \ll \ptH$. 
As a consequence, the ratio $\ptH/m_t$ provides a smooth and monotonic measure of how close a given phase-space configuration is to the kinematic domain where each approximation is expected to be valid.
Accordingly, we require that $\omega \to 1$ for $\ptH/m_t \to 0$ and $\omega \to 0$ for $\ptH/m_t \to \infty$.

Several functional forms for the weight function $\omega$ are, in principle, compatible with the required behaviour in the relevant kinematic limits. 
In this work, we adopt an exponential suppression factor to interpolate between the two regimes. This choice is motivated by the relatively narrow domain of validity of the soft-Higgs approximation, whose contribution must therefore be rapidly damped as one moves away from the very low-$\ptH$ region. At the same time, the exponential form ensures a smooth yet non-trivial transition between the two approximations, while preserving the accuracy of the massification approach in the high-$\ptH$ regime.
Explicitly, we define
\begin{equation}\label{eq:omega}
	\omega\left(\tau, \ptH/m_t\right) = \exp\left( - \tau \,\ptH/m_t \right) \,,
\end{equation}
where the parameter $\tau$ controls the sharpness of the transition between the two approximations. 
A pictorial representation of the weight function $\omega$
is given in \fig{fig:omega_plot}.
We performed a comprehensive differential study to validate the different options at one-loop level, finding that this choice yields the best agreement with the exact results.
Our findings and the corresponding validation are presented in the next subsection.

Beyond the definition of the combined approximation in \eqn{eq:Hn_CA}, a further crucial aspect of our procedure is the assessment of the systematic uncertainty associated with the approximation of the two-loop finite remainder.
A reliable uncertainty estimate is at least as important as the combination itself, since it provides the only means to assess when the approach deteriorates from the exact result in specific corners of phase space.
Taking inspiration from the {\it a posteriori} strategy adopted in \citere{Devoto:2024nhl}, we identify the following sources of systematic uncertainty:
\begin{enumerate}
	\item The variation of the scale $\muIR$ at which the combined approximation in \eqn{eq:Hn_CA} is applied. For the central values, we choose $\muIR = M$, as given in \eqn{eq:Hn_CA_default}. More generally, one can consider a scale choice of the form $\muIR = \eta M$, with $\eta$ being a dimensionless parameter of $\mathcal{O}(1)$. The corresponding finite shift from the initial scale $\eta M$ to the final scale $M$ is then determined by the RGE satisfied by the exact finite remainder in \eqn{eq:finrem-mu1mu2_relation} and can be expressed entirely in terms of known lower-loop matrix elements. 
We stress that, if the exact result were available, the dependence on $\eta$ would be absent, since the finite RGE shift would exactly compensate for the different choice of the scale $\muIR$ in the definition of $\HardVirt^{(2)}$.
\item The variation of the parameter $\tau$ entering the definition of the weight function $\omega$, whose default value is set to $\tau = 1$.
	\item The discrepancy $\delta = \bigl | \HardVirt^{(1)}_{\CA}(M)/\HardVirt^{(1)}(M) - 1 \bigr |$ 
	between the approximated and the exact hard-virtual coefficients at one-loop order. 
	This additional source of uncertainty incorporates, in a fully exclusive manner, information on the pointwise quality of the combined approximation at one-loop level, thereby ensuring a more conservative estimate of the two-loop uncertainty. 
	 In particular, if the combined approximation becomes unreliable in a given phase-space region or for a specific observable, this is reflected in an increase of $\delta$, and consequently in an enhanced two-loop uncertainty.
\end{enumerate}
The prescription we propose consists of varying the parameters $\eta$ and $\tau$ independently by a factor of two around their default values, $\eta=\tau=1$. This defines a nine-point variation band. In addition, we include two further variations obtained by replacing $\HardVirt^{(2)}_{\CA}(M)$ with $(1\pm\delta)\,\HardVirt^{(2)}_{\CA}(M)$ at each phase-space point. The envelope of these eleven predictions is then taken as our uncertainty band.

%...............................................................................
\subsection{Validation at one-loop level}
\label{sec:oneloop_validation}

In this section, we assess the quality of the soft-Higgs, the massified and the combined approximations, focusing on the virtual contribution at NLO.
We present predictions for $t\bar t H$ production in proton--proton collisions at a centre-of-mass energy of $\sqrt{s} = 13\,$TeV.
The Higgs boson and the top quark are treated as stable particles, with their pole masses set to $m_H = 125.0\,$GeV and $m_t = 173.2\,$GeV.
The top-quark Yukawa coupling is renormalised in the on-shell scheme, 
and the vacuum expectation value is determined via
$G_F=1.16637\cdot 10^{-5}\,$GeV$^{-2}$.
We use the NNPDF31\_nnlo\_as\_0118 set~\cite{Ball:2017nwa} 
with five active flavours through the LHAPDF interface~\cite{Buckley:2014ana}.
The central factorisation and renormalisation scales are set to $\muR = \muF = M$, where $M$ is the invariant mass of the $t \bar t H$ final state.
Scale uncertainties are estimated by taking the envelope of the conventional seven-point variation, which involves varying $\muR$ and $\muF$ independently by a factor of 2 
around their central value with the constraint $1/2 < \muR/\muF < 2$.

\begin{table}[t]
\begin{center}
\scalebox{0.95}{
\renewcommand{\arraystretch}{1.7}
    \begin{tabular}{ |c|c|c|c| }
   \cline{2-4}
   \multicolumn{1}{c|}{} 
    &
    LO+PS
    &
    NLO+PS (exact)
    &
    NLO+PS (CA)
\\
    \hline
    \multirow{1}{*}{$\sigma^{\ttH}_{\rm inclusive}(\sqrt{s}=$13\,TeV)} 
    &
    $0.2572(1)_{-19.6\%}^{+26.5\%}\,\text{pb}$
    &
    $0.4096(1)_{-11.6\%}^{+12.0\%}\,\text{pb}$
    &
    ${0.4035(1)_{-11.3\%[-1.9\%]}^{+11.5\%[+2.6\%]}}\,\text{pb}$
        \\
    \hline
    \end{tabular}%
}%
\vspace{0.5cm}
    \caption{\label{tab:xs_NLO} Total inclusive cross sections with corresponding scale uncertainties. The errors in round brackets represent the statistical uncertainty, while those in squared parentheses stand for the systematic one-loop uncertainty, estimated from the $\eta$ and $\tau$ dependence of the approximated hard-virtual contribution.}
\end{center}
\end{table}

In \tab{tab:xs_NLO}, we report total inclusive cross sections at
LO+PS and NLO+PS. 
The NLO+PS prediction in the third column is based on the {\it exact} calculation of the one-loop hard-virtual coefficient $\HardVirt^{(1)}$,
while the one in the fourth column has been obtained by substituting $\HardVirt^{(1)}$ with its combined approximation $\HardVirt^{(1)}_{\CA}$ defined in \eqn{eq:Hn_CA_default}.
The number in round brackets denotes the numerical error, scale
uncertainties are given in percent, while the percentages in square brackets 
represent the systematic uncertainty of the one-loop approximation obtained via a nine-point variation of the parameters $(\eta,\tau)$ (see \sct{sec:pointwise_combination}).
We observe that the central values differ by about $1.5\%$, with the approximated result being lower, but this difference is well covered by the systematic error assigned to the approximation.
Hence, the two predictions are in full agreement within the assigned systematic uncertainty, which is nearly an order of magnitude smaller than the perturbative one. 
This provides a non-trivial validation of the CA approach at the level of the total inclusive cross section and indicates that the approximation does not introduce sizeable uncontrolled effects that could become harmful at higher perturbative orders.

Comparing LO+PS and NLO+PS predictions in \tab{tab:xs_NLO}, we observe that the NLO correction increases the LO result by approximately a factor of 1.6. 
This sizeable correction is driven by the choice of the dynamical scale $\muR=\muF=M$ adopted in the present work. 
Indeed, if a fixed scale $\muR=\muF=(2m_t+m_H)/2$ is chosen as in \citere{Devoto:2024nhl}, the LO cross section is $\sim 390$ fb, with NLO corrections of roughly $+20\%$.
This comparison highlights the strong dependence of the LO cross section on the choice of the central scale. As higher perturbative orders are included, the residual sensitivity to the scale choice is expected to decrease. This is indeed observed for the NLO cross section and should further decrease at NNLO.

\begin{figure}[t!]
\begin{center}
\begin{tabular}{ccccc}
\hspace{-0.52cm}
\includegraphics[width=.35\textwidth]{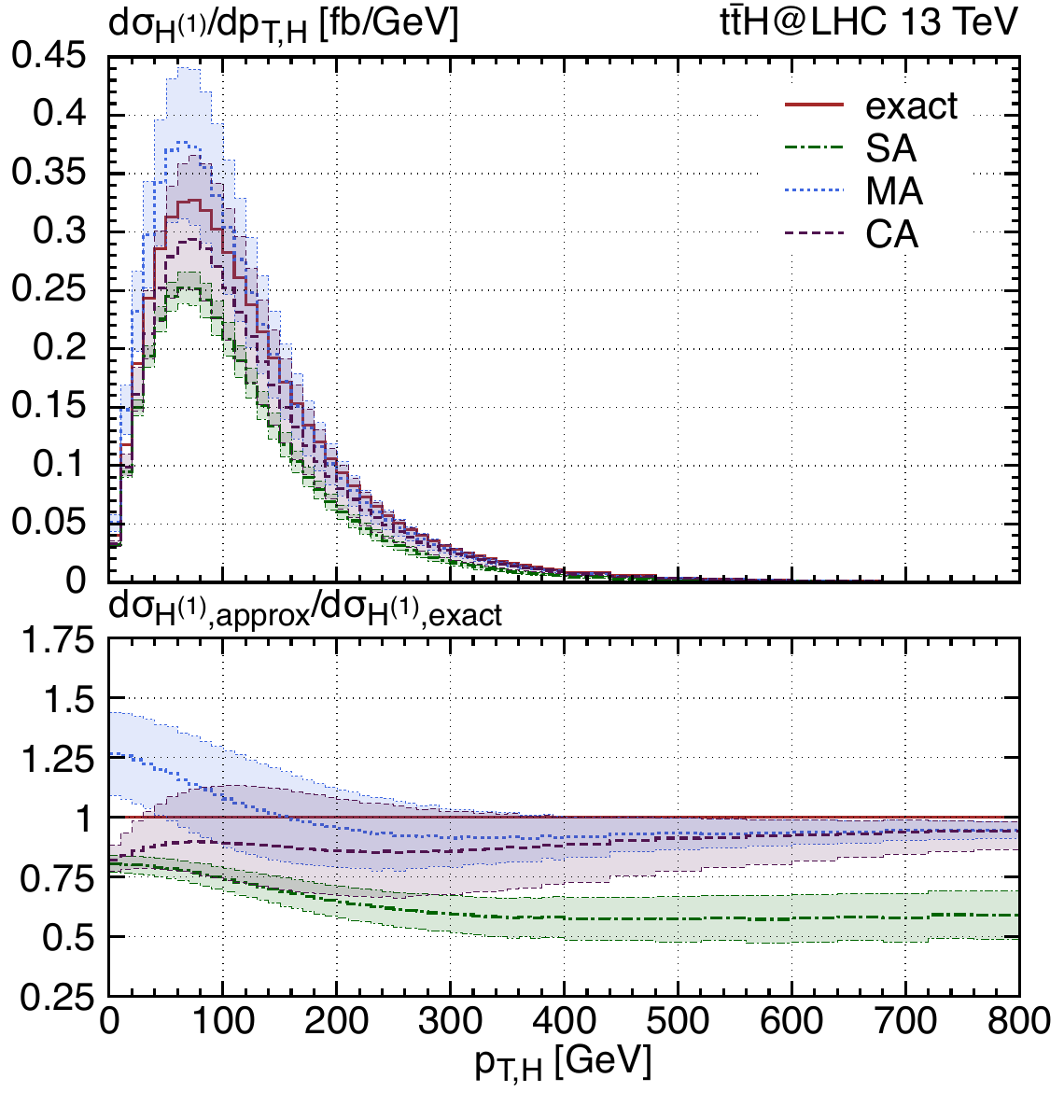}
&
\hspace{-0.73cm}
\includegraphics[width=.35\textwidth]{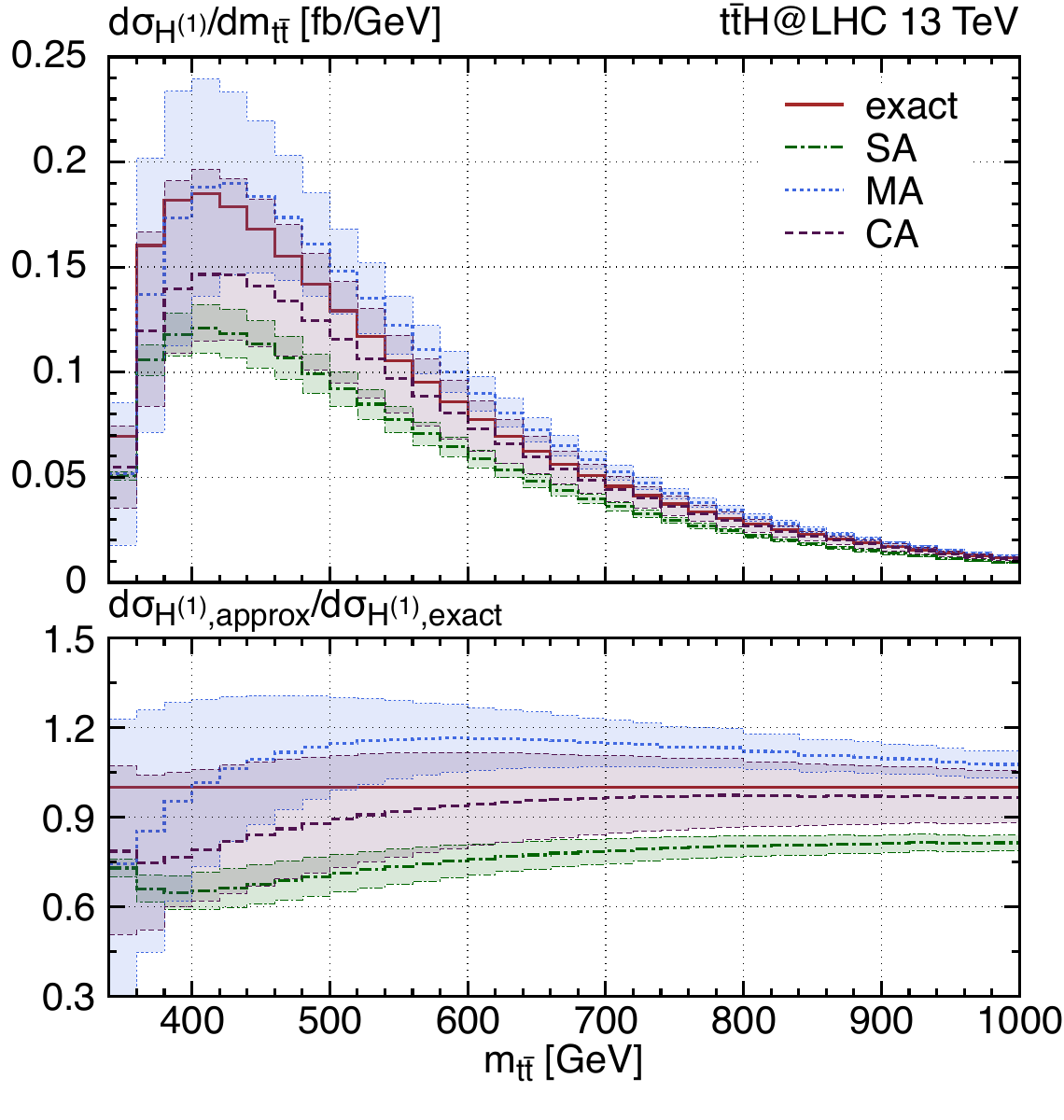}
&
\hspace{-0.73cm}
\includegraphics[width=.35\textwidth]{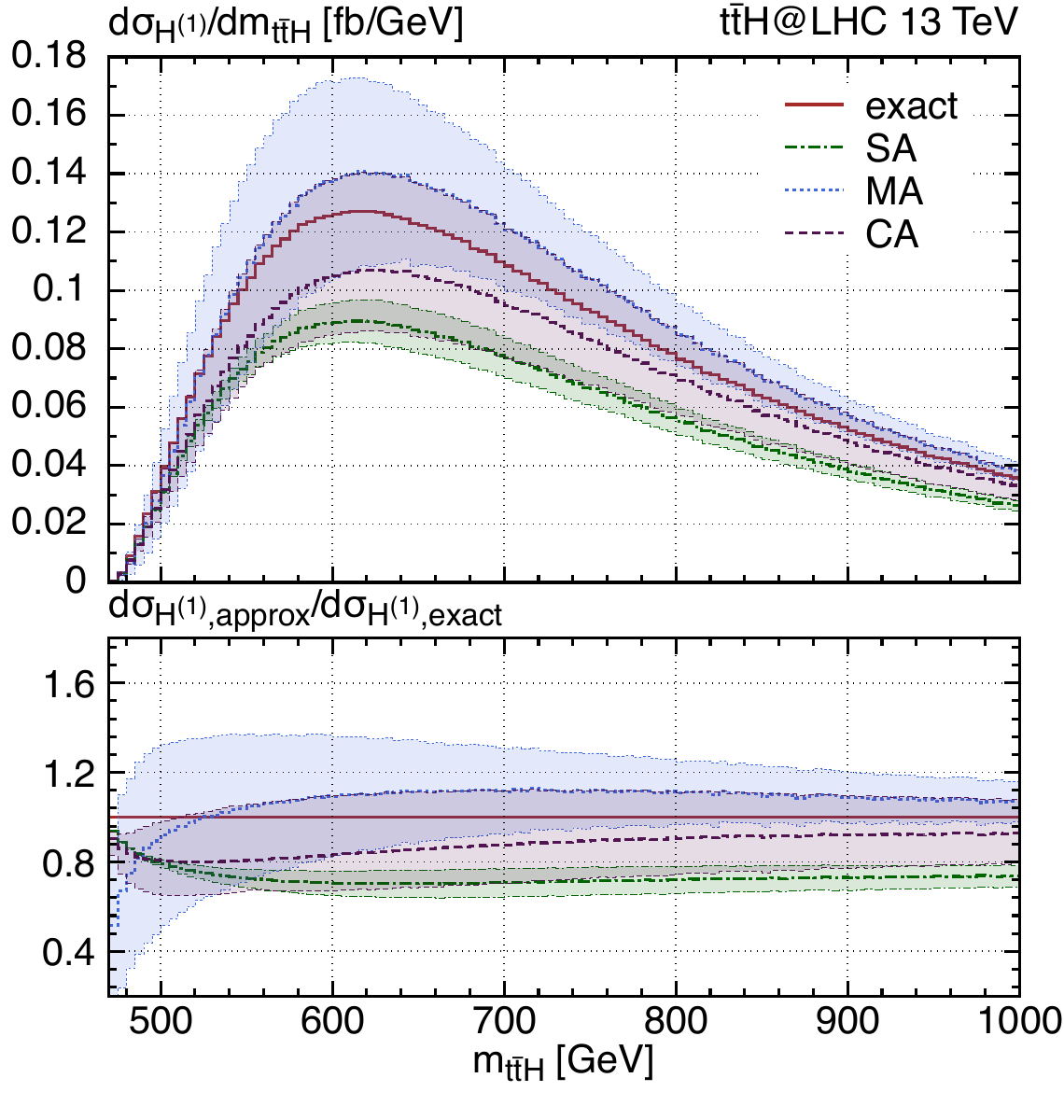}\\
\hspace{-0.52cm}
\includegraphics[width=.35\textwidth]{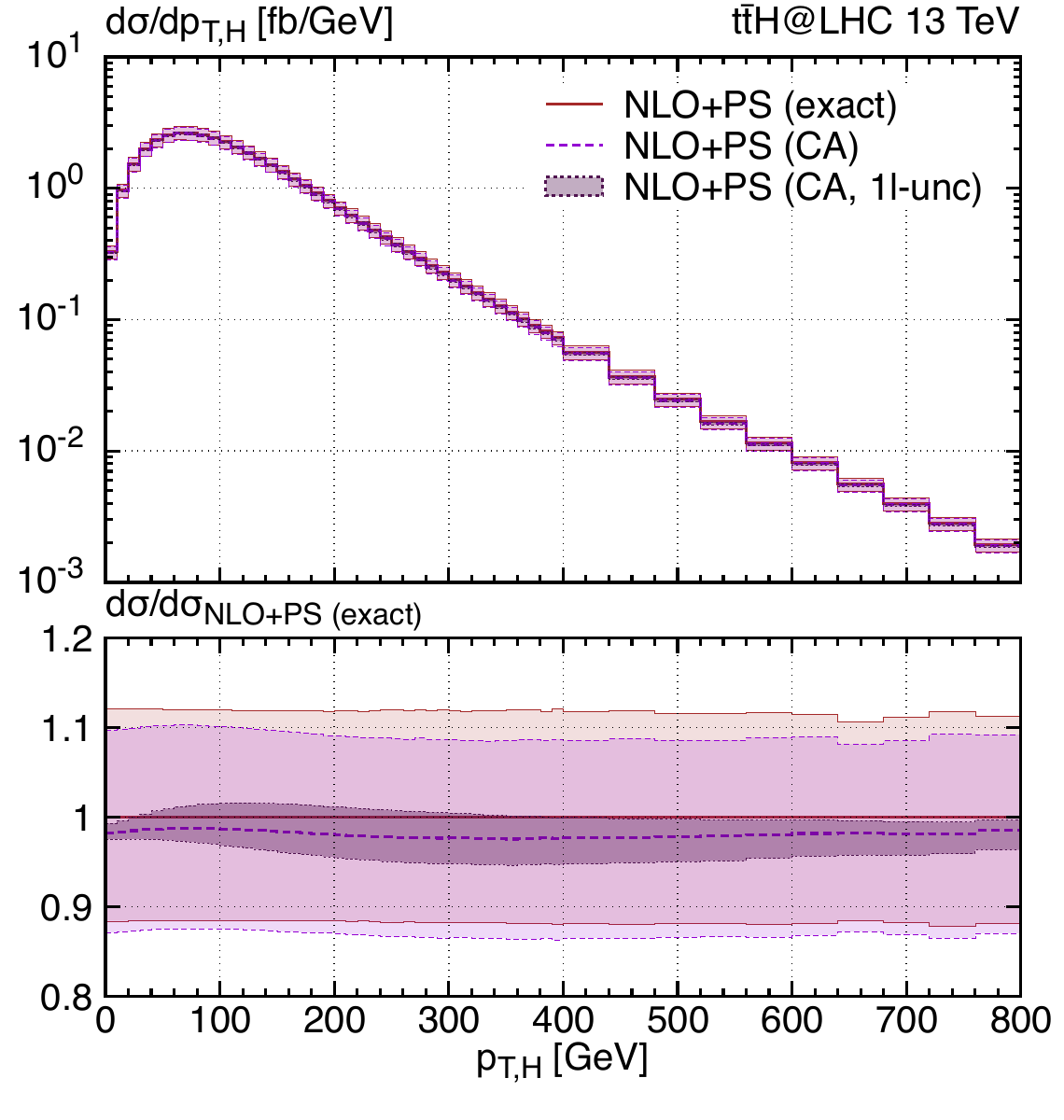}
&
\hspace{-0.73cm}
\includegraphics[width=.35\textwidth]{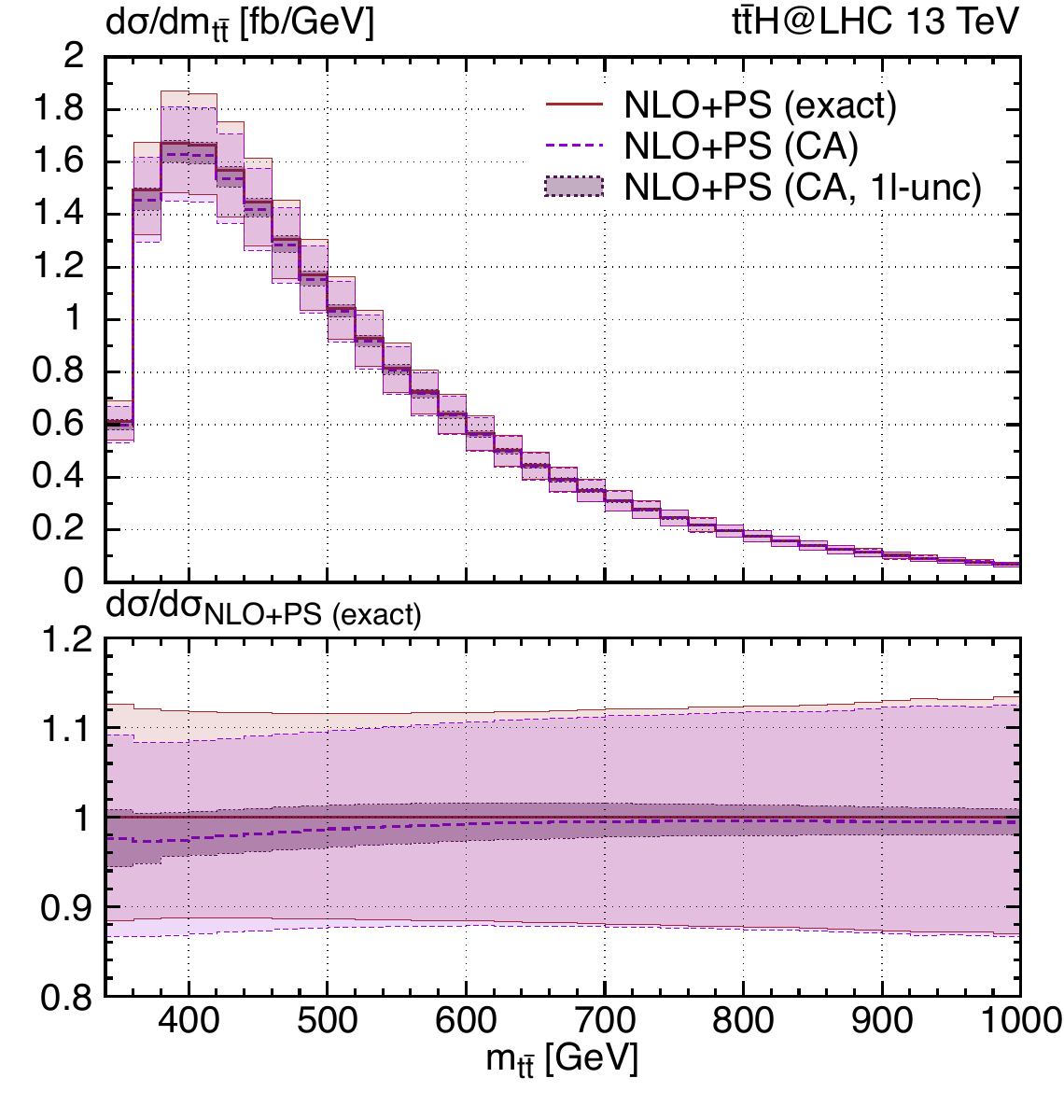}
&
\hspace{-0.73cm}
\includegraphics[width=.35\textwidth]{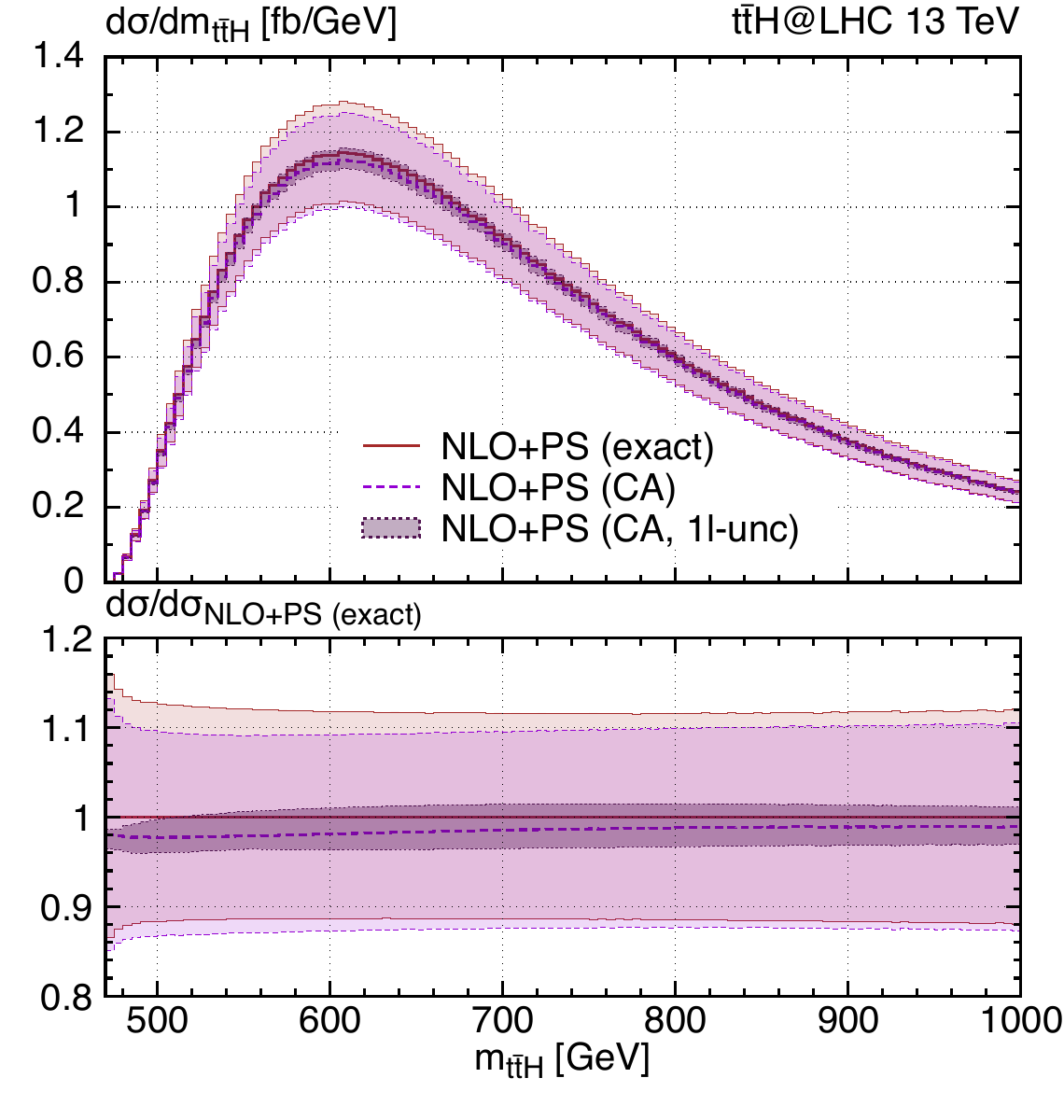}
\end{tabular}
\vspace*{1ex}
\caption{Predictions for the Higgs-boson transverse momentum $\ptH$ (left column), the invariant mass $m_{t\bar t}$ of the top-quark pair (central column), and the invariant mass $m_{t\bar t H}$ of the $t \bar t H$ final state (right column).
The upper plots show the comparison between the exact (solid brown curve), SA (dot-dashed green curve), MA (dotted blue curve), and CA (dashed violet curve) results for the integrated one-loop hard-virtual coefficient, as defined in Eq.\eqref{eq:dsigma_H1}.
The lower plots compare NLO+PS predictions based on the exact (brown solid curve) and CA (dashed violet curve) one-loop amplitude, respectively. The lighter bands display the scale uncertainties, while the darker ones represent the systematic uncertainty assigned to the CA result.}
\label{fig:one-loop_validation_fig1}
\end{center}
\end{figure}

\begin{figure}[t!]
\begin{center}
\begin{tabular}{ccccc}
\hspace{-0.73cm}
\includegraphics[width=.35\textwidth]{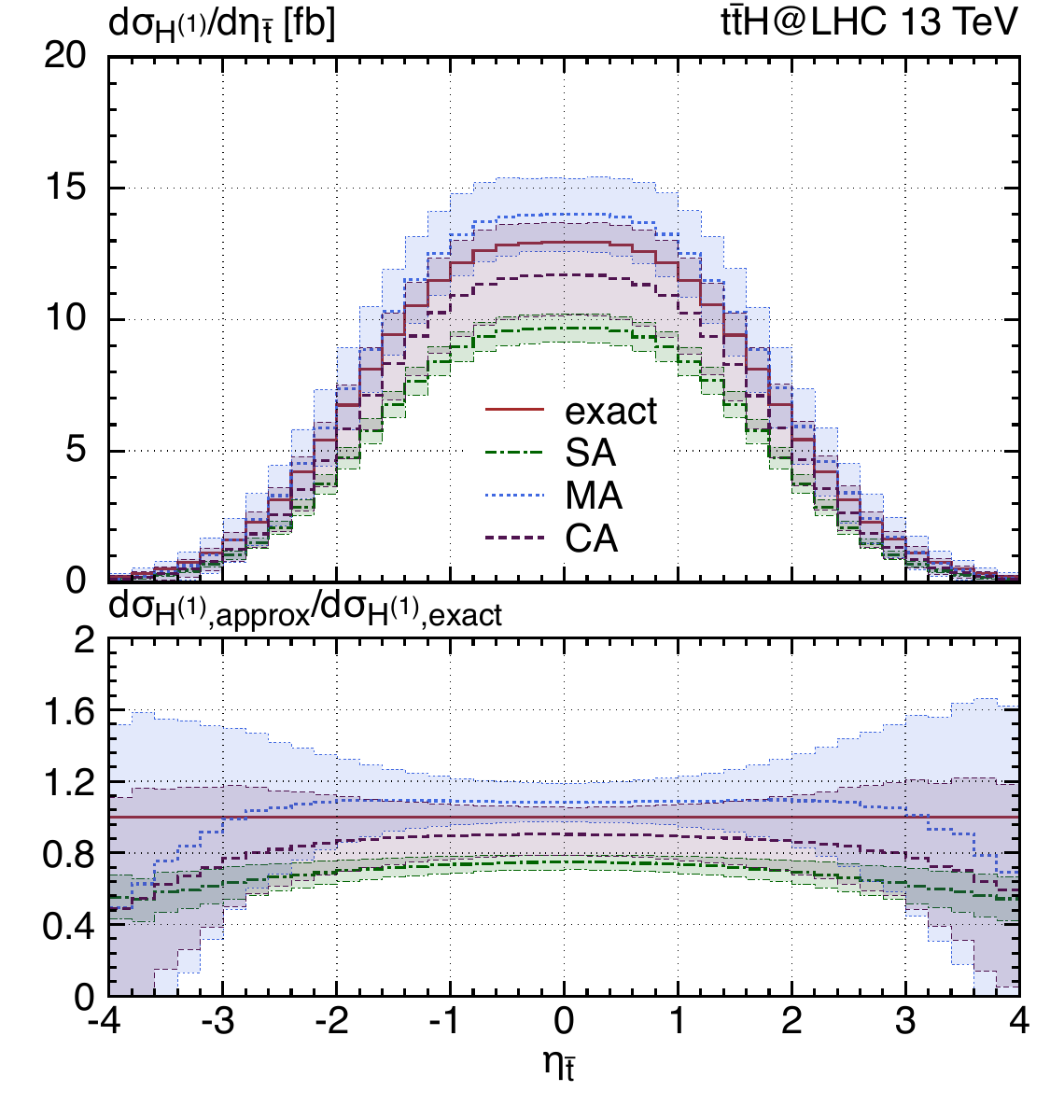}
&
\hspace{-0.73cm}
\includegraphics[width=.35\textwidth]{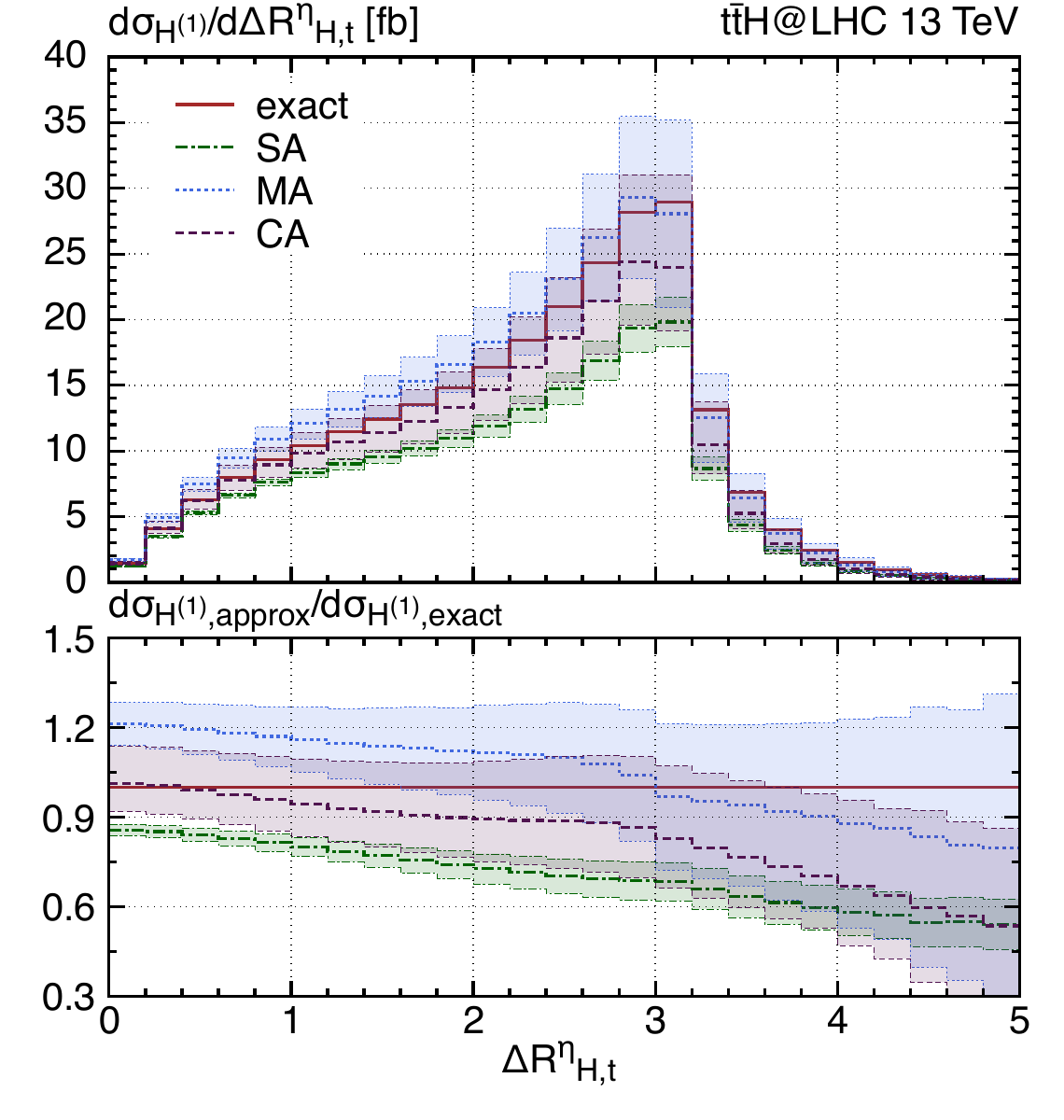}
&
\hspace{-0.73cm}
\includegraphics[width=.35\textwidth]{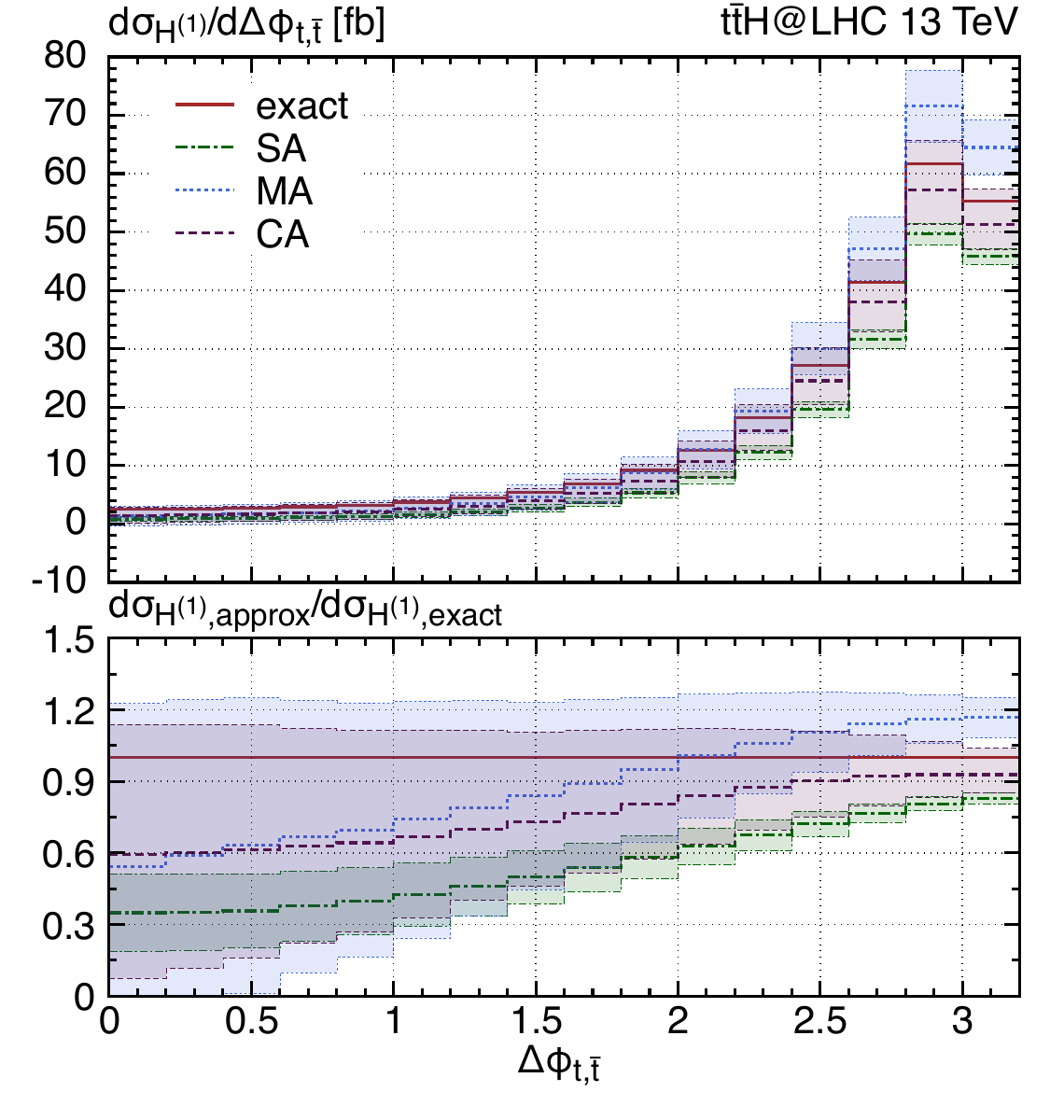}\\
\hspace{-0.73cm}
\includegraphics[width=.35\textwidth]{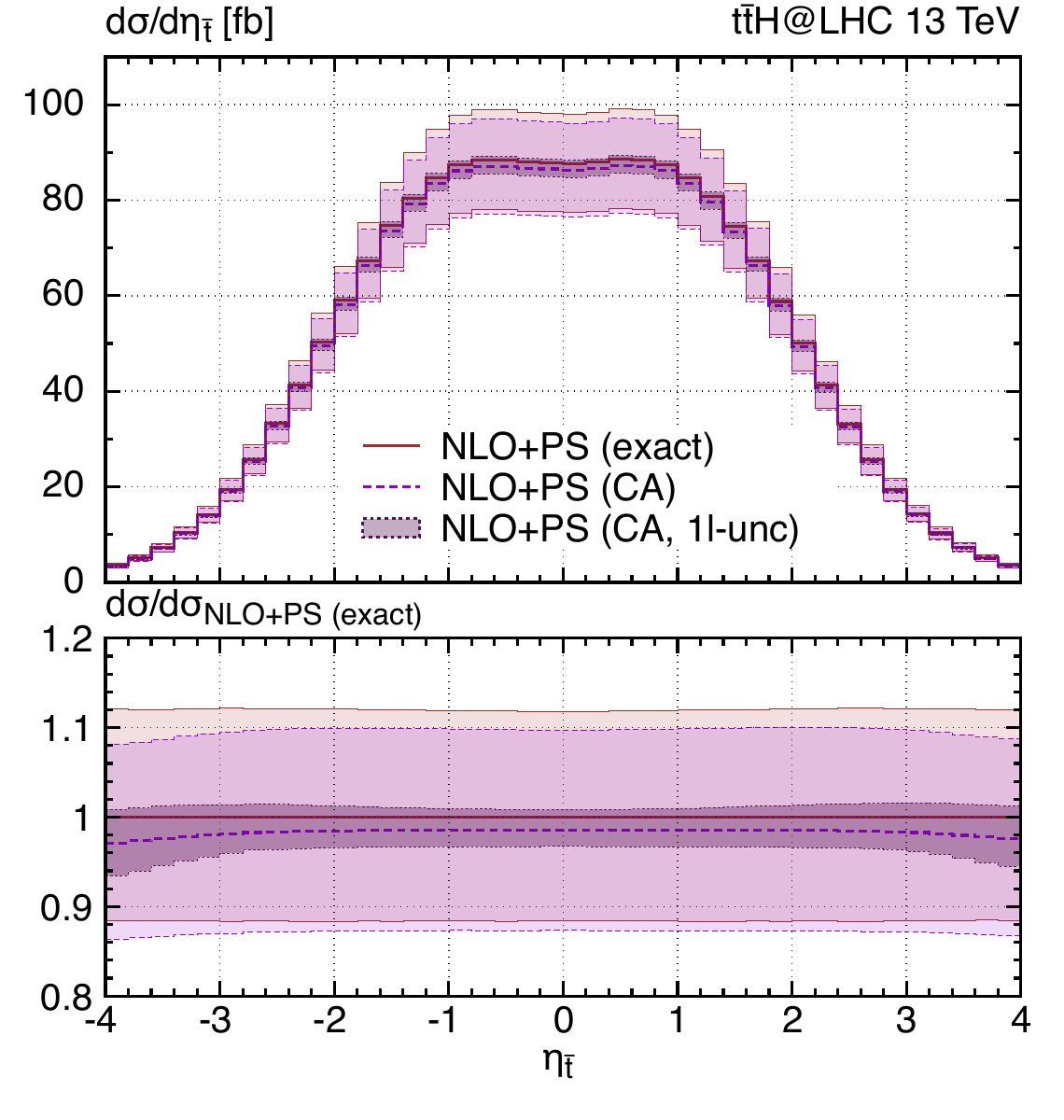}
&
\hspace{-0.73cm}
\includegraphics[width=.35\textwidth]{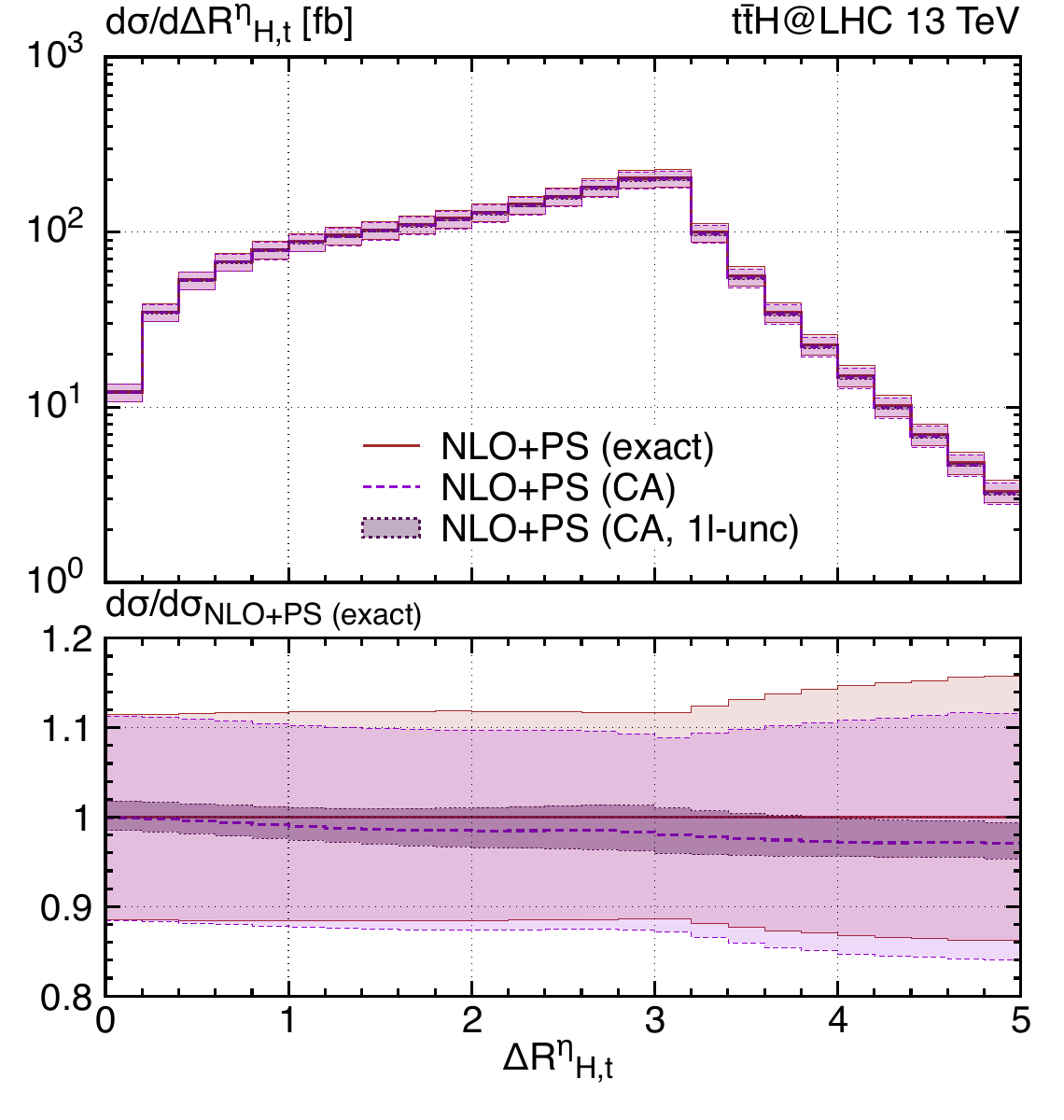}
&
\hspace{-0.73cm}
\includegraphics[width=.35\textwidth]{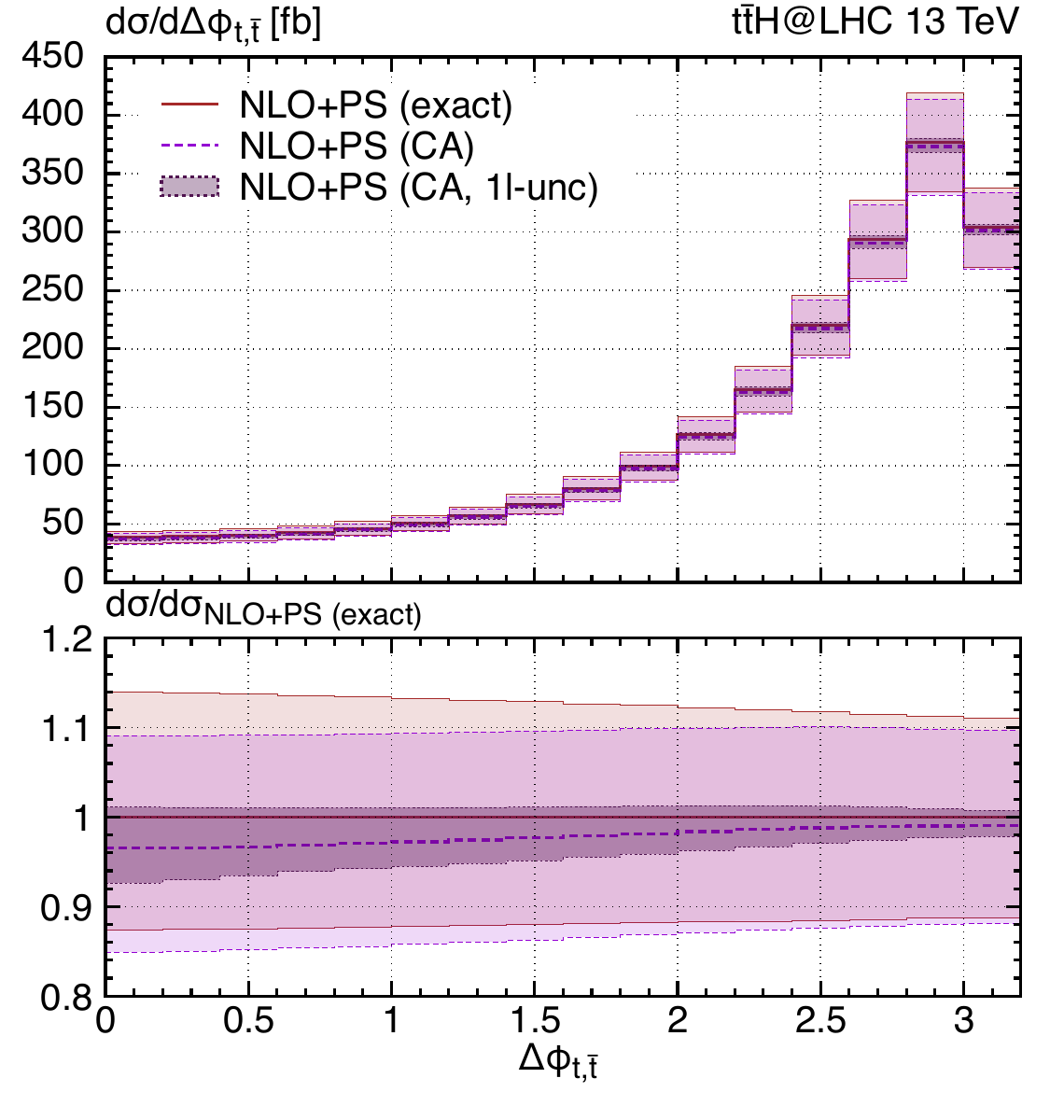}
\end{tabular}
\vspace*{1ex}
\caption{Predictions for the pseudorapidity $\eta_{\bar t}$ of the anti-top quark (left column), the separation $\Delta R^{\eta}_{H,t}$ in the $(\eta-\phi)$ plane between the Higgs boson and the top quark (central column), and the azimuthal angle separation $\Delta \phi_{t,\bar t}$ between the top and anti-top quarks (right column).
The layout is explained in the caption of figure~\ref{fig:one-loop_validation_fig1}.}
\label{fig:one-loop_validation_fig2}
\end{center}
\end{figure}

We continue our analysis by examining differential distributions in \fig{fig:one-loop_validation_fig1} and \fig{fig:one-loop_validation_fig2}. 
All plots are organised as follows: in the upper plots we study the 
different approximations of the one-loop hard-virtual coefficient function 
$\HardVirt^{(1)}$ in comparison to the exact result. 
More precisely, we display its contribution to the NLO cross
section defined as 
\begin{equation}
	{\rm d}\sigma_{\HardVirt^{(1)}} =  \frac{\alpha_s(M)}{2\pi} \, \HardVirt^{(1)}(M) \, {\rm d}\sigma_{\text{LO}} \,,
	\label{eq:dsigma_H1}
\end{equation}
where ${\rm d}\sigma_{\text{LO}}$ is the differential Born cross section.
We stress that the error band associated with the CA result is determined via a nine-point variation of the parameters $(\eta,\tau)$, while the uncertainties assigned to the SA and MA results are obtained via a three-point variation of the parameter $\eta$, that controls the choice of the scale $\muIR$ at which the approximation is applied.
The lower plots show the NLO+PS predictions, comparing the CA approach with the exact result.
The lighter bands represent the scale uncertainties. The darker band is the systematic uncertainty of the CA result.
For all plots, the main frame shows cross sections in fb, while the ratio panel shows the results normalised to the corresponding exact prediction.

% Higgs transverse momentum pT,H
We begin by discussing the two leftmost plots in \fig{fig:one-loop_validation_fig1}, which show the Higgs transverse-momentum spectrum.
This observable plays a special role, as it is directly used to define the weight function $\omega$ in \eqn{eq:omega}, which enters the definition of the CA approach in \eqn{eq:Hn_CA}. 
From the ratio in the upper plot, we observe that the CA result consistently undershoots the exact prediction across the entire $\ptH$ spectrum.
The difference amounts to $\mathcal{O}(10\text{--}15\%)$ around the peak region and gradually decreases toward larger $\ptH$.
This behaviour directly reflects the interplay between the SA and MA entering the CA prescription.
More specifically, the SA result significantly underestimates the exact prediction, with discrepancies ranging from approximately $25\%$ in the bulk region to $40\text{--}45\%$ at large $\ptH$ where the approximation is invalid.
By contrast, the MA result overestimates the exact prediction by roughly $25\%$ in the small-$\ptH$ region, while approaching the exact result in the high-$\ptH$ tail, where the accuracy of the approximation improves.
As expected from the chosen weight function $\omega$, the CA result is driven by the SA for $\ptH\to 0$ and then smoothly approaches the MA curve at large $\ptH$. 
We note, however, that the SA result is not fully reliable even at small $\ptH$, since the Higgs boson is not parametrically soft due to its sizable mass.
This explains why the SA result does not approach the exact prediction in the $\ptH\to 0$ limit and, consequently, why the CA curve remains relatively far from it in the peak region.

We now turn to the systematic uncertainty assigned to the CA result and its consistency with the exact calculation.
We observe that the CA uncertainty band fully covers the exact one-loop result over almost the entire $\ptH$ spectrum.
The only notable exception occurs in the first few bins for $\ptH\lesssim 30$\,GeV. 
In this region, the CA result is dominated by the SA contribution, which is not fully reliable and the corresponding uncertainty is visibly underestimated.
Nevertheless, we argue that the CA approach remains adequate for several reasons: the discrepancy is moderate, it affects only a tiny fraction of the cross section well below the peak region, and the one-loop contribution itself is numerically small.
This is indeed confirmed by the lower plot, where the deviation of the CA result in this limited region has a negligible numerical impact on the NLO prediction of the $\ptH$ distribution. In particular, the deviation is more than an order of magnitude smaller than the corresponding perturbative uncertainties.

Returning to the one-loop result, we find that in the intermediate $\ptH$ range, where neither SA nor MA can be expected to provide a reliable description on its own, the CA uncertainty band naturally broadens, reaching $\mathcal{O}(20\%)$.
This behaviour appropriately reflects the reduced theoretical control in the transition region.
By contrast, in the high-$\ptH$ tail, where the MA result becomes increasingly reliable, the uncertainty rapidly shrinks to the few-percent level, consistent with the good agreement between the CA and exact results.

Finally, we note that the impact of approximating the one-loop finite remainder is modest at the level of the NLO cross section, as shown in the lower plot.
Across the entire $\ptH$ spectrum, deviations remain below the $2\%$ level.
The systematic uncertainty associated with the combined prescription is well under control and remains significantly smaller than the scale-uncertainty band.
This indicates that the CA approach does not introduce any numerically significant bias in NLO-accurate predictions.

We have performed the same validation across a wide range of observables, finding consistent conclusions.
The results shown in this section include observables that exhibit the largest deviations between the CA and exact predictions. 
Before moving on, we emphasise that in the NNLO predictions presented in the next section, any one-loop discrepancy is treated as an additional systematic uncertainty on the two-loop CA result (see the third point at the end of \sct{sec:pointwise_combination}).
This provides an additional safeguard for the robustness of our NNLO+PS generator.

%...........................................

% other observables: invariant masses
Beyond the transverse momentum of the Higgs boson, \fig{fig:one-loop_validation_fig1} also shows results for two additional observables: the invariant mass of the $t\bar t$ pair (central column) and the invariant mass of the $t\bar t H$ final state (right column).
For both distributions, the ratio panels in the upper plots show that the SA result consistently underestimates the exact prediction, with an associated uncertainty band that does not cover the difference.
Its inclusion in the CA prescription is nevertheless essential to prevent the result from being dominated by the MA contribution in kinematic regions where the latter is not expected to be reliable.
This effect is particularly visible in the low-$m_{t\bar t H}$ region.
By contrast, the MA component becomes the dominant contribution to the CA result in the tails of the invariant-mass distributions, where the difference with the exact prediction reduces to a few percent. Most importantly, the systematic uncertainty of the CA approach fully covers the difference with the exact result across the entire $m_{t\bar t}$ distribution and over the $m_{t\bar t H}$ distribution, with the exception of a tiny, subdominant region at small $m_{t\bar t H}$.
For both observables, the CA approach yields an excellent description of the exact NLO+PS prediction, with the net effect of the approximation being numerically small, of $\mathcal{O}(2\%)$, and, in the case of $m_{t\bar t H}$, particularly flat across phase space.

% angular observables 
In \fig{fig:one-loop_validation_fig2}, we present results for three angular observables: the pseudorapidity $\eta_{\bar t}$ of the anti-top quark (left column), the separation $\Delta R^{\eta}_{H,t}$ in the $(\eta,\phi)$ plane between the Higgs boson and the top quark (central column), and the azimuthal separation $\Delta\phi_{t,\bar t}$ between the top and anti-top quark (right column).
The $\eta_{\bar t}$ distribution provides an example of an observable for which neither SA nor MA results are fully reliable in the bulk region, with the former undershooting it by $\mathcal{O}(30\%)$ and the latter overestimating the exact result by $\mathcal{O}(10\%)$.
Nevertheless, the combined prescription yields a reasonable description in the intermediate region, with residual shape distortions confined to the very forward region, where the cross section is small and the hard-virtual contribution is numerically suppressed.
As a result, this effect is negligible in the NLO+PS distribution. In all cases, the CA uncertainty band fully covers the exact result.

Although we found the CA approach to provide a reliable description of the exact result across phase space, for the two  angular observables presented in \fig{fig:one-loop_validation_fig2} we observe a larger impact. In particular, we notice shape distortions in the CA result in the regions $\Delta R^{\eta}_{H,t} > 3$ and $\Delta \phi_{t,\bar t}< \pi/2$. However, this affects only a very small fraction of events away from the bulk region. Moreover, the uncertainty assigned to the CA result successfully captures the differences with the exact prediction even in the low-$\Delta\phi_{t,\bar t}$ region and also in most of the $\Delta R^{\eta}_{H,t}$ range.
For the NLO+PS cross section, these shape distortions are more visible than for the other observables considered.
Nevertheless, even in the phase-space regions where the CA approach performs poorly, deviations from the exact NLO+PS result remain below $4\%$, and the associated uncertainty band increases conservatively while remaining much smaller than the scale uncertainties.
In conclusion, deviations from the exact result are substantially mitigated once the complete NLO+PS corrections are included, owing to the limited numerical impact of the hard-virtual contribution in the complete prediction.
This further supports the robustness of the CA approach, particularly in view of its application to the two-loop amplitude in our NNLO+PS predictions presented in the next section.

%============================================
\section{Phenomenological results}
\label{sec:pheno_results}
After discussing the combined approximation for the hard-virtual contribution and its one-loop validation in the previous section, we present NNLO+PS predictions for $pp$ collisions at $\sqrt{s}=13\,\text{TeV}$ at the LHC obtained with our \minnlo{} $\ttH$ generator.
Unless stated otherwise, the setup employed throughout this section is identical to that used in \sct{sec:oneloop_validation}. 
When decays are included and fiducial cuts are applied, the corresponding settings are specified in the relevant subsections.
For reproducibility of the results presented here, we list below the \minnlo{} settings adopted in this study (see also the \texttt{Manual} attached to the arXiv submission of this manuscript for a more detailed description of each parameter):
\begin{itemize}[noitemsep, topsep=0pt]
\item[]
\texttt{fixedscale    0}:  use a dynamical scale for the overall two powers of $\alpha_s$ via the flag \texttt{whichscale}. 
\item[]
\texttt{whichscale  2}:  dynamical scale choice for $\muR$, set to the invariant mass $m_{t\bar{t}H}$ of the $\ttH$ system. 
\item[]
\texttt{largeptscales      0}:  dynamical scale choice for $\muR$, $\muF$ at large-$\pt$, where the resummation effects are switched off; set to $m_{t\bar{t}H}$ to reproduce fixed-order results in that region.
\item[]
 \texttt{modlog\_p     -1d0}:  activate piecewise modified logarithms, defined as in Eq.~(4.15) of \citere{Mazzitelli:2021mmm}.
\item[]
\texttt{kappaQ          0.25d0}: parameter $\KQ$ controlling the choice of the resummation scale $Q$ as a fraction of the hard scale $m_{t\bar{t}H}$. 
\item[] 
\texttt{Q0                2d0}: parameter controlling the IR cut-off smoothly approached by the profiled scale choice in the low-$\pt$ region to avoid the Landau singularity.
\end{itemize}
The parton-shower settings in the \PYTHIA{8}\cite{Bierlich:2022pfr} interface are:
\begin{itemize}[noitemsep, topsep=0pt]
\item[]
\texttt{py8tune            21}:  Monash2013 tune.
\item[]
\texttt{nohad               1}: switch off hadronization.
\item[]
\texttt{nompi               1}:  switch off multi-parton interactions.
\item[]
 \texttt{noqed               1}:  switch off QED shower effects.
 \end{itemize}

In the following, we present a series of phenomenological results.
In \sct{sec:MATRIX_validation}, we validate the NNLO+PS predictions against fixed-order results obtained with \Matrix{}~\cite{Grazzini:2017mhc}.
In \sct{sec:pheno_on-shell_results}, we provide inclusive NNLO+PS predictions for key kinematic observables.
In \sct{sec:pheno_Higgs-decay_results} and \sct{sec:pheno_top-decay_results}, we analyse $t\bar t H$ production including Higgs-boson and top-quark decays, respectively, applying fiducial cuts inspired by recent experimental analyses~\cite{CMS:2021kom,ATLAS:2025eua}.
All results are obtained by retaining the full-colour dependence of the two-loop massive amplitude.
However, for improved computational efficiency, the two-loop amplitude with massless top quarks entering the MA approach can alternatively be evaluated in the leading-colour approximation.
This option is implemented in the $\ttH{}$ code released with this paper and can be activated via the \texttt{masslessQQHFC} flag in the input card.
Notice that even when the massless finite remainder is treated in the leading-colour approximation, the mass logarithms generated by the massification procedure are retained in full colour. 
This is essential to preserve the correct colour structure of the logarithmic terms and to avoid spurious (mis-)cancellations between real and virtual contributions at the cross-section level.
By construction, the impact of subleading-colour contributions of the massless two-loop amplitude is limited to the high-energy region and remains negligible for the total cross section.
However, their effect becomes relevant in the tails of certain differential distributions sensitive to the MA approach---such as the transverse-momentum spectrum of the top-quark pair---where negative subleading-colour corrections of a few percent appear at the NNLO+PS cross-section level.

%...............................................................................
\subsection{Validation against fixed-order predictions}
\label{sec:MATRIX_validation}

In this section, we compare NNLO+PS predictions obtained with our \minnlo{} generator to NNLO-accurate fixed-order results~\cite{Devoto:2024nhl} computed within the \Matrix{} framework~\cite{Grazzini:2017mhc}.
For this validation, the two-loop hard-virtual contribution $\mathbb{H}^{(2)}$ is set to zero in both codes.

$\ttH$ production represents the first $\QQF$ process studied within the \minnlo{} framework for which a fixed-order calculation at NNLO accuracy is available.
The present comparison thus provides a crucial benchmark for the extension of the \minnlo{} methodology to $\QQF$ processes developed in \citere{Mazzitelli:2024ura}, allowing us to assess its validity and NNLO accuracy.
As previously observed for heavy-quark pair production~\cite{Mazzitelli:2020jio,Mazzitelli:2021mmm}, the large numerical value of the top-quark mass leads to improved perturbative convergence compared to bottom-quark pair production~\cite{Mazzitelli:2023znt}.
In this respect, the $t\bar t H$ process represents an ideal testing ground, offering a non-trivial validation of the process-independent ingredients of the \minnlo{} framework.
These ingredients also enter the massive $b\bar b Z$~\cite{Mazzitelli:2024ura} and $b\bar b H$~\cite{Biello:2024pgo} implementations, for which fixed-order predictions are not yet available.

Within the \minnlo{} formulation, the organisation of the perturbative series differs from the fixed-order expansion by terms of formally higher order.
It is therefore essential to verify that the numerical impact of those terms remains small, within the quoted perturbative uncertainties, and does not spoil the formal NNLO accuracy.
We emphasise that this comparison constitutes a key validation step of the \minnlo{} approach.
As we will show, differences with the fixed-order results are very moderate, and the impact of formally higher-order terms remains small for inclusive observables, even at the fully differential level.

\begin{table}[b]
\begin{center}
\scalebox{0.95}{
\renewcommand{\arraystretch}{1.7}
    \begin{tabular}{ |c|c|c|c| }
   \cline{2-4}
   \multicolumn{1}{c|}{} 
   &
    \minlo{}
    &
    \minnlo{} ($\HardVirt^{(2)}=0$)
   &
    \Matrix{} ($\HardVirt^{(2)}=0$)
 \\
    \hline
    \multirow{1}{*}{$\sigma^{\ttH}_{\rm inclusive}(\sqrt{s}=$13\,TeV)}
    &
    $0.3864(2)_{-16.2\%}^{+19.2\%}\,\text{pb}$
    &
    $0.4734(2)_{-5.9\%}^{+5.4\%}\,\text{pb}$
    &
    $0.4742(9)_{-6.0\%}^{+4.7\%}\,\text{pb}$
        \\
    \hline
    \end{tabular}%
}%
\vspace{0.5cm}
    \caption{\label{tab:xs_MATRIXvsMiNNLO} Total inclusive cross sections for \ttH{} production at 13\,TeV. Percentages represent perturbative uncertainties, and statistical errors are in round brackets.}
\end{center}
\end{table}

We first comment on the predictions for the total cross section reported in \tab{tab:xs_MATRIXvsMiNNLO}.
The second and third columns show the \minlo{} and \minnlo{} predictions, respectively, while the last column reports the fixed-order result obtained with \Matrix{}.
Both the \minnlo{} and fixed-order results are NNLO-accurate, except for the hard-virtual contribution, which has been set to zero in this comparison.
We observe that the two central values are fully compatible within the statistical error, highlighting the remarkable agreement between the NNLO+PS and fixed-order predictions.
This confirms that resummation and parton-shower effects have a negligible impact on the total cross section in a fully inclusive setup, as expected.
Even the scale uncertainties, despite being treated differently in the two calculations, are very similar in size, further demonstrating that the \minnlo{} approach yields genuinely NNLO-accurate predictions.
Comparing the \minnlo{} and \minlo{} results, it is evident that the inclusion of the $D$-terms, see \eqn{eq:minnlo_cross_section}, which are absent in the \minlo{} formulation, is essential to achieve NNLO accuracy.
The \minlo{} prediction undershoots the NNLO cross section by $\mathcal{O}(18\%)$, well within its quoted scale uncertainty, which is more than a factor of three larger than that of \minnlo{}.

% MATRIX vs MiNNLO comparison
\begin{figure}[t]
\begin{center}
\begin{tabular}{ccc}
\hspace{-0.52cm}
\hspace{-0.73cm}
\includegraphics[width=.35\textwidth]{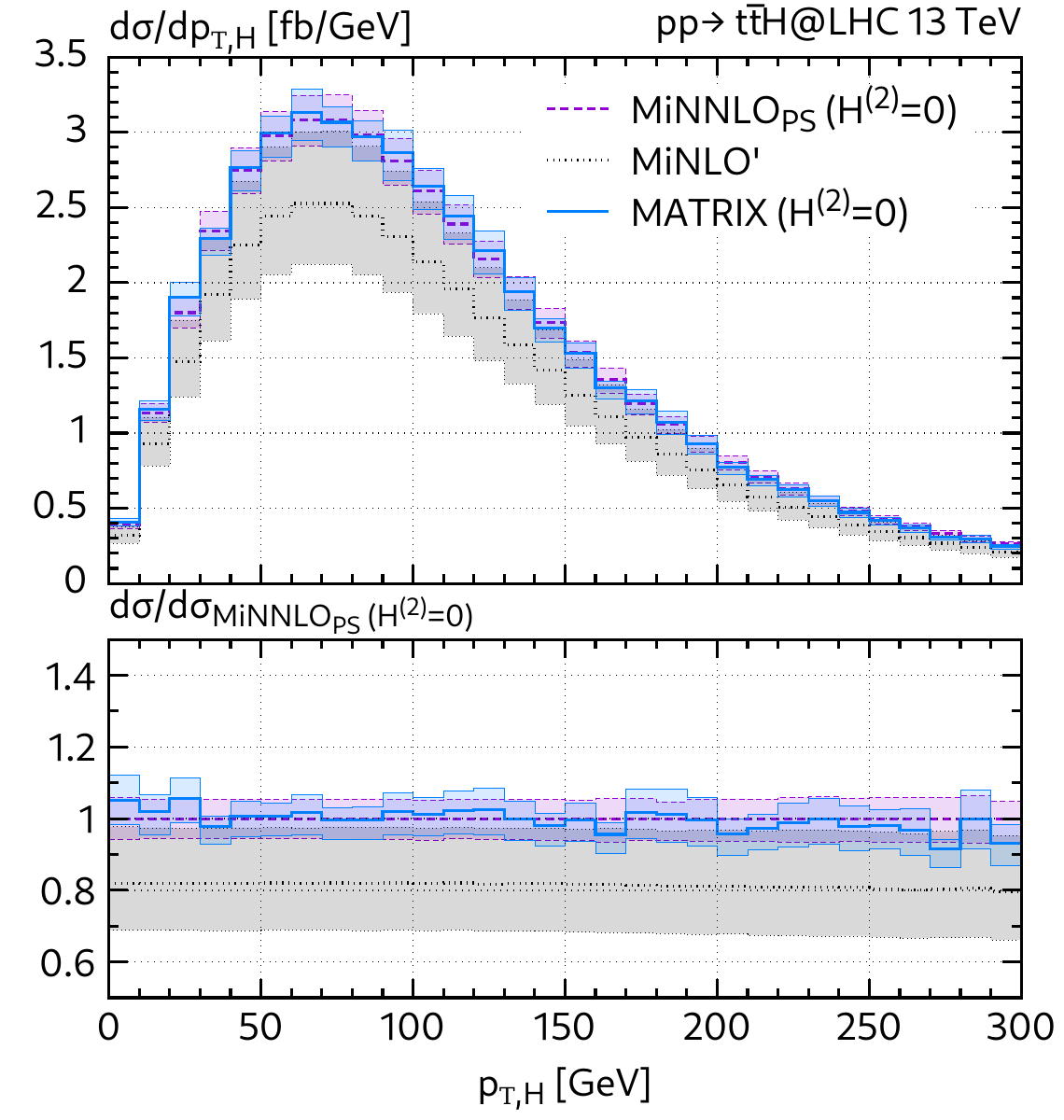}
&
\hspace{-0.73cm}
\includegraphics[width=.35\textwidth]{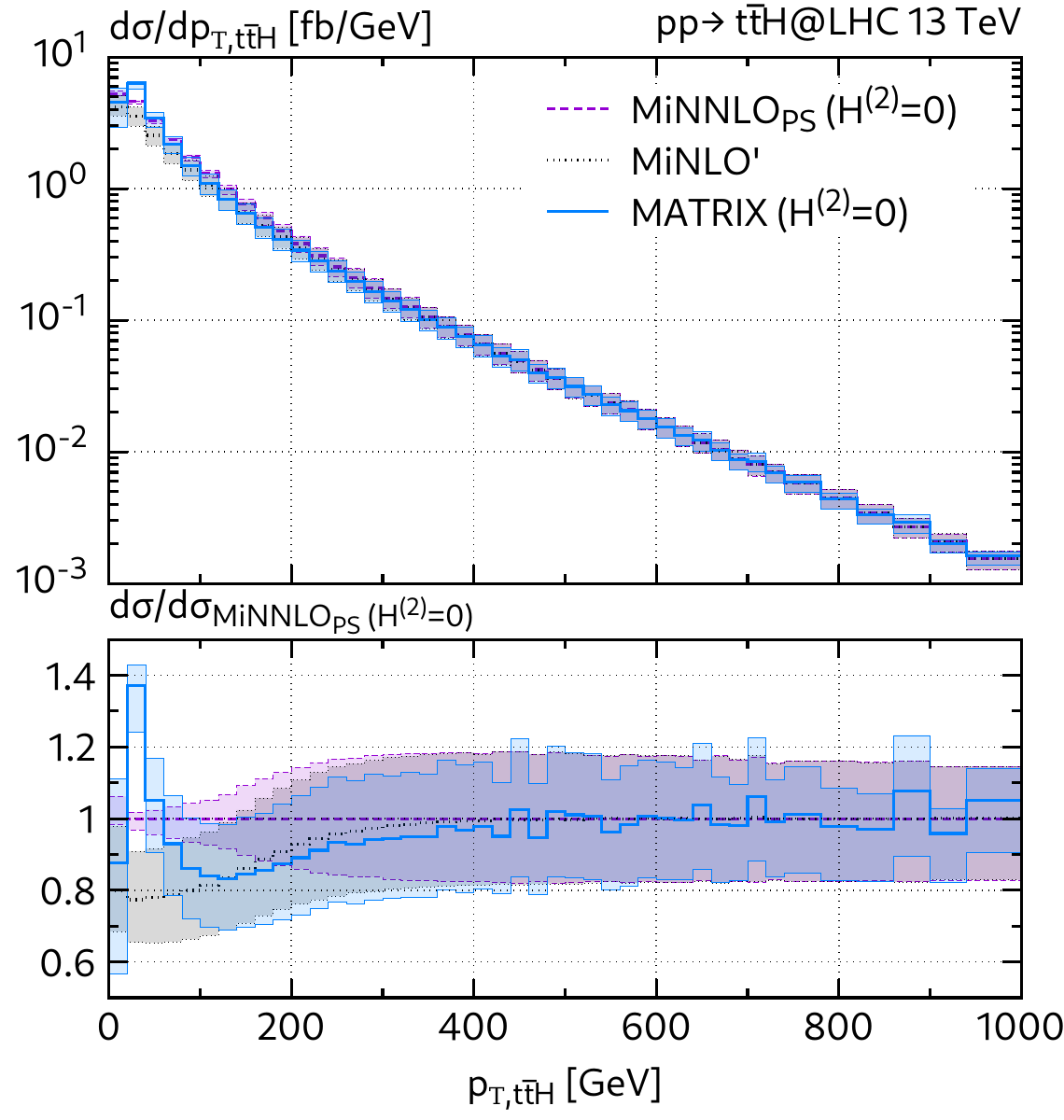}
&
\hspace{-0.73cm}
\includegraphics[width=.35\textwidth]{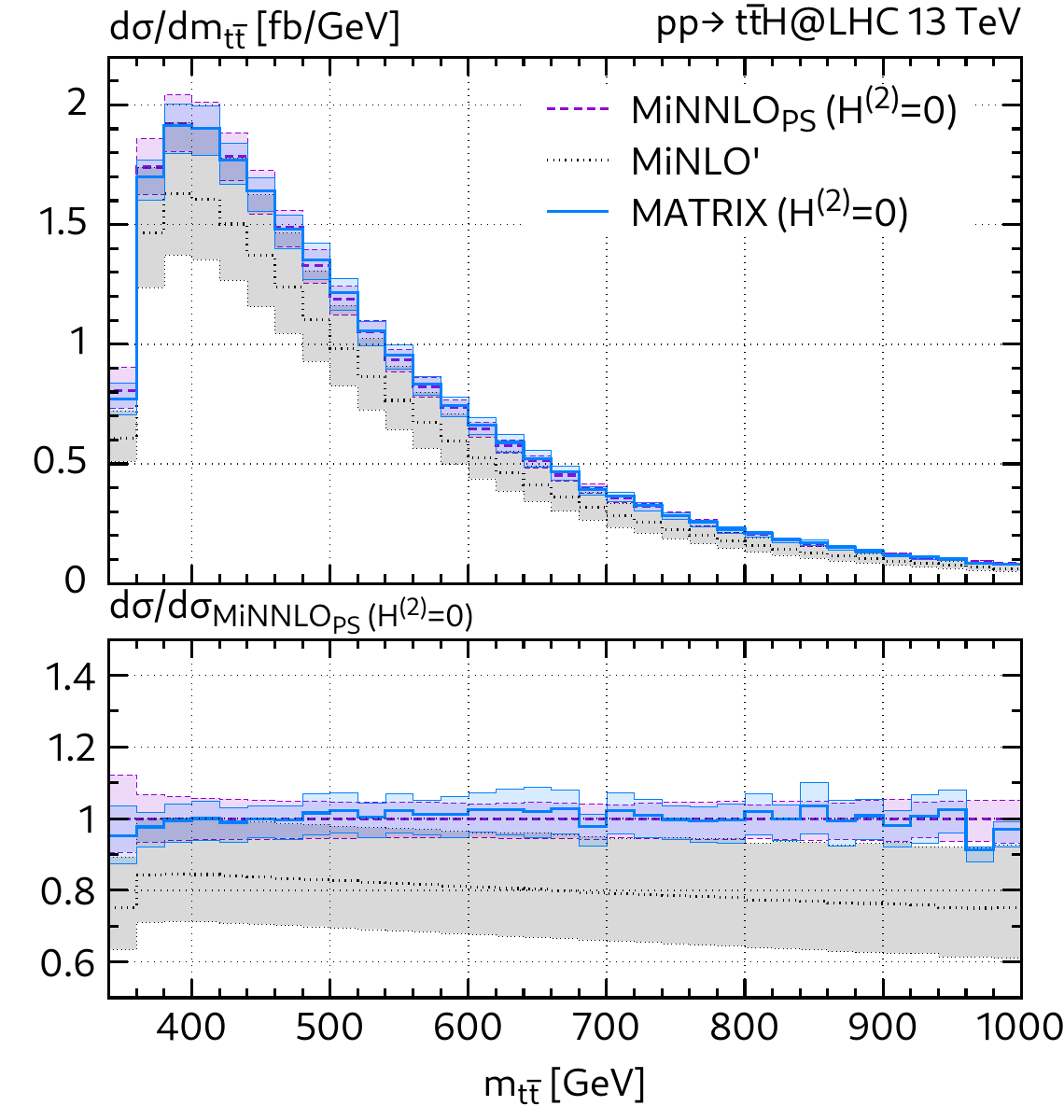}\\
\hspace{-0.73cm}
\includegraphics[width=.35\textwidth]{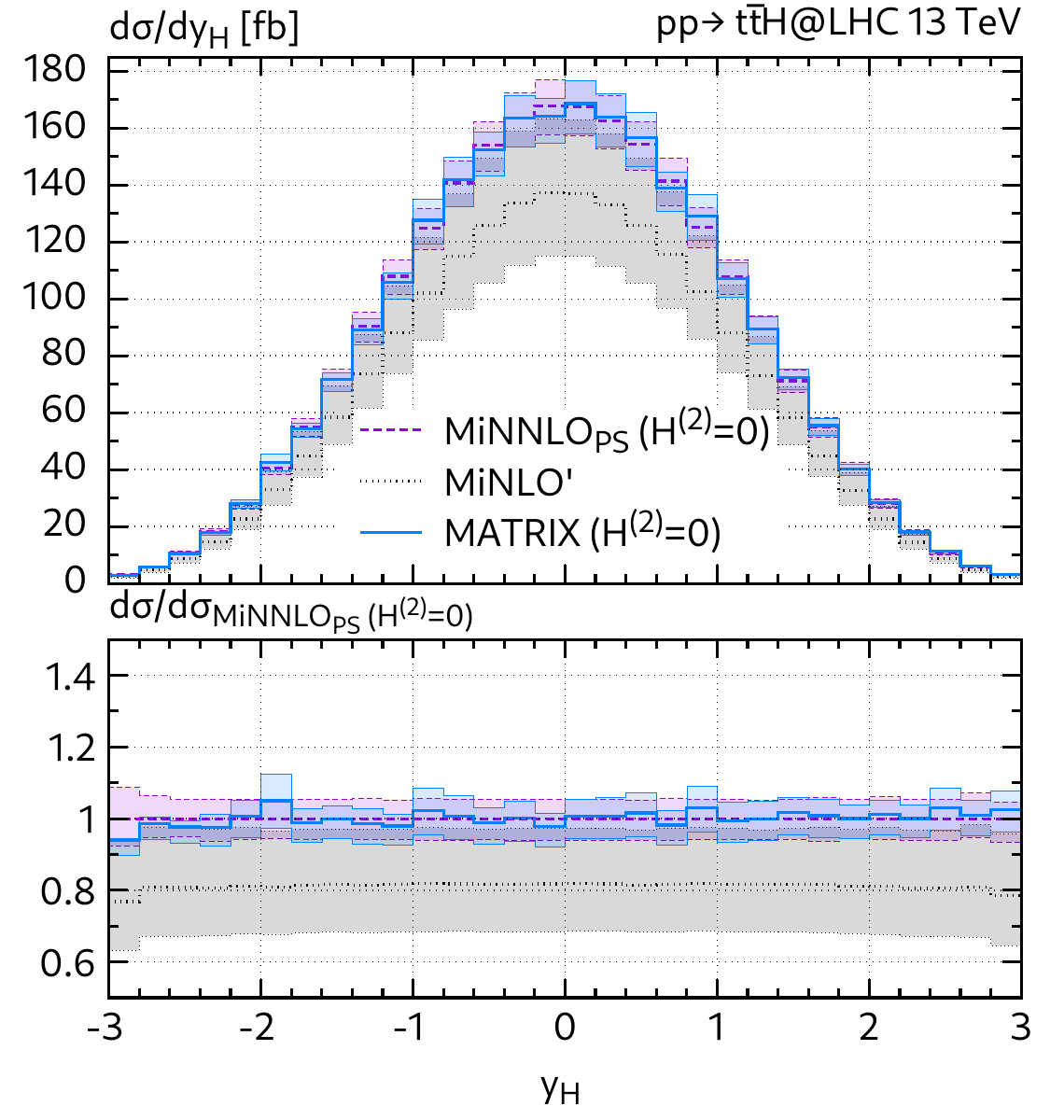}
&
\hspace{-0.73cm}
\includegraphics[width=.35\textwidth]{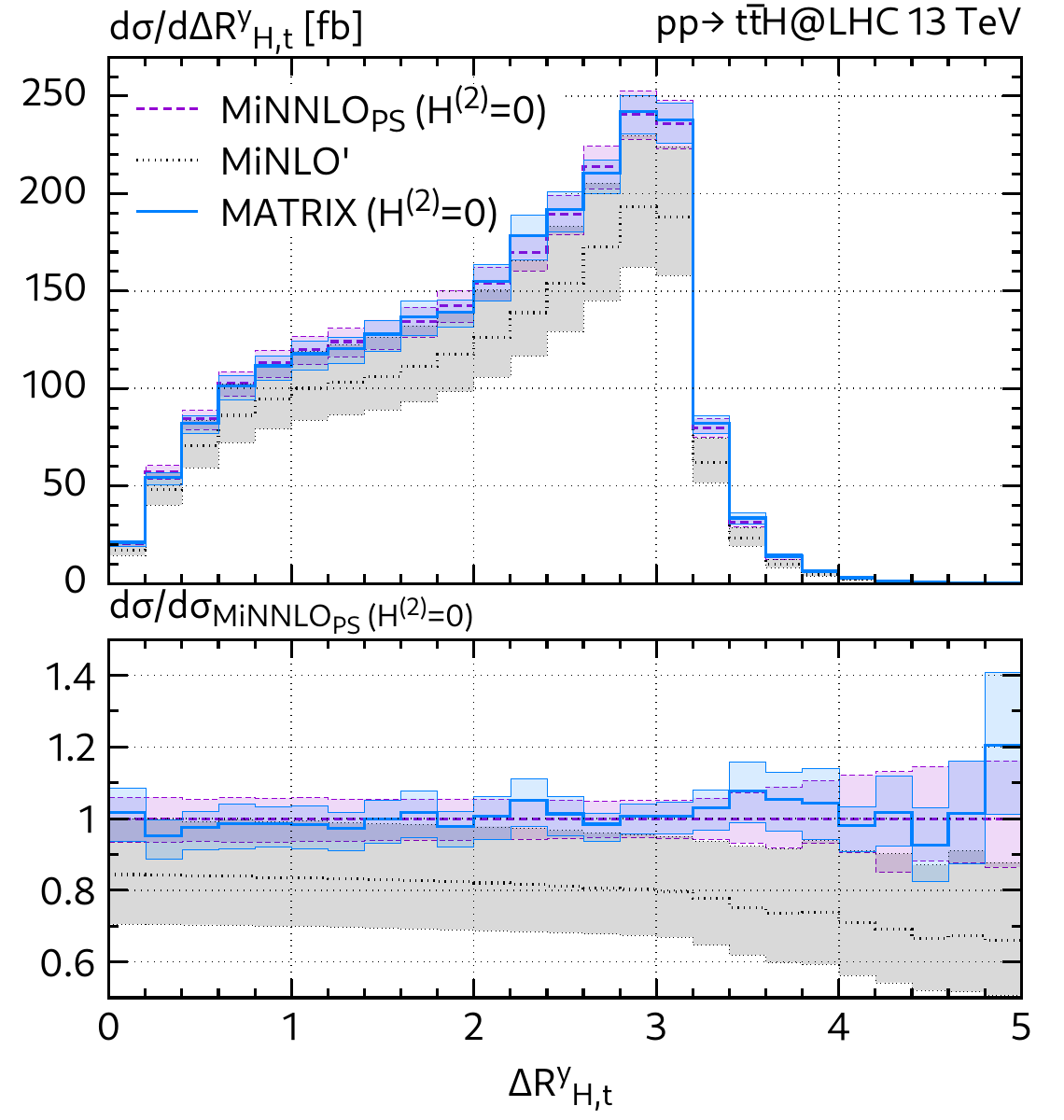}
&
\hspace{-0.73cm}
\includegraphics[width=.35\textwidth]{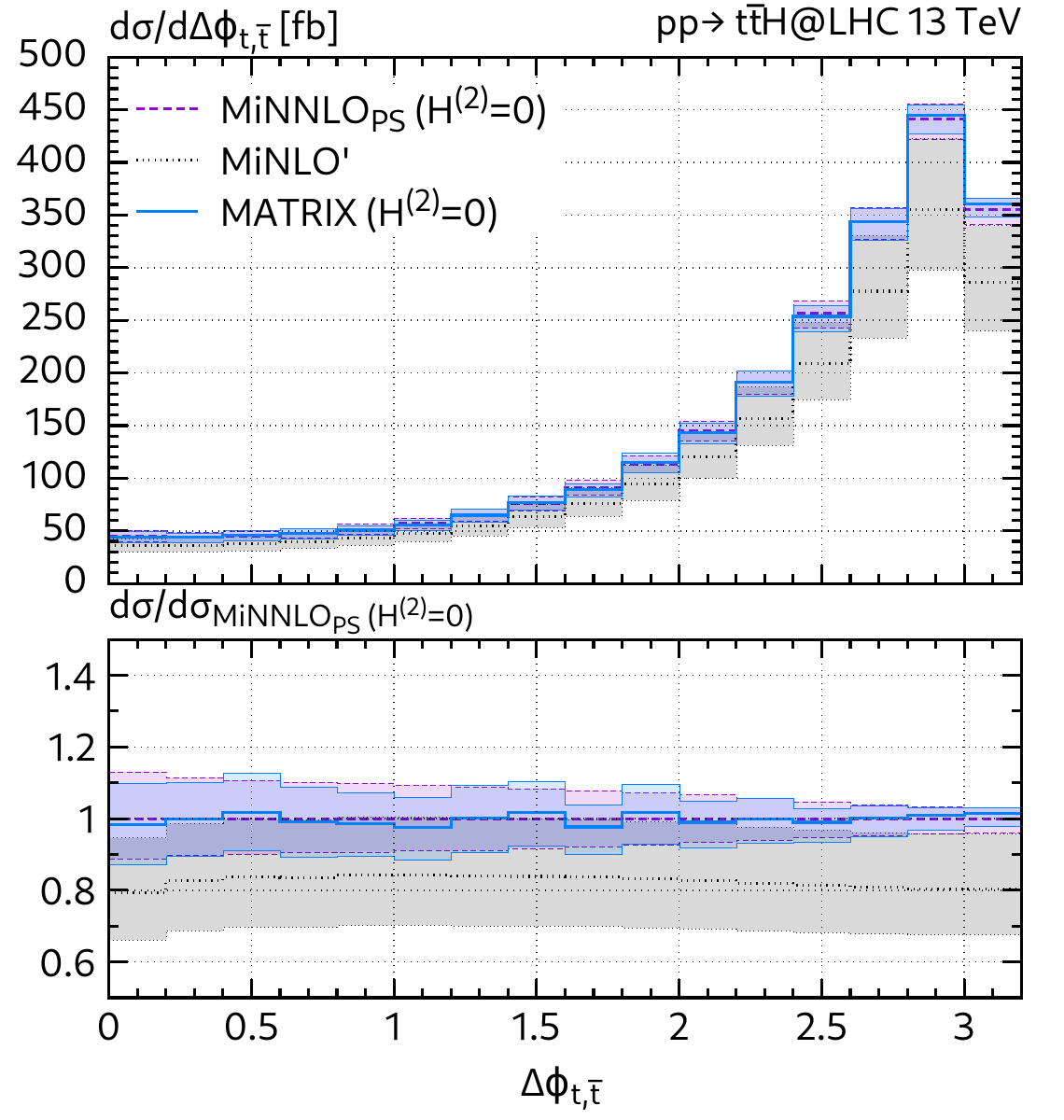}\\
\end{tabular}
\vspace*{1ex}
\caption{\label{fig:MiNNLO_vs_MATRIX} Comparison of \minnlo{}, \minlo{} and NNLO QCD predictions. See text for details.}
\end{center}
\end{figure}
%.............................

We now turn to the comparison of differential distributions.
In \fig{fig:MiNNLO_vs_MATRIX}, we present results for six observables: the transverse momentum $\ptH$ and the rapidity $\yH$ of the Higgs boson (left column); the transverse momentum $\ptTTH$ of the $\ttH{}$ system and the separation $\Delta R^{y}_{H,t}$ in the $(y,\phi)$ plane between the Higgs boson and the top quark (central column); and the invariant mass $m_{t\bar t}$ of the top-quark pair together with their azimuthal-angle separation $\Delta \phi_{t,\bar t}$ (right column).
In all plots, the upper panel compares the \minnlo{} prediction (dashed violet curve) with the \minlo{} result (dotted grey curve) and the fixed-order \Matrix{} prediction (solid blue curve). 
The lower panel shows the ratio to the \minnlo{} prediction, which illustrates the impact of the NNLO corrections absent in \minlo{}, as well as the resummation effects by the parton shower absent in the fixed-order \Matrix{} result. 
The bands correspond to the scale uncertainties. 

Our findings can be summarised as follows:
\begin{itemize}
\item[i)] The inclusion of the $D$-terms in the \minnlo{} predictions increases the cross section by $\mathcal{O}(20\%)$ with respect to the \minlo{} results.
This enhancement is largely flat across phase space, with the exception of $\ptTTH$ and $\Delta R^{y}_{H,t}$, where noticeable shape distortions are observed.
The effect is particularly pronounced in the small-$\ptTTH$ region, where the $D$-terms are redistributed to $\ptTTH>0$\,GeV through the spreading function $F^{\text{corr}}$, see \eqn{eq:Bbar}.
The inclusion of NNLO corrections also leads to a substantial reduction of the scale-uncertainty bands. 
\item[ii)] The \minnlo{} predictions are in excellent agreement with the fixed-order results, up to statistical fluctuations, for observables inclusive over radiation, where resummation effects are negligible.
This agreement holds for both the central values and the relative scale uncertainties.
In contrast, in regions where resummation effects become relevant, notably at small $\ptTTH$, the fixed-order prediction becomes unreliable, and parton-shower matching in the \minnlo{} framework is essential.
\end{itemize}
We conclude that the excellent agreement observed for all inclusive observables confirms that the \minnlo{} approach achieves NNLO accuracy while consistently incorporating resummation and parton-shower effects at the fully differential level.

%...............................................................................
\subsection[Results for on-shell $\ttH{}$ production]{Results for on-shell \boldmath{$\ttH{}$} production}
\label{sec:pheno_on-shell_results}

\begin{table}[b]
\begin{center}
\scalebox{0.9}{
\renewcommand{\arraystretch}{1.7}
    \begin{tabular}{ |c|c|c|c| }
   \cline{2-4}
   \multicolumn{1}{c|}{} 
    &
    NLO+PS (exact)
    &
    \minnlo{} ($\HardVirt^{(2)}=0$)
    &
    \minnlo{} (CA)
\\
\hline
    \multirow{1}{*}{$\sigma^{\ttH}_{\rm inclusive}(\sqrt{s}=$13\,TeV)}
    &
    $0.4096(1)_{-11.6\%}^{+12.0\%}\,\text{pb}$
    &
    $0.4734(2)_{-5.9\%}^{+5.4\%}\,\text{pb}$
    &
    $0.4778(5)_{-6.3\%[-0.57\%]}^{+6.1\%[+1.04\%]}\,\text{pb}$
        \\
    \hline
    \end{tabular}%
}%
\vspace{0.5cm}
    \caption{\label{tab:xs_onshell_final} Total inclusive cross sections for \ttH{} production at 13\,TeV. Percentages represent perturbative uncertainties, statistical errors are in round brackets, and percentages in squared brackets are the systematic uncertainties assigned to the CA of the two-loop amplitude discussed in \sct{sec:pointwise_combination}.}
\end{center}
\end{table}

%...................................
\begin{figure}[t!]
\begin{center}
\begin{tabular}{ccc}
\hspace{-0.75cm}
\hspace{-0.75cm}
\includegraphics[width=.34\textwidth]{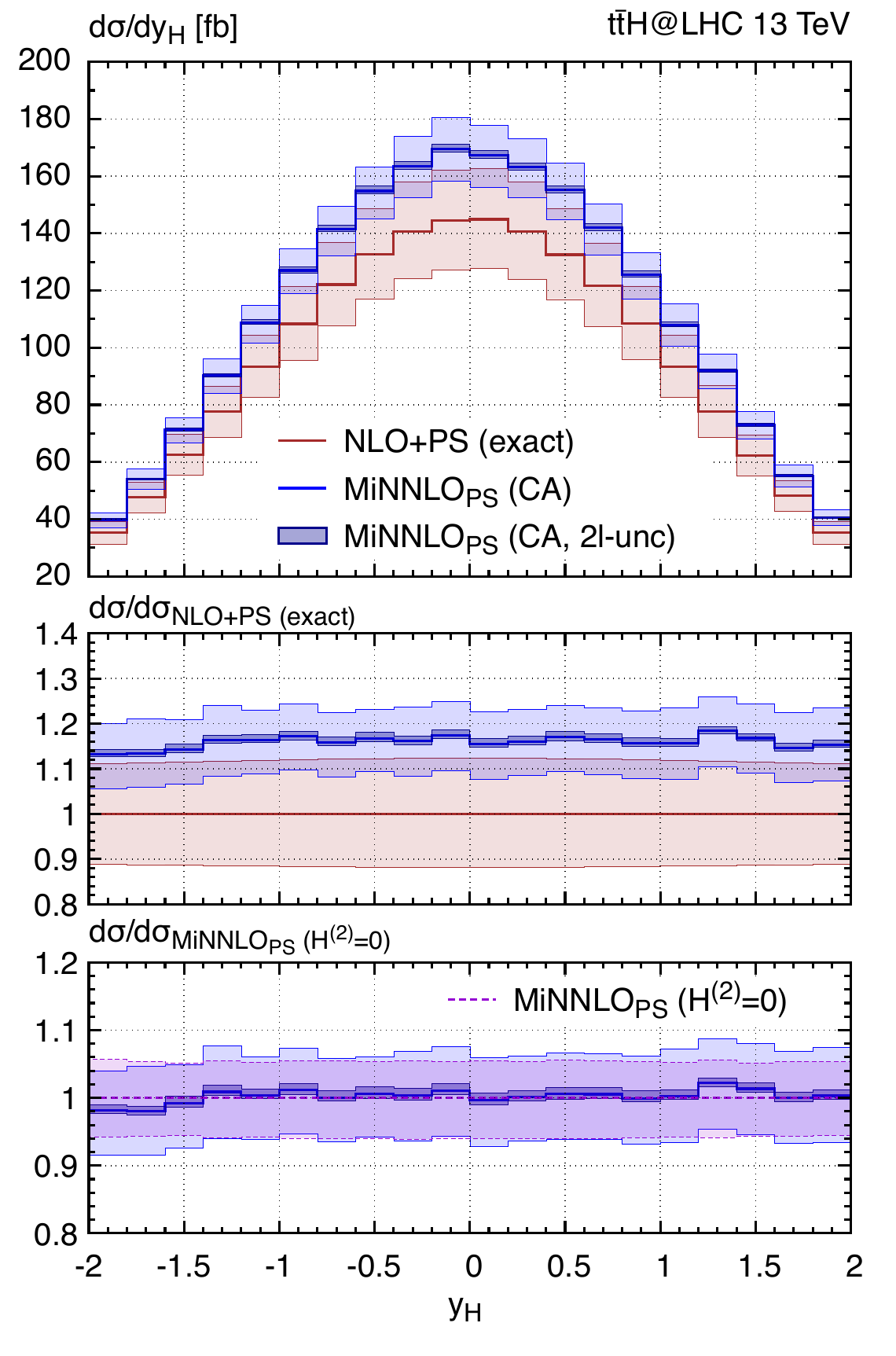}
\hspace{-0.75cm}
&
\hspace{-0.75cm}
\includegraphics[width=.34\textwidth]{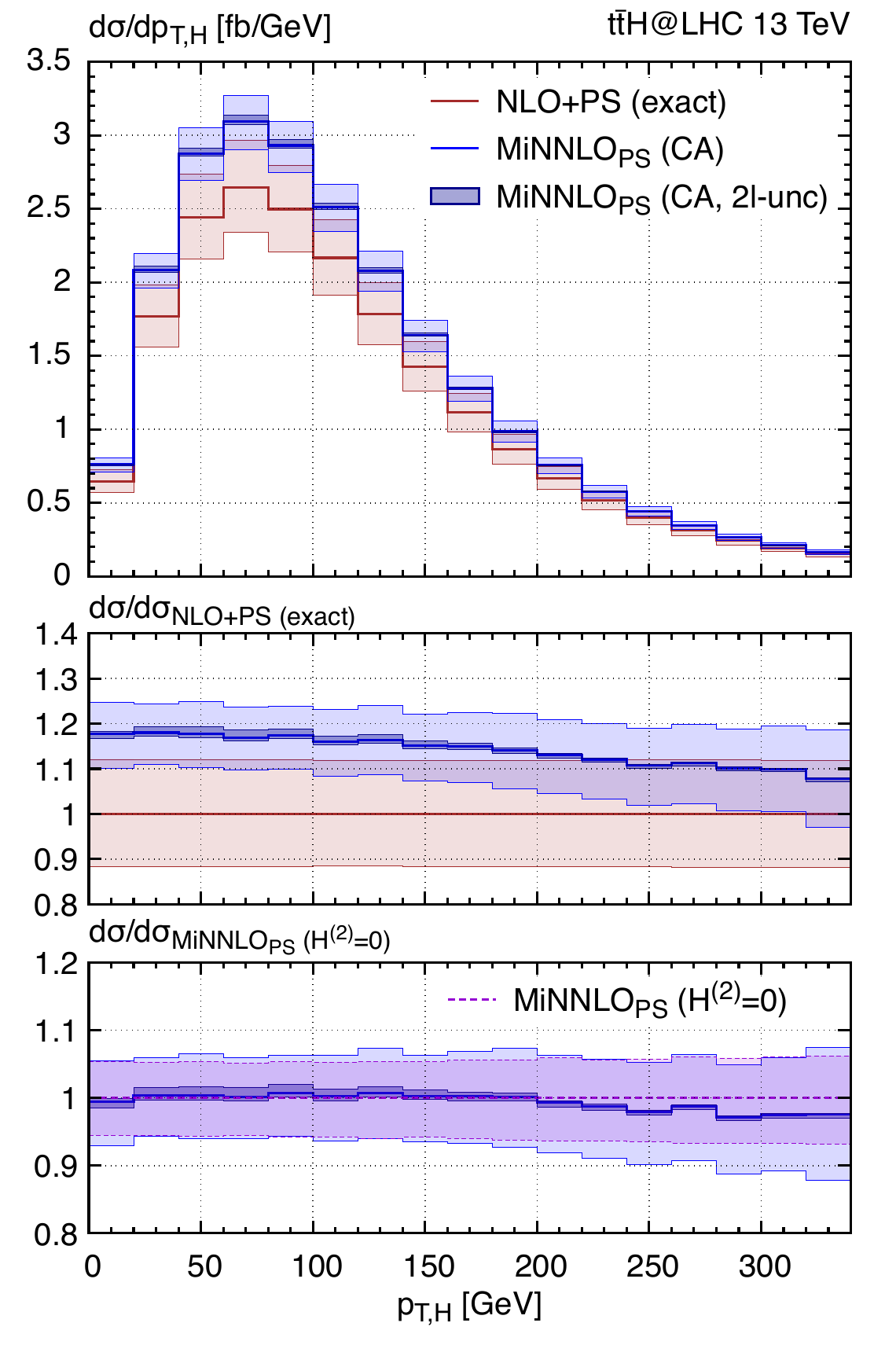}
&
\hspace{-0.75cm}
\includegraphics[width=.34\textwidth]{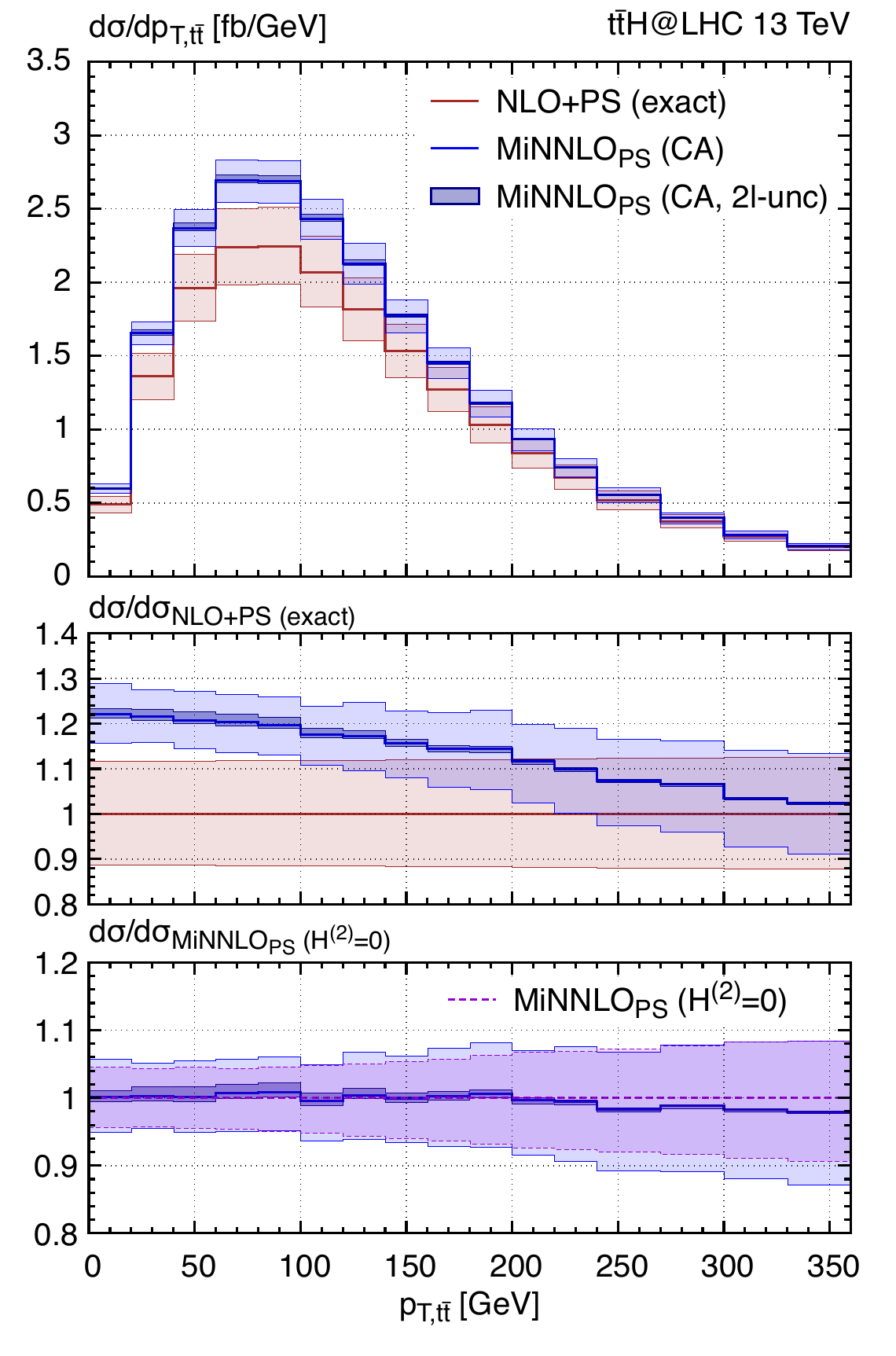}\\
\hspace{-0.75cm}
\includegraphics[width=.34\textwidth]{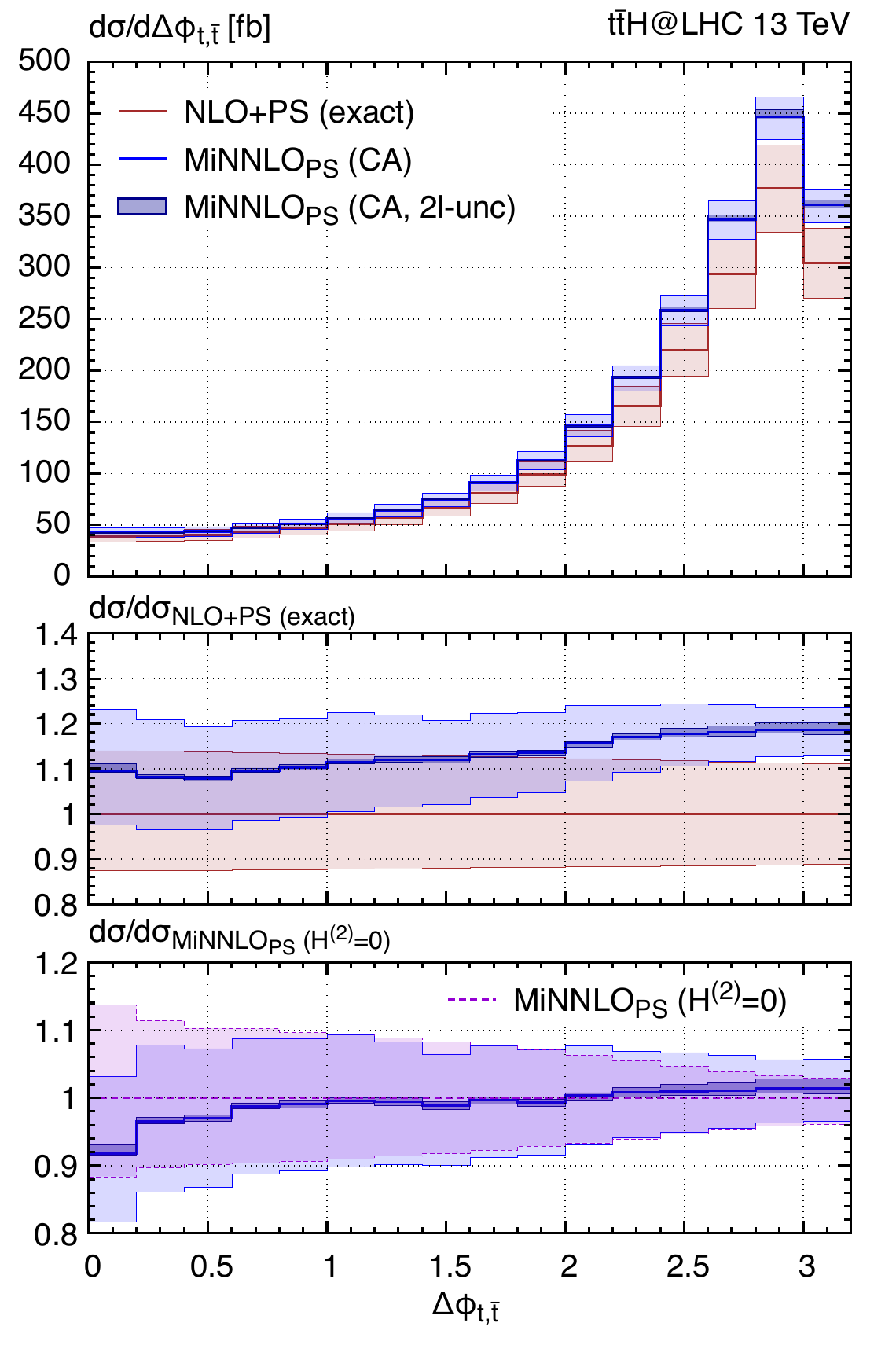}
&
\hspace{-0.75cm}
\includegraphics[width=.34\textwidth]{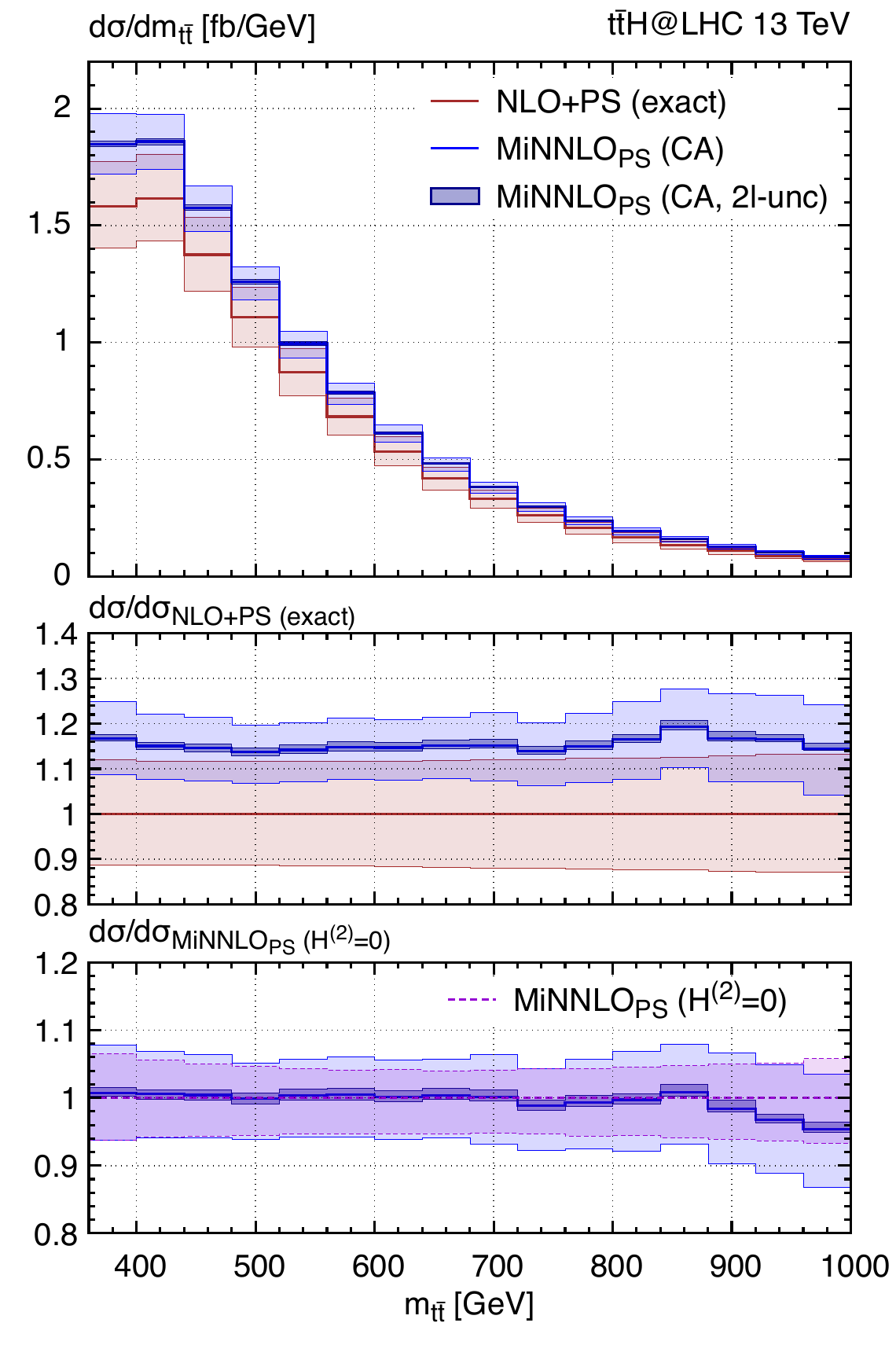}
&
\hspace{-0.75cm}
\includegraphics[width=.34\textwidth]{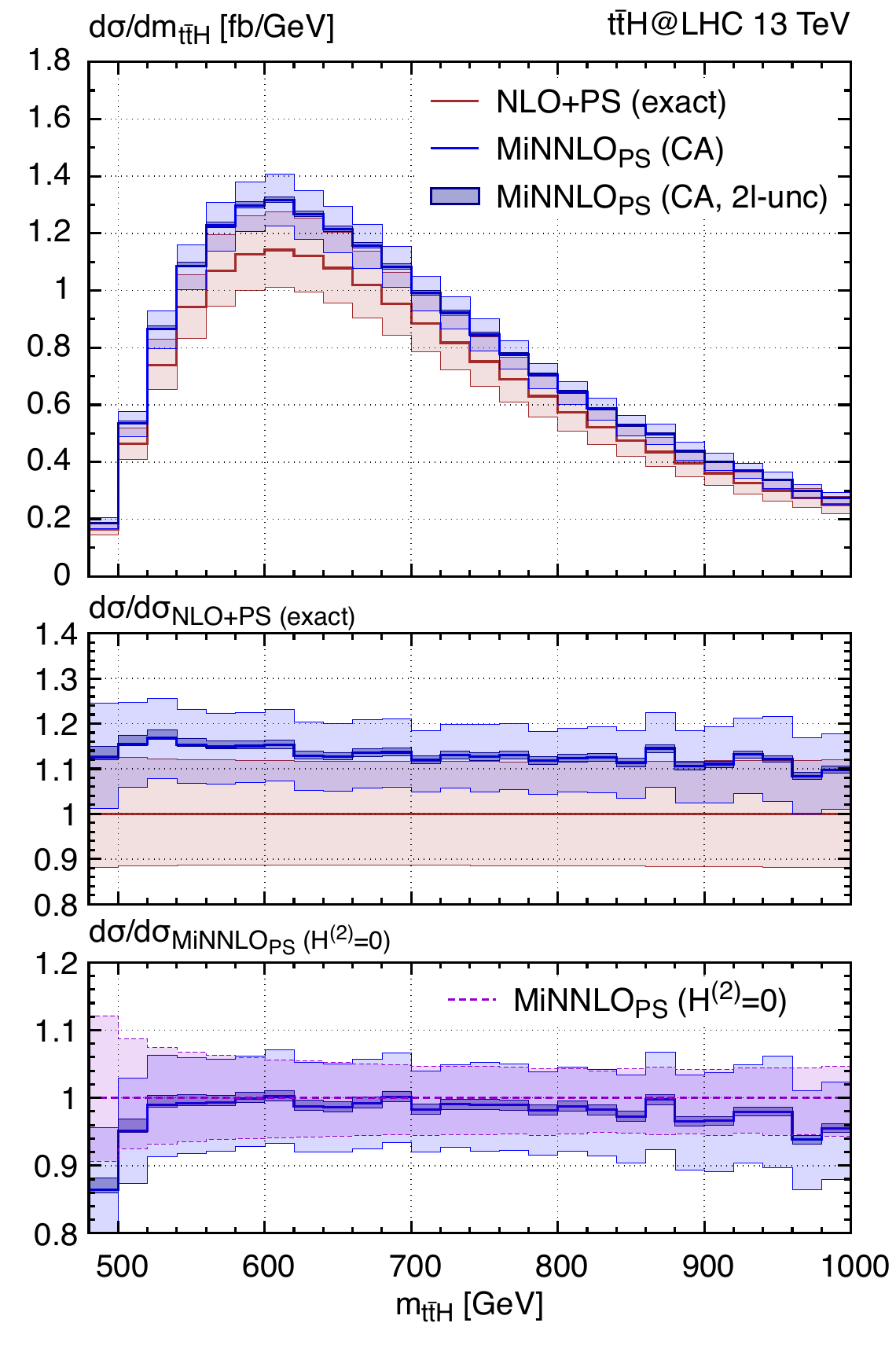}
\\
\end{tabular}
\vspace*{1ex}
\caption{\label{fig:final_results_on-shell} Comparison of NLO+PS and \minnlo{} predictions. The latter is based on the CA prescription of the two-loop finite remainder, whose numerical impact is visible in the lowest panel of each plot, which shows the ratio to the \minnlo{} result with $\mathbb{H}^{(2)}=0$. The central panel shows the size of the NNLO corrections relative to the NLO+PS result. The lighter bands display the scale uncertainties, while the darker ones represent the systematic uncertainty assigned to the CA result.}
\end{center}
\end{figure}
%...................................

In this section, we present results for $t\bar t H$ production with a stable Higgs boson and stable top quarks.
We recall that the NNLO+PS predictions considered here are complete, except for the two-loop hard-virtual coefficient, which is approximated using the CA procedure described in \sct{sec:pointwise_combination}.

Before presenting differential results, we comment on the predictions for the total cross section reported in \tab{tab:xs_onshell_final}.
The second column reports the exact NLO+PS result, while the last two columns display the \minnlo{} predictions with the two-loop hard-virtual contribution either set to zero (third column) or estimated using the CA prescription (fourth column).
Uncertainties quoted as percentages originate from scale variation, whereas those shown in square brackets represent the systematic uncertainty associated with the two-loop approximation.
The latter is estimated through a nine-point variation of the $\eta$ and $\tau$ parameters combined with the one-loop discrepancy, as discussed in \sct{sec:pointwise_combination}.
We observe that the complete \minnlo{} prediction in the fourth column is $\mathcal{O}(15\%)$ higher than the NLO+PS result in the second column, with scale uncertainties reduced by roughly a factor of two.
The numerical impact of the two-loop hard-virtual contribution on the NNLO+PS cross section is below $1\%$, and the associated CA systematic uncertainty remains much smaller than the scale uncertainty.

In \fig{fig:final_results_on-shell}, we compare \minnlo{} and NLO+PS predictions for six observables: the rapidity $\yH$ and transverse momentum $\ptH$ of the Higgs boson, and the transverse momentum $\ptTT$ of the top-quark pair (first row); the azimuthal-angle separation $\Delta\phi_{t,\bar t}$ between the top and anti-top quarks, and the invariant masses $m_{t\bar t}$ and $m_{t\bar t H}$ of the top-quark pair and of the Born-level system, respectively (second row).
For each observable, three panels are shown.
In the upper panel, the \minnlo{} predictions (solid blue curve) are compared to the NLO+PS results (solid brown curve).
The impact of NNLO corrections is illustrated in the central panel, which shows the ratio to the NLO+PS prediction.
The lower panel shows the ratio between the complete \minnlo{} prediction, based on the CA estimate of the two-loop amplitude, and the result obtained by setting the two-loop hard-virtual coefficient $\mathbb{H}^{(2)}$ to zero.
The darker uncertainty band represents the CA systematic uncertainty of the two-loop amplitude, while the lighter band corresponds to the scale uncertainty.

First, we highlight features common to all distributions shown in \fig{fig:final_results_on-shell}.
By and large, a clear reduction of the perturbative uncertainty is observed when moving from NLO+PS to \minnlo{}, as evident from the central panel of each plot.
Moreover, the CA systematic uncertainty associated with the two-loop amplitude is well under control, remaining at the level of about $1\text{--}2\%$ and thus well below the perturbative uncertainty across the entire phase space.
Finally, the numerical impact of the two-loop coefficient $\HardVirt^{(2)}_{\CA}$ is generally small, as illustrated in the third panel of each plot, with the exception of the first bins of the $\Delta\phi_{t,\bar t}$ distribution and the low-$m_{t\bar t H}$ region, where effects of $\mathcal{O}(10\%)$ are observed.

We now turn to a more detailed discussion of the behaviour of individual observables.
For the $\yH$, $m_{t\bar t}$, and $m_{t\bar t H}$ distributions, the NNLO corrections are found to be remarkably flat relative to the NLO+PS predictions.
In these cases, the scale uncertainty is reduced from about $10\%$ at NLO+PS to approximately $5\%$ in the \minnlo{} predictions.
This reduction is largely uniform across the full range of these observables and mirrors the behaviour observed for the total inclusive cross section.
Moreover, the NNLO+PS uncertainty bands overlap with those at NLO+PS, although the corresponding central values are generally not contained within the NLO+PS bands, or only marginally so.

A qualitatively different behaviour is observed for the remaining three observables, namely the $\ptH$, $\ptTT$, and $\Delta\phi_{t,\bar t}$ distributions, for which NNLO corrections induce non-trivial shape distortions relative to the NLO+PS predictions.
In particular, for the transverse-momentum spectra, NNLO corrections enhance the cross section by $\mathcal{O}(20\%)$ in the low-$\pt$ region.
This enhancement gradually decreases at higher transverse momenta, reaching $\mathcal{O}(+10\%)$ at $\ptH = 250\,$GeV and only a few percent in the $\ptTT$ tail above 300\,GeV. 
We further observe that, in the low-$\ptTT$ region, the \minnlo{} scale-variation bands do not overlap with those of NLO+PS, clearly underscoring the importance of NNLO effects.
For the azimuthal-angle separation $\Delta\phi_{t,\bar t}$, the trend is reversed: NNLO corrections amount to about $\mathcal{O}(+10\%)$ in the region $\Delta\phi_{t,\bar t} < \pi/2$ and increase monotonically, reaching $\mathcal{O}(+20\%)$ as $\Delta\phi_{t,\bar t} \to \pi$, with only marginal overlap of the uncertainty bands in that region.
For the transverse-momentum distributions, we also observe a visible reduction of the CA systematic uncertainty in the high-$\pt$ tails.
These regions are dominated by the massification procedure within the CA approach and are therefore expected to exhibit very good agreement with the exact two-loop result.

%...............................................................................
\subsection{Simulations with Higgs decay into photons}
\label{sec:pheno_Higgs-decay_results}

In this section, we include the Higgs-boson decay into a photon pair, $H \to \gamma \gamma$.
Although forbidden at tree level in the SM, this decay proceeds via loop diagrams involving fermions and $W$ bosons and has a branching ratio of $\mathrm{BR}(H \to \gamma \gamma) = 0.00227$~\cite{LHCHiggsCrossSectionWorkingGroup:2016ypw}.
The decay is modelled in the narrow-width approximation using \PYTHIA{8}~\cite{Bierlich:2022pfr}.
This is possible thanks to the fully exclusive nature of our simulation: each on-shell event generated by the \minnlo{} implementation is passed to the parton shower, which subsequently performs the decay into a photon pair according to a Breit–Wigner distribution and isotropic angular configurations in the Higgs rest frame.
Since the decay products are colourless, this procedure preserves the QCD accuracy of the calculation.

Experimentally, this decay channel benefits from the excellent invariant-mass resolution of electromagnetic calorimeters, which allows the Higgs signal to be efficiently separated from the continuum background arising from diphoton, photon--jet, and multijet events with misidentified photons.
This leads to precise determinations of the Higgs mass, which is of particular importance as it is tightly connected to other electroweak parameters of the SM  through radiative corrections, such as the weak mixing angle and the masses of the $Z$ and $W$ bosons and the top quark.

We apply the following fiducial cuts
\begin{equation}
p_{\text{\scalefont{0.77}T},\gamma_1}/m_{\gamma\gamma} > 1/3   \; ,
\quad 
p_{\text{\scalefont{0.77}T},\gamma_2}/m_{\gamma \gamma} > 1/4 \;, 
\quad
|\eta_{\gamma_{1,2}}| < 2.5 \; ,  
\end{equation}
where $m_{\gamma\gamma}$ is the invariant mass of the photon pair\footnote{In our case, we consider the on-shell Higgs boson decay. Consequently, the invariant mass of the photon pair equals the Higgs boson mass.}, while $p_{\text{\scalefont{0.77}T},\gamma_i}$ and $\eta_{\gamma_{i}}$ are the transverse momentum and rapidity of the leading ($i=1$) and subleading ($i=2$) photons, ordered in transverse momentum. 

%.................................................
\begin{figure}[h!]
\begin{center}
\begin{tabular}{ccc}
\hspace{-0.53cm}
&
\hspace{-0.45cm}
\includegraphics[width=.35\textwidth]{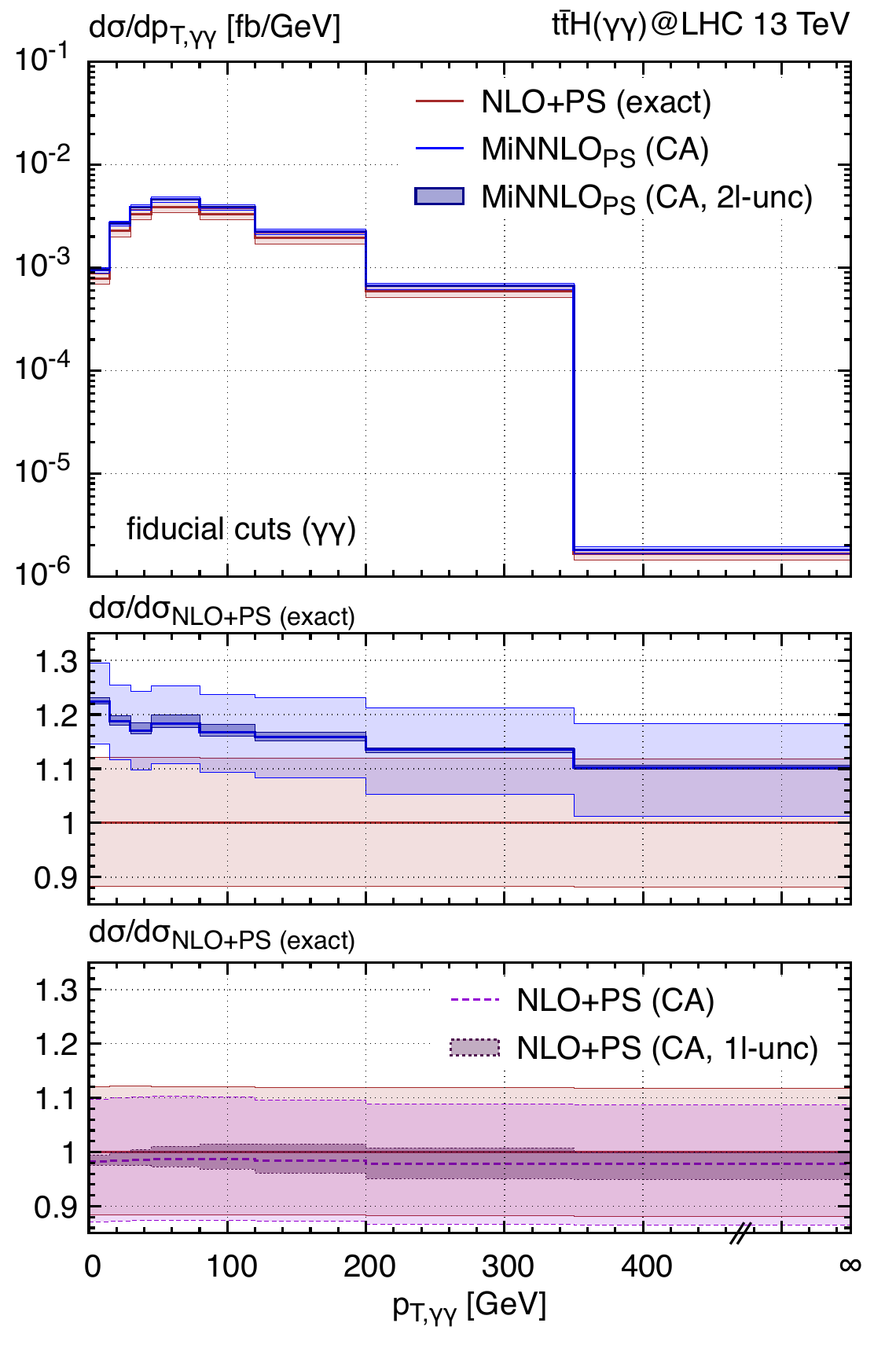}
\hspace{-0.45cm}
\includegraphics[width=.35\textwidth]{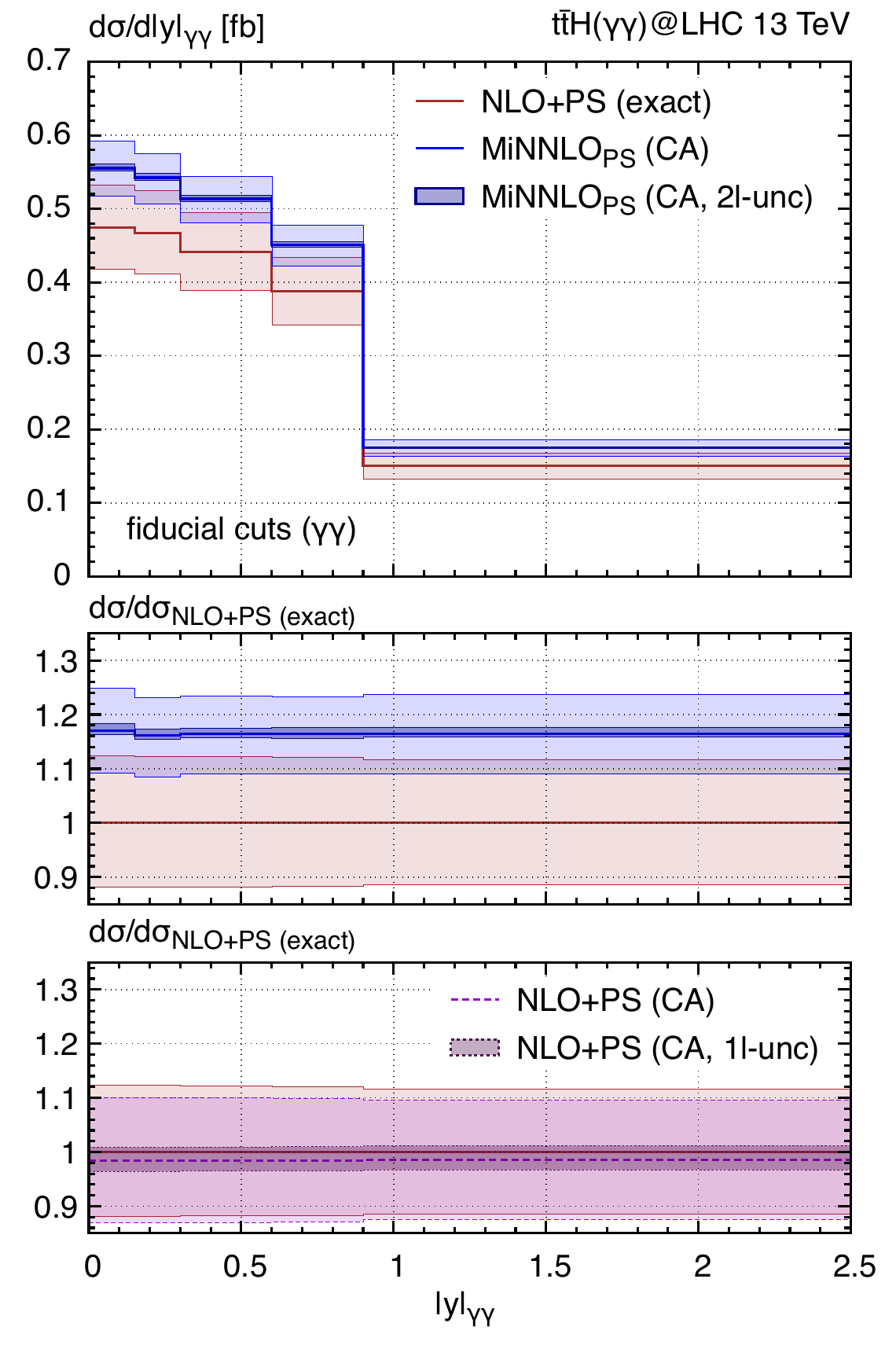}
\hspace{-0.45cm}
\includegraphics[width=.35\textwidth]{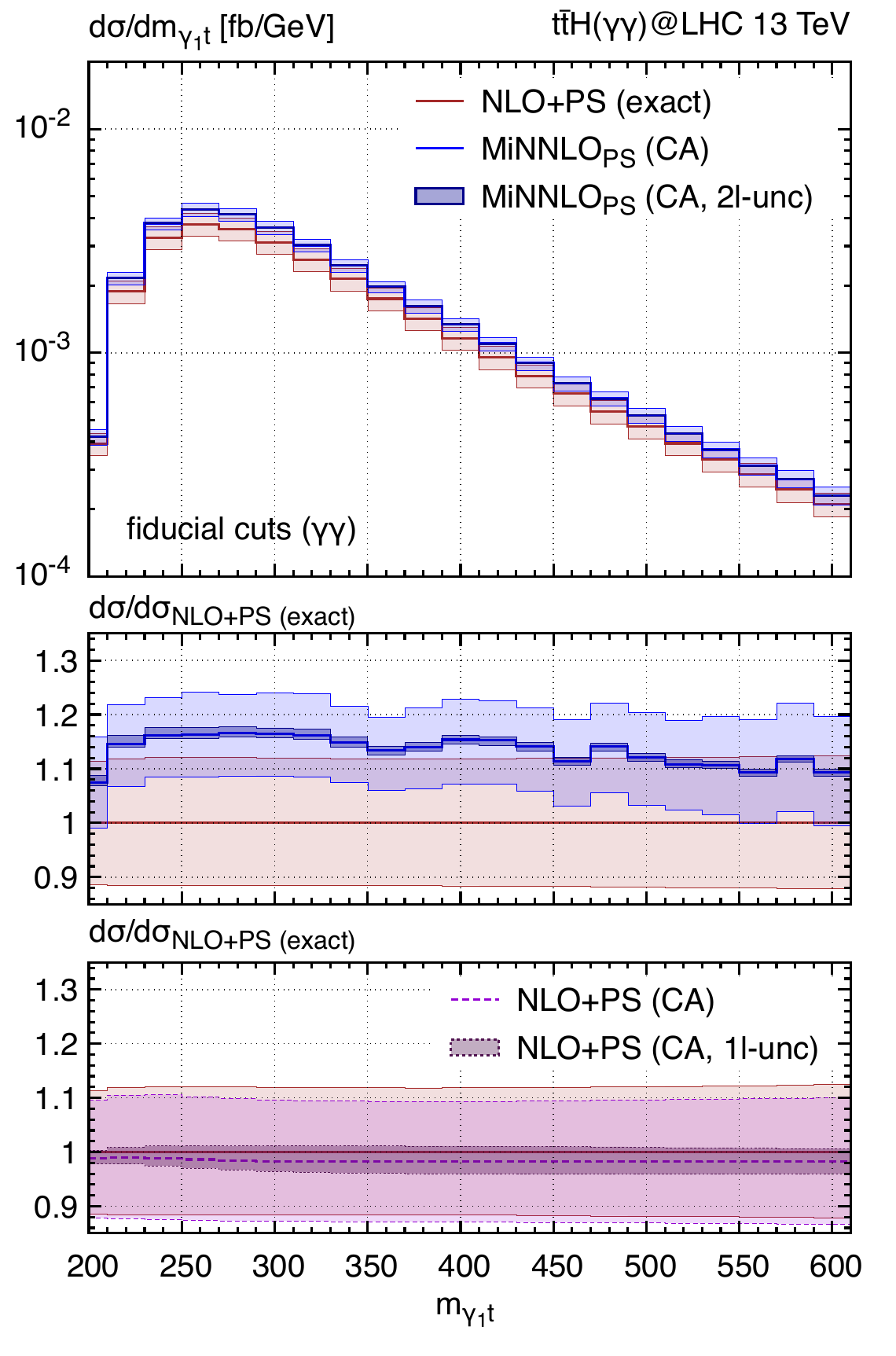}\\
\end{tabular}
\vspace*{1ex}
\caption{\label{fig:final_results_Higgs-decay} 
Comparison of NLO+PS and \minnlo{} predictions for $t \bar t H$ production with $H \to \gamma \gamma$ decay in the fiducial region defined in the main text. The lighter bands display the scale uncertainties, while the darker ones represent the systematic uncertainty assigned to the CA result.}
\end{center}
\end{figure}
%...............................

The left and central plots in \fig{fig:final_results_Higgs-decay} show two observables of the reconstructed Higgs boson, with the binning inspired by \citere{CMS:2025fwn}: the transverse momentum $p_{\text{\scalefont{0.77}T},\gamma\gamma}$ (left) and the absolute value of the rapidity $|y_{\gamma\gamma}|$ (central) of the diphoton system. The rightmost plot displays the invariant mass $m_{\gamma_1 t}$ of the leading photon and top quark. 
The latter serves as a representative example of observables used to discriminate the $t\bar{t}H$ signal from the $t\bar{t}\gamma$ background.
For each distribution, three panels are shown.
In the upper panel, the \minnlo{} cross section (solid blue curve), based on the combined prescription (CA) for the two-loop contribution, is compared to the NLO+PS one (solid brown curve).
The impact of NNLO corrections can be appreciated from the central panel, which displays the ratio to the NLO+PS result. 
In the lower panel, we instead show the NLO+PS predictions obtained using the CA approach (dashed violet curve) and the exact calculation (solid brown curve), normalised to the exact result.
The lighter shaded bands represent the scale uncertainties, while the darker bands indicate the uncertainty assigned to the approximated two-loop (upper and central panels) and one-loop (lower panels) contributions.

We begin by discussing the impact of the NNLO corrections.
For the rapidity distribution, the \minnlo{} prediction shows an almost flat enhancement of about $+18\%$ with respect to the NLO+PS result. 
A different pattern is observed for the $p_{\text{\scalefont{0.77}T},\gamma\gamma}$ ($m_{\gamma_1 t}$) distribution, where the NNLO correction is not uniform: it amounts to roughly $20\%$ ($15\%$) in the bulk of the spectrum and gradually decreases to about $10\%$ for $p_{\text{\scalefont{0.77}T},\gamma\gamma} \gtrsim 350$~GeV ($m_{\gamma_1 t} \gtrsim 500$ GeV). These sizeable, non-flat effects highlight the phenomenological importance of NNLO corrections and the need for a fully differential NNLO+PS description. As expected, we also observe a significant reduction in the perturbative uncertainties at NNLO+PS.

Moreover, the systematic uncertainties of the CA predictions are consistently contained within the scale-uncertainty bands and are roughly an order of magnitude smaller. This further corroborates the robustness of the CA results in fiducial setups, extending the conclusions previously drawn for inclusive observables in \fig{fig:final_results_on-shell}.

Finally, we comment on the quality of the CA approach at one-loop order,  illustrated in the lower panels. 
As in the inclusive setup, the central predictions differ by up to $2\%$, and the two NLO results are fully consistent within their respective perturbative uncertainties. 
Moreover, the exact NLO+PS prediction lies entirely within the systematic uncertainty assigned to the combined one-loop approximation, which remains roughly an order of magnitude smaller than the standard scale-variation band throughout the entire phase-space region.

%...............................................................................
\subsection{Simulations with top-quark decays}
\label{sec:pheno_top-decay_results}

For top-quark pair production, the consistent inclusion of off-shell, non-resonant, and spin-correlation effects in multi-purpose Monte Carlo generators is well established at NLO accuracy. 
In particular, several generators~\cite{Frederix:2012ps, Hoeche:2014qda, Cormier:2018tog, Jezo:2015aia, Jezo:2016ujg, Frederix:2016rdc, Jezo:2023rht} provide NLO corrections included consistently in the fully decayed process.
However, performing such calculation at NNLO QCD, i.e.\ achieving NNLO corrections to the fully decayed process with off-shell, non-resonant and spin-correlation effects, remains an open problem.

For this reason, in the present work, we retain NNLO accuracy in the $\ttH$ production process, while incorporating off-shell effects and spin correlations in the top-quark decays at leading order, restricted to double-resonant topologies. 
Previous studies have shown that this approximation provides a reliable description of spin-sensitive observables when higher-order QCD corrections are included in the production stage.
The decays are implemented using an algorithm developed in the \texttt{POWHEG-hvq} generator \cite{Frixione:2007nw}, which we adapt here for $t\bar tH$ production. Applied to double-resonant configurations, it delivers a consistent leading-order description, including spin correlations, suitable for both exclusive final states and observables inclusive over QCD radiation. A closely related implementation for $t\bar t$ production was presented in \citere{Mazzitelli:2021mmm} and validated against \MADSPIN{}~\cite{Artoisenet:2012st}, confirming the robustness of this approach for phenomenological applications.

Within our framework, \minnlo{} first generates NNLO-accurate weights in the on-shell approximation for the top quarks.\footnote{In practice, the on-shell weights are initially produced with $\mathbb{H}^{(2)}=0$. After the off-shell event construction, the hard two-loop contribution is included via an {\it a posteriori} reweighting.} 
Each on-shell configuration is then projected onto off-shell kinematics according to the procedure detailed in Appendix B of \citere{Mazzitelli:2021mmm}, and its weight is multiplied by a correction factor, namely the branching ratio, to account for the enlarged off-shell phase space.
The final Les Houches events (LHEs) are generated at the last stage via a hit-or-miss procedure based on the ratio of the off-shell and on-shell tree-level matrix elements evaluated at the generated kinematics. 
Due to the underlying \POWHEG{} \ttHJ{} process, \ttH{} events are produced with one or two additional QCD partons, depending on whether the hardest \POWHEG{} radiation is vetoed. Consequently, the off-shell description requires tree-level amplitudes for configurations with up to two extra QCD emissions, which are obtained via \Recola{2}~\cite{Denner:2017vms,Denner:2017wsf}.

We consider two distinct setups: the {\it dilepton channel}, which corresponds to the fully leptonic top-quark decay mode with one electron (positron) and one (anti)muon in the final state, and the {\it semileptonic channel} defined by requiring one top quark to decay leptonically (into either an electron or a muon) and the other hadronically, thus resulting in one charged lepton and two light quarks among the decay products.

Initially, the top quarks and intermediate $W$-bosons are treated in the narrow-width approximation. 
On the contrary, for the evaluation of the tree-level off-shell matrix element, we set $\Gamma_t=1.35$\,GeV, $\Gamma_W=2.0854$\,GeV and $\Gamma_Z=2.4952$\,GeV. These values are controlled from the input card by the flags \texttt{twidth\_rcl}, \texttt{wwidth\_rcl} and \texttt{zwidth\_rcl}, respectively. 
For the EW parameters, we adopt the $G_\mu$ scheme, such that the
relevant coupling and mixing angle are computed via the relations
$\alpha_{G_\mu} = \sqrt{2}/\pi G_F m_W^2 \sin^2\theta_W$ and
$\cos^2\theta_W= m_W^2/m_Z^2$, with $m_W=80.385$\,GeV and $m_Z=91.1876$\,GeV. 
The Higgs boson is treated as stable, and the remaining settings are those described in \sct{sec:pheno_results}.

%%%%%%%%%%%%%%%%%%%%%%%%%%%
\subsubsection{Dilepton setup}
\label{sec:dilept}
In this section we present the results of the simulations in the dilepton channel. 
By setting the input-card options \texttt{topdecaymode 11000} and \texttt{semileptonic 0}, we select events with one electron and one muon originating from the decay of the two intermediate $W$ bosons. The corresponding event weights are rescaled by the branching ratio ${\rm BR}(t \to b \ell \nu_\ell)=0.108$ for each top decay.

Jets are reconstructed using the anti-$k_T$ algorithm~\cite{Cacciari:2008gp} with a radius $R=0.4$, requiring transverse momentum $p_{\text{\scalefont{0.77}T},\text{jet}} > 25~\mathrm{GeV}$ and pseudorapidity $|\eta_{\text{jet}}| < 2.5$. A jet is flavour-tagged as $b$-jet if at least one bottom quark is found among its constituents, and only events with at least two identified $b$-jets are retained.
Events are required to contain exactly one electron and one muon, each satisfying $p_{\text{\scalefont{0.77}T},\ell} > 10~\mathrm{GeV}$ and $|\eta_\ell| < 2.5$. Lepton isolation is imposed through a standard cone-based criterion: for each lepton $\ell \in \{e,\mu\}$, we consider all final-state particles $i$, excluding leptons, within a cone of radius
\begin{equation}
\Delta R(i,\ell) = \sqrt{(\Delta \eta)^2 + (\Delta \phi)^2} < 0.4 \, ,
\end{equation}
and we require
\begin{equation}
\sum_{i \neq \ell,\, \Delta R(i,\ell)<0.4} \!\!\!\!p_{\text{\scalefont{0.77}T},i} 
< 0.35 \, p_{\text{\scalefont{0.77}T},\ell} \, .
\end{equation}
If this condition is not fulfilled, the lepton is classified as \textit{non-isolated}, and the event is rejected. 
The isolation criterion is applied independently to both leptons.
We observe that this requirement leads to a reduction of the fiducial cross section after parton showering compared to the LHE level. Additional radiation from the shower increases hadronic activity around the leptons, particularly in boosted configurations where the leptons are emitted close to the direction of the parent top quark, thereby reducing the probability that the isolation condition is satisfied.
Finally, we impose a cut $p_{\text{\scalefont{0.77}T},\ell_1} > 15~\mathrm{GeV}$ on the leading lepton, defined as the lepton with the highest transverse momentum in the event.

\begin{figure}[h!]
\begin{center}
\begin{tabular}{cc}
\hspace{-0.52cm}
\hspace{-0.73cm}
\includegraphics[width=.4\textwidth]{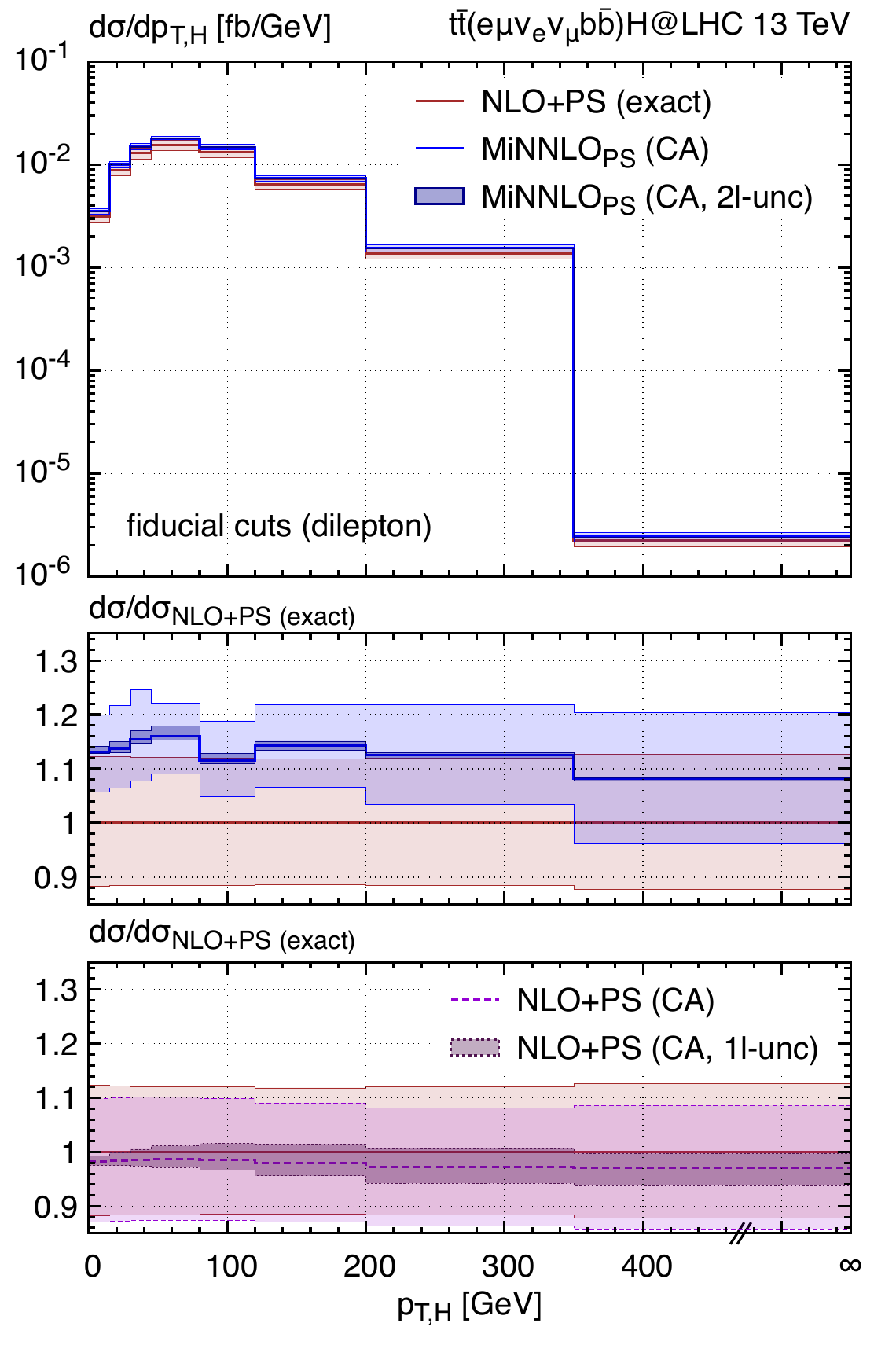}
\includegraphics[width=.4\textwidth]{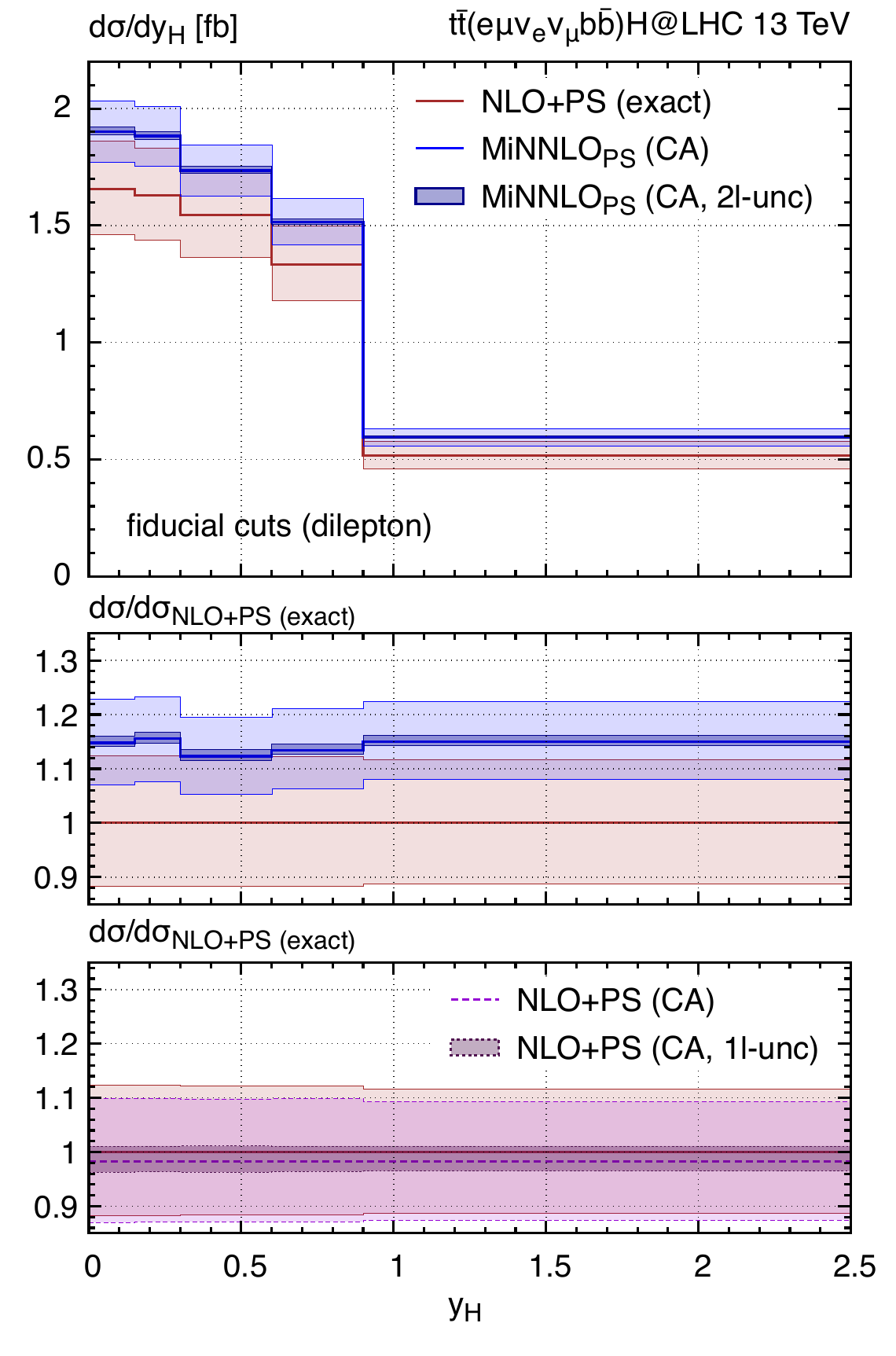}
\end{tabular}
\vspace*{1ex}
\caption{Comparison of NLO+PS and \minnlo{} predictions for Higgs-related observables in $t \bar t H$ production with leptonic decays of the top quarks in the fiducial region defined in the main text. The lighter bands display the scale uncertainties, while the darker ones represent the systematic uncertainty assigned to the CA result.}
\label{fig:decay_fully_lep_Higgs}
\end{center}
\end{figure}

In \fig{fig:decay_fully_lep_Higgs}, we study the transverse-momentum $\ptH$ (left column) and rapidity $\yH$ (right column) distributions of the Higgs boson in the fiducial volume. 
The style of the plots is analogous to that of \fig{fig:final_results_Higgs-decay}.
The rapidity distribution receives NNLO corrections that are nearly flat across the full kinematic range, corresponding to an increase of approximately $15\%$ relative to the NLO+PS prediction (central panel). For this observable, the effect of the combined approximation at one-loop order is negligible and essentially flat, resulting in deviations of roughly $-2\%$ relative to the exact NLO+PS result. Moreover, the uncertainty band of the CA prediction fully covers the exact result (lower panel).
By and large, also the NNLO corrections in the $\ptH$ distribution are flat, with a slight decrease in the high-$\ptH$ tail. Their impact reaches approximately $+16\%$ around the peak and gradually decreases to about $+10\%$ at larger values of $\ptH$. The central \minnlo{} prediction is only marginally covered by the NLO+PS scale uncertainties, but the NNLO+PS and NLO+PS uncertainty bands fully overlap.
The combined approximation of the one-loop amplitude induces a small distortion of the $\ptH$ shape relative to the exact prediction, with differences reaching about $-3\%$ in the high-$\ptH$ region. These deviations remain well within the estimated perturbative uncertainties across the full kinematic range and therefore do not affect the overall theoretical accuracy of the prediction (lower panel). 
For both Higgs-boson observables, the systematic uncertainty (darker blue band) assigned to the approximated two-loop contribution at NNLO+PS is well under control and significantly smaller than the scale uncertainties (lighter shaded blue band).

%.................................
%.................................
\begin{figure}[h!]
\begin{center}
\begin{tabular}{ccc}
\hspace{-0.53cm}
&
\hspace{-0.45cm}
\includegraphics[width=.35\textwidth]{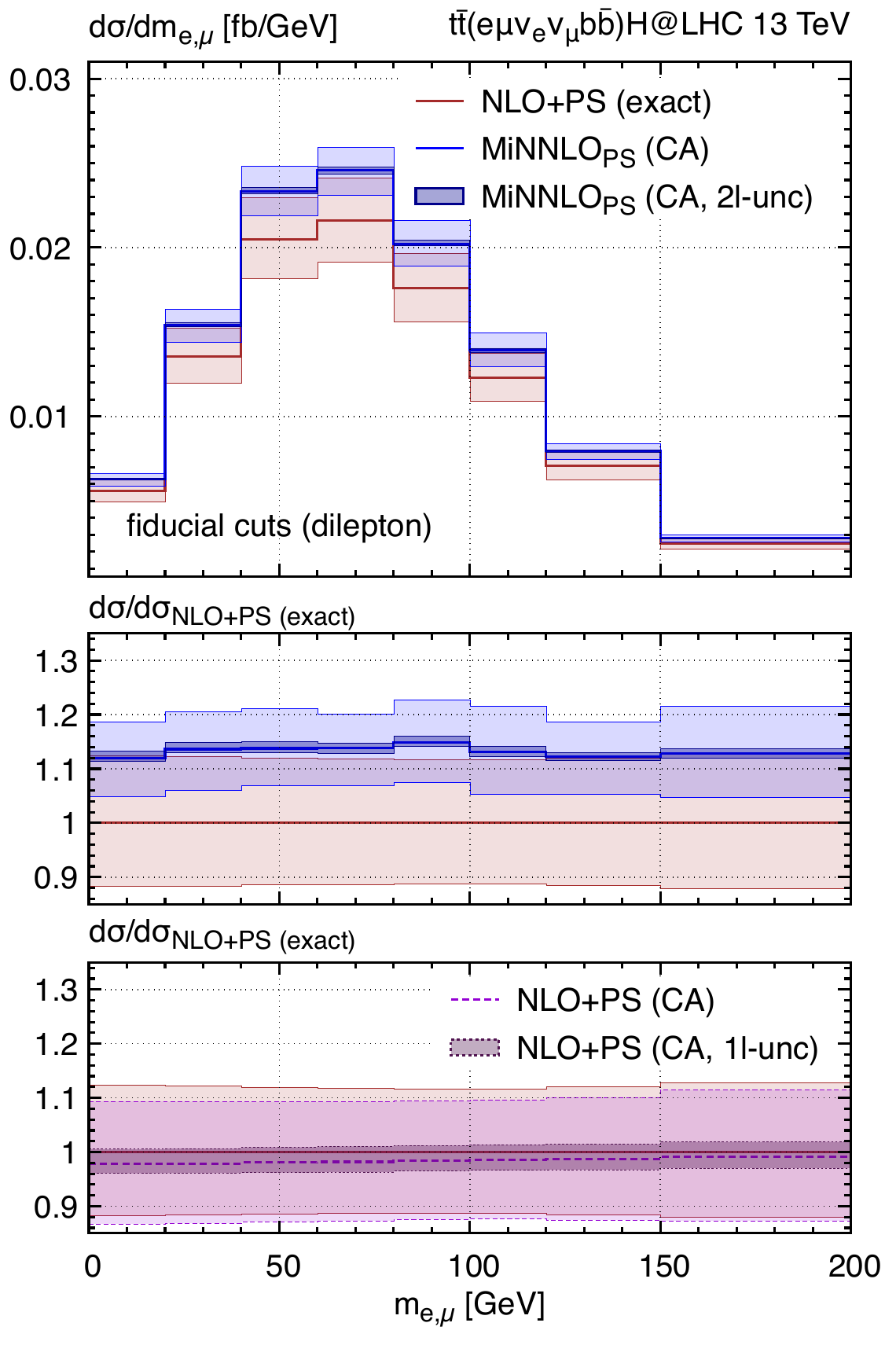}
\hspace{-0.45cm}
\includegraphics[width=.35\textwidth]{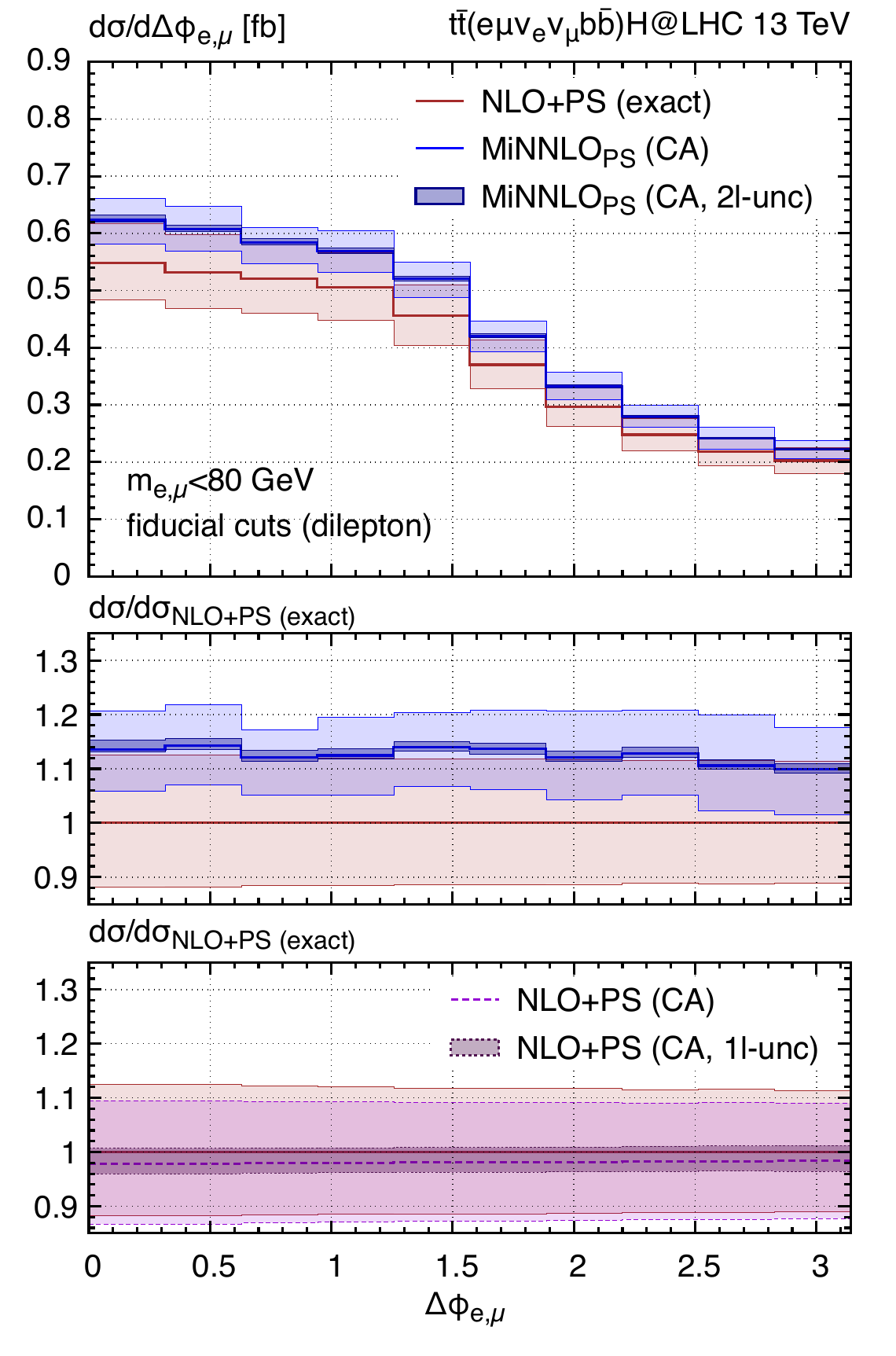}
\hspace{-0.45cm}
\includegraphics[width=.35\textwidth]{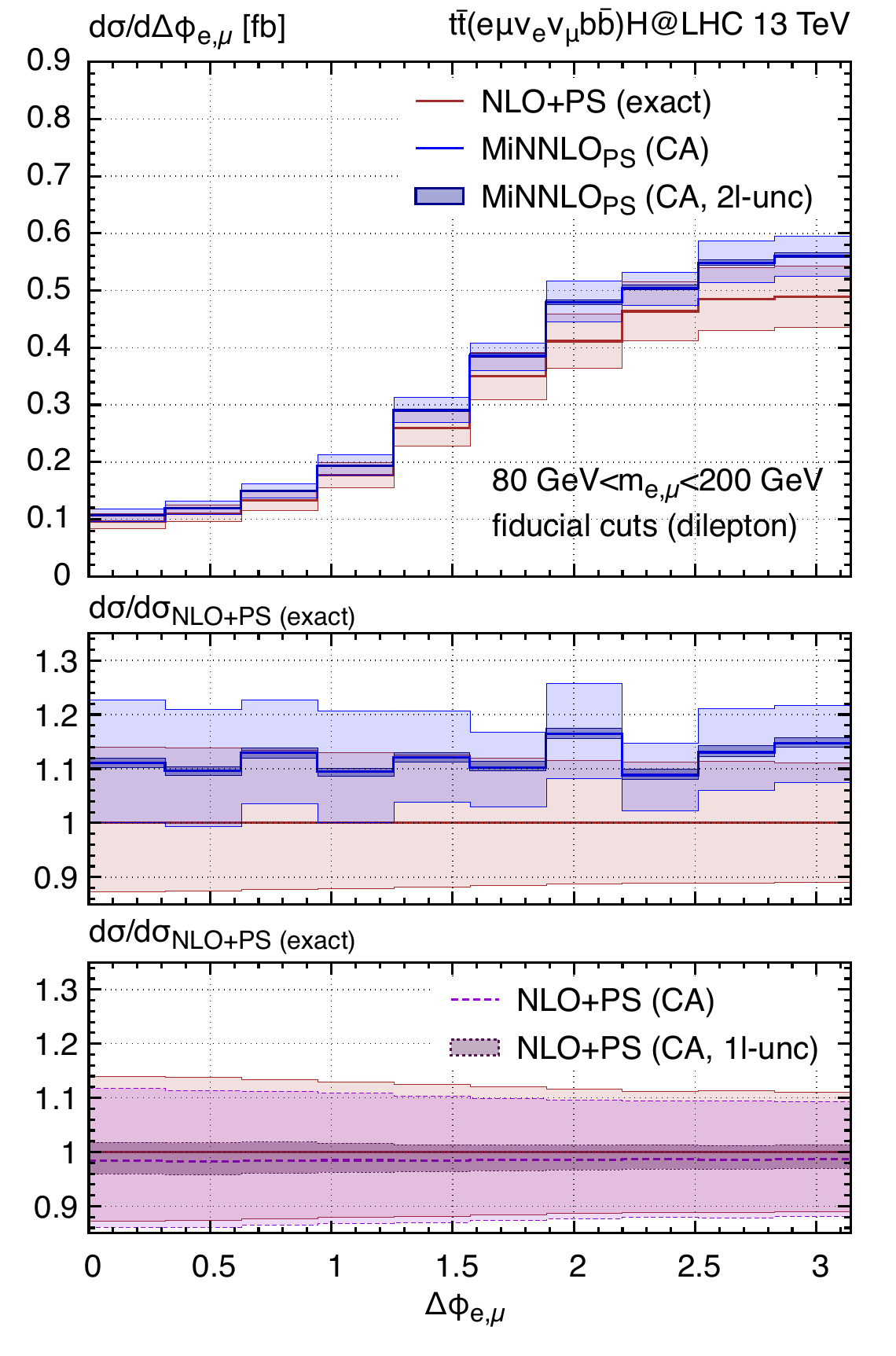}
\end{tabular}
\vspace*{1ex}
\caption{Same as \fig{fig:decay_fully_lep_Higgs}, but for observables sensitive to spin-correlation effects.}
\label{fig:decay_fully_lep_spincorr}
\end{center}
\end{figure}

We now discuss the spin-correlation effects captured by our simulations of the top-quark decays.
Since top quarks decay before hadronising, their spin information is largely preserved and remains unaffected by non-perturbative QCD effects that could otherwise depolarise them prior to decay.
As a result, the angular and kinematic distributions of their decay products retain a strong sensitivity to the underlying spin configuration of the top-antitop system.
In dilepton events, observables built from the two charged leptons, such as the azimuthal angle separation $\Delta\phi_{e\mu}$ and the dilepton invariant mass $m_{e\mu}$, shown in \fig{fig:decay_fully_lep_spincorr}, provide clean probes of these effects. 
The invariant mass $m_{e\mu}$ effectively acts as a spin filter: events below the peak ($m_{e\mu}\lesssim 80$\,GeV) tend to populate the region of small $\Delta\phi_{e\mu}$, see central plot of \fig{fig:decay_fully_lep_spincorr}, whereas events in the tail of the $m_{e\mu}$ distribution (e.g.\ in the range $80$-$200$\,GeV, or higher) tend more towards back-to-back configurations of the leptons, see rightmost plot in \fig{fig:decay_fully_lep_spincorr}.
This behaviour can be traced back to the electroweak nature of the top-quark decay, which makes the charged lepton a very efficient spin analyser, being preferentially emitted along the direction of the parent top-quark spin. 
Consequently, the low-$m_{e\mu}$ region, where the two leptons are typically close to each other in phase space, is predominantly populated by configurations in which the reconstructed top and anti-top spins are aligned, while larger dilepton invariant masses are more often associated with opposite spin configurations.
In \ttH{} production, the recoil induced by the Higgs boson partly smears these correlations, weakening the direct relation between the top-quark spin direction and the charged-lepton direction compared to the simpler $t\bar t$ case, where the top quarks are emitted back-to-back at the Born level. Despite this smearing, spin correlations play a crucial role in \ttH{} simulations, as neglecting tree-level spin information would lead to significantly flatter angular distributions.

The relevance of spin-correlation effects motivates their inclusion in our NNLO+PS simulations at tree level.
As shown in \fig{fig:decay_fully_lep_spincorr}, \minnlo{} predictions increase the NLO+PS results by $\mathcal{O}(10\text{--}15\%)$ for all three distributions. These corrections are largely uniform across the kinematic range considered.
Also the impact of the combined approximation of one-loop amplitude with respect to the exact NLO+PS result (lower panels) is quite flat, with deviations below $2\%$, which are fully covered by the assigned uncertainty.
As for Higgs-boson observables in \fig{fig:decay_fully_lep_Higgs}, the systematic uncertainty associated with the approximated two-loop contribution is well under control.

%%%%%%%%%%%%%%%%%%%%%%%%%%%
\subsubsection{Semileptonic setup}
\label{sec:semilept}
In this section we present simulations in the semileptonic channel. 
By setting the input-card options \texttt{topdecaymode 22022} and \texttt{semileptonic 1}, we select events that feature one charged lepton $\ell \in \{e,\mu\}$ and a light quark $q \in \{u,d,c,s\}$ in the final state.
The corresponding event weights are rescaled by the leptonic branching ratio, specified in \sct{sec:dilept}, and by the hadronic branching ratio BR$(t \rightarrow b q\bar{q}')=0.676$.
%.................................
\begin{figure}[h!]
\begin{center}
\begin{tabular}{ccc}
\hspace{-0.53cm}
&
\hspace{-0.45cm}
\includegraphics[width=.35\textwidth]{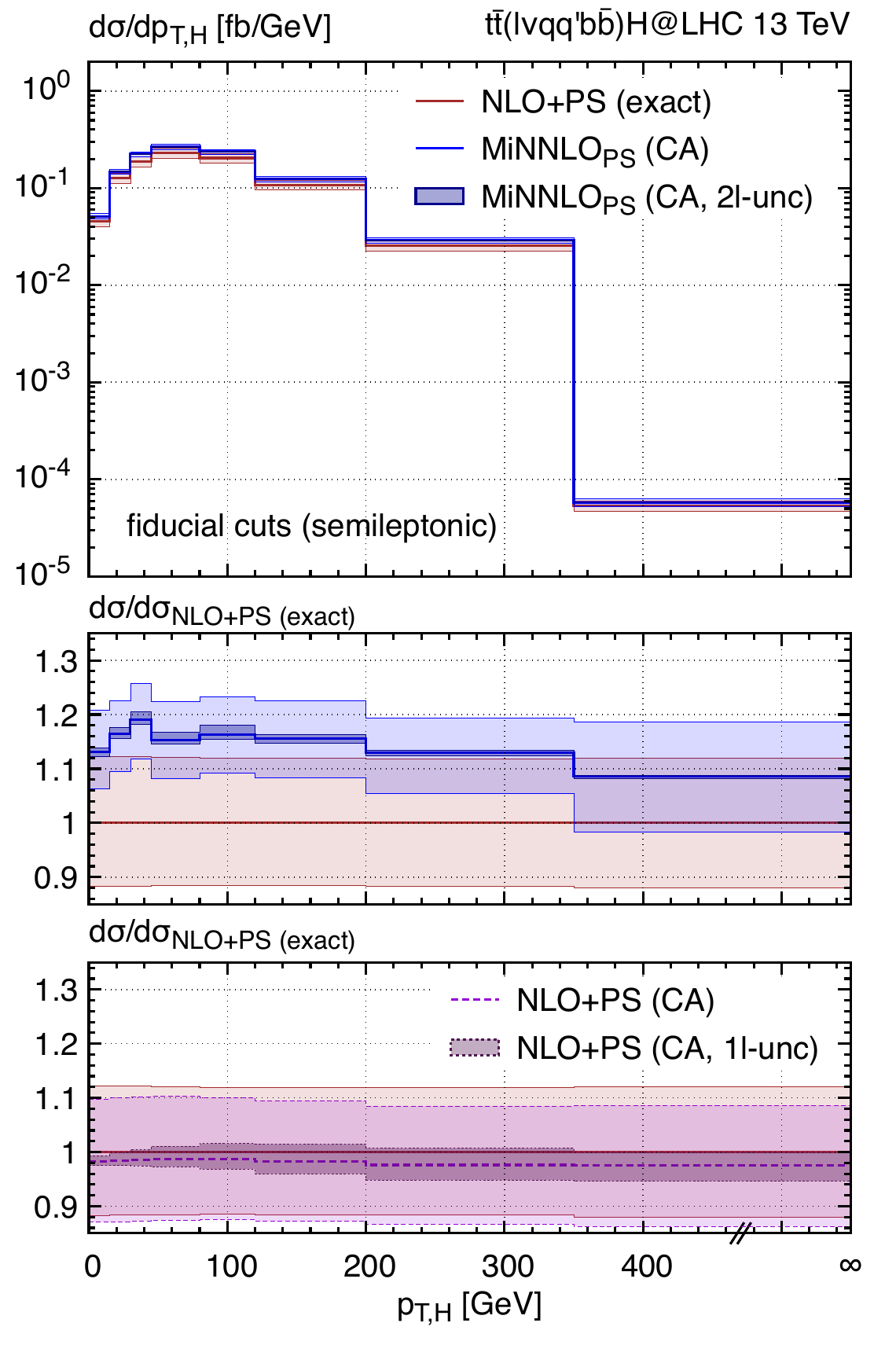}
\hspace{-0.45cm}
\includegraphics[width=.35\textwidth]{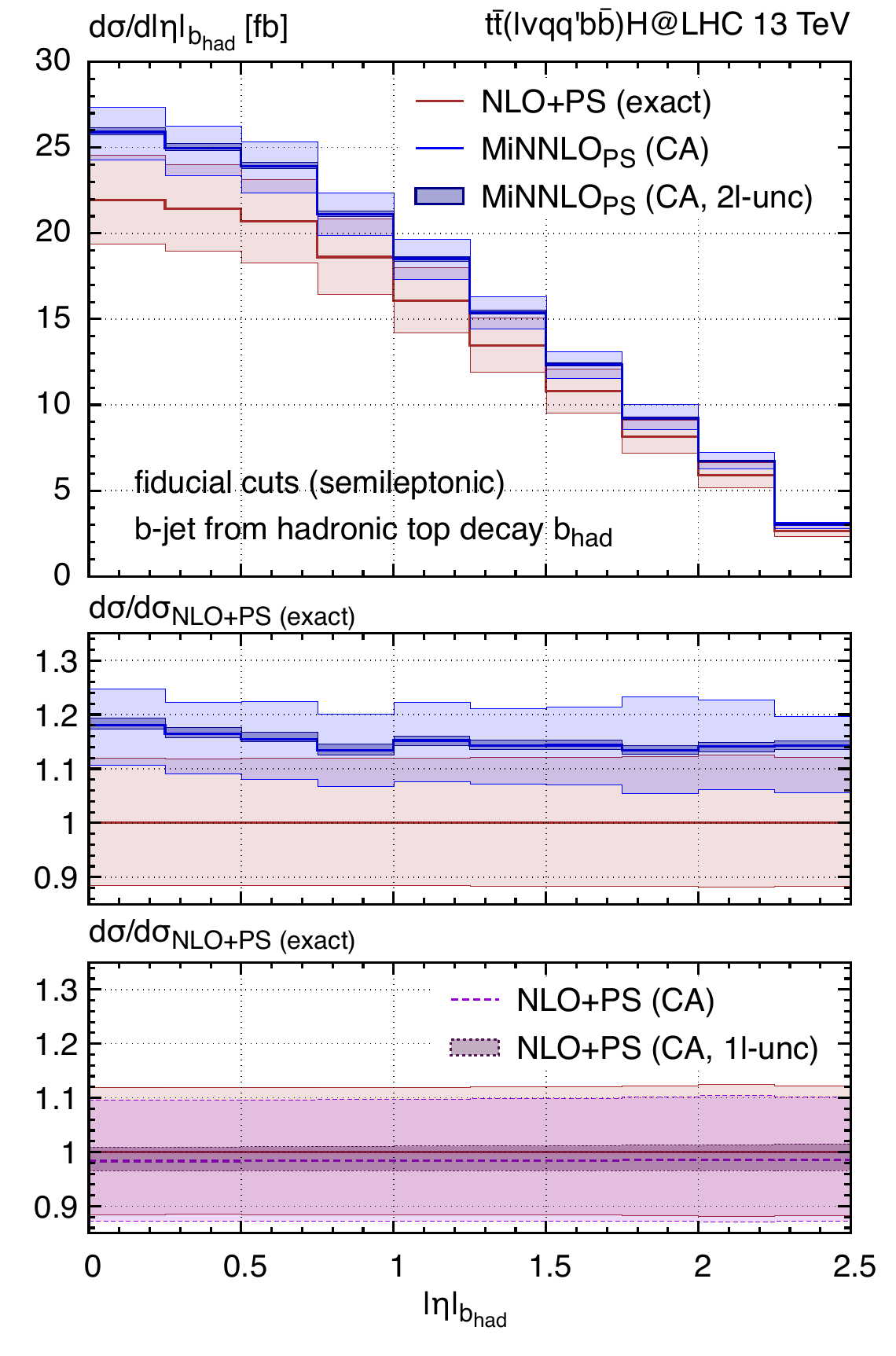}
\hspace{-0.45cm}
\includegraphics[width=.35\textwidth]{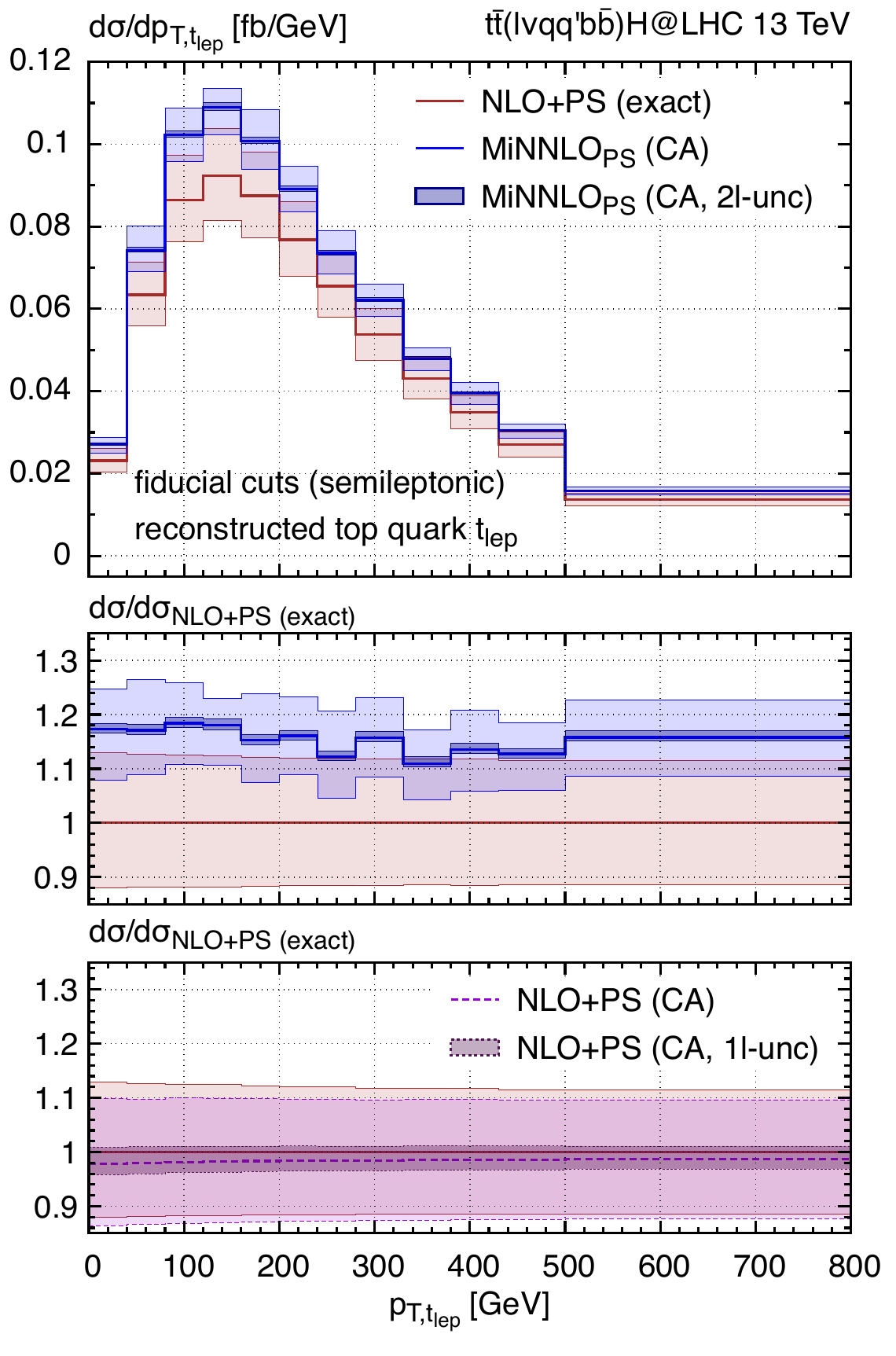}
\end{tabular}
\vspace*{1ex}
\caption{Same as \fig{fig:decay_fully_lep_Higgs}, but for the Higgs transverse momentum, the rapidity of the reconstructed $b$-jet, and the transverse momentum of the reconstructed top quark in the semileptonic top-decay channel.}
\label{fig:decay_semi_lep}
\end{center}
\end{figure}
%%%%%%%%%%%%%%%%%%%%%%%%%%%%%%%%%%%%%%%%%%%%

For the definition of the fiducial region we follow the analysis of \citere{CMS:2018htd}. 
We accept events that feature one charged lepton with $p_{\text{\scalefont{0.77}T},\ell} > 30~\mathrm{GeV}$ and $|\eta_\ell| < 2.4$. 
We reconstruct jets via the anti-$k_T$ algorithm~\cite{Cacciari:2008gp} with $R=0.4$, requiring transverse momentum $p_{\text{\scalefont{0.77}T},\text{jet}} > 25~\mathrm{GeV}$ and pseudorapidity $|\eta_{\text{jet}}| < 2.5$, as in the dilepton case.
Jets are tagged as $b$-jets when they contain at least one bottom quark. We then select events with at least two $b$-jets and two light jets. 
The reconstruction of the top-quark momenta is not unique and we follow the procedure used in top-quark analyses~\cite{CMS:2016oae,CMS:2018htd}. 
More precisely, for each set of momenta related to the missing energy component ($p_{\rm miss}$), the charged lepton ($p_\ell$), two $b$-jets ($p_{b_1},p_{b_2}$) and two light-jets ($p_{j_1},p_{j_2}$), we compute the following quantity, known as \textit{distance-to-the-resonances}:\footnote{Notice that we reconstruct $p_{\rm miss}$ directly from the neutrino momentum present in our generated events, while in the experimental analysis of \citere{CMS:2018htd} it is reconstructed by imposing on-shell conditions on the top quark and the $W$ boson.}
\begin{align}
	\delta_r=((p_{\rm miss}+p_{\rm \ell}+p_{b_1})^2-m_t^2)^2+((p_{j_1}+p_{j_2})^2-m_W^2)^2+((p_{j_1}+p_{j_2}+p_{b_2})^2-m_t^2)^2\,.
\end{align}
The set of momenta that minimises $\delta_r$ is selected, and the corresponding assignment of jets and leptons is kept to define the momenta of the hadronically and leptonically decaying top quarks.

As representative distributions, \fig{fig:decay_semi_lep} shows the transverse momentum of the Higgs boson ($\ptH$), the pseudorapidity of the $b$-jet from the hadronic top decay ($\eta_{\rm b_{\rm had}}$), and the transverse momentum of the top quark associated with the leptonically decaying $W$ boson ($p_{\text{\scalefont{0.77}T,$t_{\rm lep}$}}$), defined by the reconstruction algorithm. The overall behaviour is similar to that observed in the dilepton channel in \sct{sec:dilept}.
NNLO corrections modify the shapes of the distributions only marginally, being $\mathcal{O}(+15\%)$ around the peak and $\mathcal{O}(+10\%)$ for transverse momenta above 350\,GeV. The qualitative behaviour is analogous to that in the left plot of \fig{fig:decay_fully_lep_Higgs}, with the only difference being that the cross section is larger by about a factor of six due to the hadronic branching ratio compared to the leptonic one.
Also for the other two observables, related to the reconstructed $b$-jets and top quarks, the NNLO corrections are relatively uniform with an effect of about $+20\%$. 
As in the dilepton channel, the systematic uncertainty (darker blue band) associated with the approximated two-loop contribution is well under control for all considered observables and much smaller than the perturbative uncertainties (lighter blue band) across the whole kinematic range.
Moreover, the CA result at NLO+PS deviates by less than 2\% from the exact one, which is fully covered by the corresponding systematic uncertainty, thereby confirming the robustness of our combined approximation in realistic experimental setups.

%============================================
\section{Conclusions}
\label{sec:conclusions}

We have presented the first fully exclusive NNLO+PS simulation for hadronic $t\bar t H$ production. 
Our calculation consistently combines NNLO QCD corrections with parton-shower effects within the \minnlo{} framework and is exact up to the two-loop amplitude, for which a novel approximation is employed.

A central aspect of the present study is the development of an approximation suitable for fully differential Monte Carlo event generators, since the exact two-loop amplitude is currently beyond reach. 
We propose a pointwise combination of two complementary approximations previously used in fixed-order calculations~\cite{Devoto:2024nhl}, each valid in a distinct kinematic regime: the soft Higgs-boson limit and the high-energy limit where finite top-quark mass effects can be neglected. When combined appropriately, they capture the dominant contributions across the full phase space, as verified at one-loop order.

The proposed combined approximation (CA) is constructed as a weighted average of the two aforementioned approximations and applied on an event-by-event basis. 
To our knowledge, this represents the first implementation of such a pointwise approximation at the fully differential level for a process like $t\bar t H$ production.
We validated the CA by applying it to the one-loop amplitude at NLO+PS and comparing the resulting predictions with the exact NLO+PS calculation across a broad set of observables. The CA accurately reproduces the shapes of all considered distributions, with the ratio to the exact result remarkably flat and deviations typically at the few-percent level.

Quantifying the uncertainty associated with the CA approach is as important as constructing the approximation itself. 
We therefore introduced a dedicated systematic uncertainty that accounts for both the IR-scale dependence of the approximated finite remainder and variations of the weight function $\omega$. 
At NLO, deviations of the CA prediction from the exact result are largely covered by this uncertainty estimate, which is nearly an order of magnitude smaller than the perturbative uncertainty from seven-point scale variation. 
This demonstrates the robustness of our approach. 
For the two-loop amplitude entering the NNLO+PS prediction, we identified an additional source of uncertainty by evaluating, point by point in phase space, the difference between the one-loop CA and exact finite remainders. 
This procedure ensures that, for any observable, limitations of the one-loop approximation are systematically propagated to the two-loop level through a conservative uncertainty. The resulting uncertainty remains at the sub-percent level and is therefore well below the perturbative uncertainty from NNLO scale variation.

As a first and essential step, we validated our \minnlo{} implementation against NNLO-accurate fixed-order predictions~\cite{Devoto:2024nhl} obtained within the \Matrix{} framework~\cite{Grazzini:2017mhc}, setting the two-loop contribution to zero in both calculations.
The excellent agreement observed at the fully differential level provides a crucial and non-trivial validation of the \minnlo{} method for the associated production of heavy quarks with a colour singlet.
It confirms that the framework reliably achieves NNLO accuracy and can successfully handle complex multi-scale processes such as $\ttH$ production.
Moreover, it demonstrates that any additional higher-order contributions inherent to the Monte Carlo simulation are formally beyond NNLO accuracy and numerically subleading.

Having established the formal accuracy of the method, we included the two-loop finite remainder using the CA prescription and consistently incorporated resummation and parton-shower effects.
The resulting NNLO+PS predictions exhibit corrections of $\mathcal{O}(15\%)$ relative to standard NLO+PS results.
These corrections are largely uniform across kinematic distributions, although they can become more pronounced in specific phase-space regions.
The perturbative scale uncertainty is significantly reduced at NNLO, yet it remains an order of magnitude larger than the systematic uncertainty associated with the two-loop approximation.
This hierarchy of uncertainties indicates that the CA approach does not introduce uncontrolled effects beyond the formal perturbative accuracy of the calculation.

Beyond differential results for stable top quarks and a stable Higgs boson, we also considered Higgs-boson and top-quark decays.
In particular, we presented predictions for $H \to \gamma\gamma$ measurements and for top-quark decays in the dilepton and semileptonic channels.
By incorporating tree-level spin correlations, we obtained reliable predictions for spin-sensitive observables constructed from the top-quark decay products.
For all distributions studied, NNLO corrections have a significant impact and reduce perturbative uncertainties.
Moreover, by comparing exact and approximate NLO+PS predictions in the same fiducial regions, we demonstrated that the CA remains reliable in realistic experimental setups.

The present work establishes a new state-of-the-art benchmark for fully differential $\ttH$ predictions, and we expect our publicly available Monte Carlo code to provide valuable theoretical input for future measurements involving $\ttH$ final states, both as signal and as background.
The calculation of the exact two-loop amplitude for $\ttH$ production remains an open challenge towards complete NNLO QCD accuracy without relying on approximations.
Our strategy provides a quantitatively controlled determination of the missing contribution and enables fully exclusive NNLO+PS predictions that already satisfy current experimental precision requirements.
Nevertheless, the exact two-loop calculation will ultimately be necessary to eliminate the residual systematic uncertainty associated with the approximate NNLO predictions.
Our \minnlo{} implementation is designed to incorporate the exact two-loop contribution in a straightforward manner as soon as it becomes available.
In particular, the fully exclusive nature of our event generation allows even numerically demanding two-loop results to be included via reweighting techniques, while reusing already generated event samples.

The computational framework and validation strategy developed in this work are not specific to $\ttH$ production and can be extended to other $t\bar{t}F$ processes with a colour-singlet final state $F$, such as $t\bar{t}Z$ production.
In these cases, where the two-loop amplitude remains the primary bottleneck, a similar combination of controlled approximations and systematic one-loop validations provides a viable path towards NNLO+PS predictions.
\\

%%%%%%%%%%%%%%%%%%%%%%%%%%%%%%%%%%%%%%%%%%%%%%%%%%%%%%%%%%%%%
\noindent {\bf Public release of the code.} \\
Together with this manuscript, we release the \mbox{{\tt POWHEG-BOX-RES/ttHJ}} code, referred to as the \texttt{ttH-MiNNLO} event generator, making it publicly available for experimental analyses and phenomenological studies.
The code can be downloaded by following the instructions on the \POWHEGBOX{} website: \url{http://powhegbox.mib.infn.it}.
To obtain the implementation, navigate to the \texttt{POWHEG-BOX-RES} git directory and execute
\lstset{basicstyle=\small, frame=none}
{\tt
\begin{lstlisting}[language=bash]
  $ git submodule update --remote --init ttHJ
\end{lstlisting}
}%$
\vspace{-0.3cm}
The \texttt{ttH-MiNNLO} generator provides NLO+PS predictions for $t\bar t HJ$ observables and NNLO+PS simulations for $t\bar t H$ production based on the combined two-loop approximation described in this work. 
All required dependencies are documented in the manual located at \\\mbox{\texttt{ttHJ/Docs/manual\_ttH-MiNNLO.pdf}}.
Section\,3 explains how to download and compile the dependencies and build the executable for event generation.
Section\,4 describes the event-generation workflow and the inclusion of the two-loop contribution via the \POWHEG{} reweighting facilities.
Section\,5 details the most relevant input flags and available features, including the treatment of off-shell top-quark decays in the production of LHEs.
Section\,6 presents the native analysis modules, which can be used directly within \POWHEG{} using the LHE files or in conjunction with the \PYTHIA{8} parton shower.

%%%%%%%%%%%%%%%%%%%%%%%%%%%%%%%%%%%%%%%%%%%%%%%%%%%%%%%%%%%%%

\noindent {\bf Acknowledgements.}
We would like to thank Massimiliano Grazzini and Giulia Zanderighi for comments on the manuscript.
C.SS.\ is indebted to Vitaly Magerya for valuable discussions. 
C.B.\ thanks CERN TH Department for hospitality while this research was partially carried out. 
C.B.\ acknowledges funding from the Swiss National Science Foundation, grant 10001706.
The work of C.S.\ is partly supported by the Excellence Cluster ORIGINS, funded by the
Deutsche Forschungsgemeinschaft (DFG, German Research Foundation) under
Germany's Excellence Strategy --- EXC-2094-390783311.
We have used the Max Planck Computing and Data Facility (MPCDF) in
Garching to carry out all simulations presented here.

\bibliography{MiNNLO}

\end{document}